\pdfoutput=1
\documentclass[
  aps,
  prx,
  twocolumn,
  superscriptaddress,
  nofootinbib,
  showkeys,
  longbibliography
]{revtex4-2}

\usepackage{graphicx}
\usepackage{hyperref}
\usepackage{amsmath,amssymb,amsfonts} % <-- add this line
\usepackage{mathtools}
\usepackage{float}
\usepackage{booktabs}
\usepackage{xcolor}\usepackage{color,graphicx,hyperref,amsmath}
\usepackage{mathtools}
\usepackage{graphicx}
\usepackage{float}
\usepackage{booktabs}



\begin{document}
\title{Nonlinear Model Reduction of Complex Networks via Spectral Submanifolds}

\author{Kaviya Bhaskaran}
\affiliation{Department of Mechanics and Aerospace Engineering, Southern University of Science and Technology, Shenzhen 518055, China}

\author{Shobhit Jain}
\affiliation{Delft Institute of Applied Mathematics, TU Delft, Mekelweg 4, 2628 CD Delft, The Netherlands}

\author{Mingwu Li}
\affiliation{Department of Mechanics and Aerospace Engineering, Southern University of Science and Technology, Shenzhen 518055, China}

%\date{\today}
\date{\today}
\begin{abstract}
Complex networked systems are prevalent in biology, engineering, and the social sciences, yet their high-dimensional, nonlinear dynamics pose major challenges for analysis and prediction. A mathematically rigorous route to simplification is to represent system behavior on a low-dimensional, smooth invariant manifold known as a spectral submanifold (SSM). Here we present a comprehensive SSM reduction framework and its globalized extension (gSSM) for dimensionality reduction in large-scale nonlinear networks. Our approach yields accurate global and node-level predictions across synthetic and real networks, including highly heterogeneous topologies and systems with higher-order interactions. Crucially, SSM is a robust tipping-point predictor: even at low truncation order (e.g., $O(2)$) it reliably identifies the onset of sustained activity, while higher orders and gSSM capture post-onset amplitudes and saturation.  Consistently, the reduction collapses the full network dynamics to a one-dimensional system, offering clarity and efficiency. Across all the realizations, SSM/gSSM consistently outperform classical spectral and mean-field methods in modeling critical transitions at both microscopic and macroscopic scales, establishing SSM-based reduction as a robust, interpretable tool for nonlinear networked systems with broad applicability to epidemiology, ecology, and engineered networks.
\end{abstract}

\maketitle

\section{Introduction}

Complex networked systems are fundamental to key processes in ecology, epidemiology, power engineering, and neuroscience \cite{pascual2006,anderson1991,dobson2007,rumelhart1986}. Their interactions are high-dimensional and nonlinear, rendering direct analysis and large-scale simulations computationally demanding and often intractable. Dimensionality reduction is, therefore, indispensable for prediction, control, and decision-making while preserving the essential mechanisms that govern system behavior \cite{gao2016,landi2018}.

Existing approaches to reduce network dynamics range from mean-field and spectral projections to broader analytical and data-driven methods. Mean-field schemes provide coarse-grained, low-dimensional descriptions of system-level resilience and may capture tipping-point behavior in favorable settings, but they typically lose node-level fidelity~\cite{gao2016,jiang2018} and do not uniformly predict critical thresholds accurately, especially in heterogeneous networks. Spectral reductions project the dynamics onto dominant eigenmodes and can reveal bifurcations and collective trends; refinements that incorporate sub-dominant modes or modular structure can enhance accuracy in heterogeneous networks ~\cite{laurence2019,masuda2022,vegue2023}. General analytical frameworks further condense high-dimensional dynamics into a small set of effective variables, including treatments for heterogeneous, discrete-time, and stochastic regimes \cite{tu2021,tu2022discrete,tu2023stochastic}. Complementary strategies combine spectral ideas with coarse-graining and adaptivity \cite{thibeault2020}, use entropy-based compression for resilience \cite{wu2023entropy}, or employ nature-inspired optimization and data-driven prediction with limited topological information \cite{mohammadi2019,prasse2022pnas,prasse2021chaos,ding2024arxiv}. Recent theory also highlights low-rank structure as a foundation for reduction in complex systems \cite{thibeault2024natphys}.

Despite this progress, many reductions struggle with strongly nonlinear, heterogeneous, or node-resolved behaviors. Mean-field models compress the dynamics over global coordinates, obscuring heterogeneity and potentially misidentifying critical thresholds \cite{gao2016,jiang2018}. Spectral projections rely on the sufficiency of a few linear modes. Indeed, when the leading spectral gap $\Delta=\operatorname{Re}\lambda_1-\operatorname{Re}\lambda_2$ is small, where $\lambda_1$ and $\lambda_2$ denote the eigenvalues with the two largest real parts, or when the dominant eigenvector is localized on high-degree hubs rather than distributed across the network, trajectories may depart from a one-mode reduced subspace and node-level errors can increase \cite{laurence2019,masuda2022,vegue2023}. Hybrid, entropy-based, and data-driven methods can fit aggregate behavior but often lack mathematical guarantees and may generalize poorly beyond training regimes \cite{thibeault2020,wu2023entropy,mohammadi2019,prasse2022pnas,prasse2021chaos,ding2024arxiv}. These limitations motivate a nonlinear, invariant, and globalizable reduction that preserves fidelity across regimes \cite{kaszas2025globalizing,tu2021,tu2022discrete,tu2023stochastic}.

To address these limitations, we introduce a reduction framework based on spectral submanifolds (SSMs) and their globalized extension (gSSM). Given an equilibrium $\mathbf{x}^\ast=\mathbf{0}$ of $\dot{\mathbf{x}}=f(\mathbf{x})$ and a spectral subspace $E$ of the linearized operator $L=Df(\mathbf{0})$, a spectral submanifold (SSM) is the smoothest invariant manifold $W(E)$ tangent to $E$ at $\mathbf{x}^\ast$. The dynamics on this manifold are governed by a reduced equation $\dot{\eta}=R(\eta)$, while the full network state is reconstructed through the lifting map $\mathbf{x}=W(\eta)$. In this work, the globalized SSM (gSSM) refers to a Pad\'e-type rational continuation of the local Taylor-series SSM reduction, introduced to extend the validity of the reduced model beyond the convergence region of its polynomial approximation. This approach is rooted in rigorous theoretical foundations and leverages scalable computational techniques for constructing invariant manifolds and their reduced dynamics \cite{haller2016,jain2022}. When local Taylor-series parameterizations of an SSM yield only a limited domain of convergence, we employ Padé-type rational approximations to extend the validity domain of the reduced models significantly~\cite{kaszas2025globalizing}. Recent advances in the SSM literature include explicit steady-state analyses and prediction of bifurcation structure in resonant systems \cite{breunung2018,ponsioen2020,li2022i,li2022ii}, as well as data-driven extensions that recover SSMs and their reduced dynamics directly from trajectory data, enabling robust model reduction even for non-linearizable or chaotic systems~\cite{cenedese2022nc,liu2024chaos}. Our main goal is to develop an equation-driven SSM/gSSM workflow that preserves the nonlinear structure of network dynamics across heterogeneous topologies, while delivering accurate node-level and macroscopic forecasts.

\begin{figure*}[t]
  \centering
  \includegraphics[width=\textwidth]{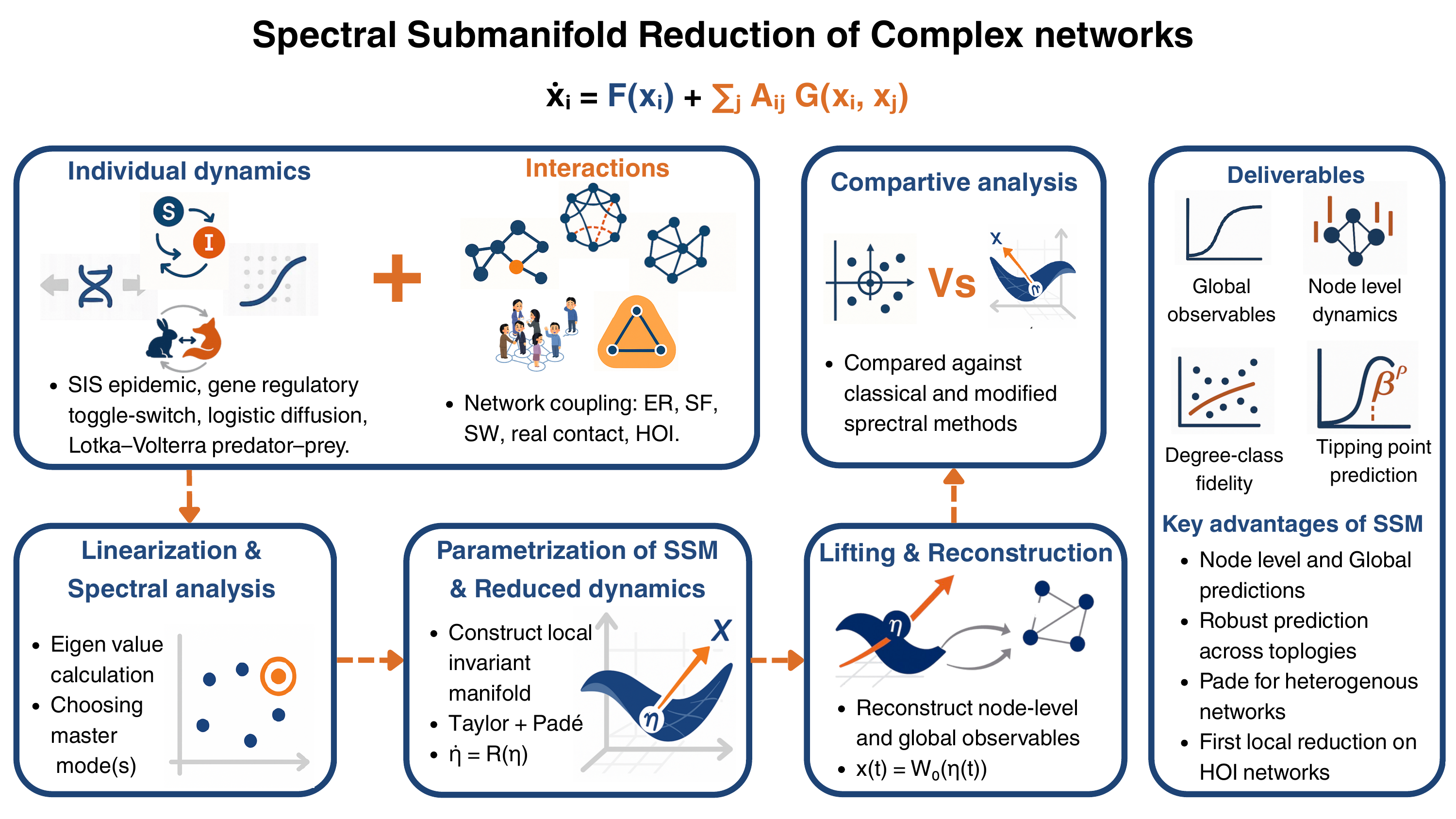}
  \caption{Schematic of the SSM reduction pipeline for complex networks. Starting from node dynamics and topology, a spectral analysis selects a master spectral mode (more generally, a master spectral subspace) of the linearized system and constructs a low-dimensional SSM and its autonomous reduced dynamics. The lifting map reconstructs node-level trajectories and global observables, enabling robust prediction of dynamics, tipping points, and degree-class behavior across heterogeneous and higher-order networks.}
  \label{fig:schematic}
\end{figure*}

Our results demonstrate that SSM/gSSM-based reductions substantially enhance predictive accuracy for both node-level trajectories and global observables across synthetic and empirical contact networks. Here, by a global observable we mean any aggregate quantity $\mathbf{y}(t)=\mathcal{O}(\mathbf{x}(t))$, where $\mathcal{O}:\mathbb{R}^N\to\mathbb{R}^q$ maps the full node-state vector to a low-dimensional summary of collective behavior; in the SIS examples below, the principal case is the mean prevalence $\langle I\rangle(t)=N^{-1}\sum_{i=1}^N x_i(t)$.  These SSM/gSSM reductions outperform classical spectral and mean-field reductions, especially in heterogeneous settings and near critical thresholds, and remain effective in the presence of higher-order interactions (HOI). A key advantage of our approach is its parsimony: for the SIS cases studied here, the full node-level dynamics are effectively captured by a single scalar reduced coordinate on the dominant spectral submanifold, from which both global and node-wise behaviors can be reconstructed. Notably, SSM provides a robust tipping-point predictor: even at low truncation orders it accurately identifies the onset of sustained activity, while higher orders and gSSM recover post-onset amplitudes and saturation. This framework is interpretable because the reduction yields a single latent coordinate whose evolution captures the dominant collective dynamics, while the corresponding lifting map reconstructs every node’s behavior from this low-dimensional description. In this way, the model offers a transparent connection between individual node interactions and macroscopic network outcomes, supporting reliable forecasting and control of high-dimensional nonlinear systems in epidemiology, ecology, and engineered infrastructures.

The remainder of the paper is organized as follows: we first formulate the network dynamics and governing equations; next, we develop the SSM framework and its globalized variant (gSSM); we then apply the reduction to epidemic networks by the Susceptible--Infected--Susceptible (SIS) model and benchmark it against alternative methods; we extend the study to higher–order SIS dynamics with triadic interactions; and we conclude with key findings and future directions. Additional applications and full diagnostic analyses are provided in the Supplementary Material (SM)~\cite{sup_ssm}.

\section{Network Dynamics and Model Reduction}

\subsection{Network dynamics formulation}

We consider a dynamical system defined on a network with adjacency matrix $\mathbf{A}$,
\begin{equation}
\label{eq:full}
    \dot{x}_i \;=\; F(x_i)+\sum_{j} A_{ij}\,G(x_i,x_j), 
    \qquad 1\le i\le N,
\end{equation}
where $F$ encodes intrinsic node dynamics, $G$ describes pairwise interactions, and $A_{ij}$ is the $(i,j)$ entry of $\mathbf{A}$. Without loss of generality, we shift an equilibrium so that $x_1=\cdots=x_N=0$ is an attracting fixed point of~\eqref{eq:full}. For compactness, we focus on undirected graphs and assume $\mathbf{A}$ is symmetric (extensions to directed/weighted graphs are straightforward). This compact form covers a broad class of networked systems (e.g., SIS-type epidemic, ecological, and logistic diffusion models) while keeping notation minimal for the reduction that follows. We next construct a one-dimensional reduced-order model using spectral submanifolds (SSMs) and their globalized rational extension (gSSM). For clarity and reproducibility, we summarize the general SSM/gSSM methodology in Sec.~\ref{sec:ssm_method_main}, while the network-specific coefficient derivations and higher-order recursions are provided in Appendix~\ref{sec:s1-methods}.

\subsection{SSM/gSSM reduction method}
\label{sec:ssm_method_main}

In all examples considered in this paper, the system admits multiple equilibria, but the reduction is constructed only about the trivial equilibrium. Let $\mathbf{f}:\mathbb{R}^N\to\mathbb{R}^N$ denote the network vector field and let $\mathbf{L}=D\mathbf{f}(0)$ be its linearization at the selected fixed point. In all these examples, the resulting one-dimensional SSM recovers the heteroclinic orbit connecting to the other fixed point, so the relevant nonlinear evolution is obtained from this single local construction. We compute the eigenpairs $(\lambda_k,\mathbf{v}_k)$ of $\mathbf{L}$ and select a master spectral subspace
\[
E=\mathrm{span}\{\mathbf{v}_1,\dots,\mathbf{v}_m\}
\]
associated with the eigenvalue(s) having the largest real part (slowest decay).

 In the present study, this choice is appropriate because our objective is to approximate the slow departure from the attracting  trivial equilibrium and the subsequent approach along the heteroclinic connection toward the other fixed point. When the leading eigenvalue is real and spectrally separated, this choice gives a one-dimensional master subspace. Other choices of master subspace are meaningful for different dynamical questions. For instance, in a Laplacian-coupled system near a synchronous state, if we restrict that the perturbation from the synchronous state is orthogonal to the null space of the Laplacian,  an SSM constructed to the study of approaching the synchronous state would naturally be built over the weakly stable or critical transverse Laplacian modes. Likewise, near-degenerate leading spectra, complex conjugate pairs, or symmetry-breaking modes would require a higher-dimensional master subspace. This clarifies that the master-mode selection is context-dependent and chosen to capture the dominant slow dynamics relevant to the examples studied in this manuscript.
 
 Under standard nonresonance and spectral-quotient conditions, there exists a $C^r$ spectral submanifold (SSM) $\mathbf{W}:E\to\mathbb{R}^N$ tangent to $E$ at the origin that carries autonomous reduced dynamics $\dot{\boldsymbol{\eta}}=\mathbf{R}(\boldsymbol{\eta})$~\cite{haller2016}. The pair $(\mathbf{W},\mathbf{R})$ is defined by the invariance equation
\begin{equation}
D\mathbf{W}(\boldsymbol{\eta})\,\mathbf{R}(\boldsymbol{\eta})=\mathbf{f}\!\big(\mathbf{W}(\boldsymbol{\eta})\big).
\end{equation}

We compute $(\mathbf{W},\mathbf{R})$ via the parameterization method~\cite{jain2022}, using Taylor series expansions
\begin{equation}
\mathbf{W}(\boldsymbol{\eta})=\sum_{|\boldsymbol{\alpha}|\ge1}\mathbf{w}_{\boldsymbol{\alpha}}\,\boldsymbol{\eta}^{\boldsymbol{\alpha}},
\qquad
\mathbf{R}(\boldsymbol{\eta})=\boldsymbol{\Lambda}\boldsymbol{\eta}+\sum_{|\boldsymbol{\alpha}|\ge2}\mathbf{r}_{\boldsymbol{\alpha}}\,\boldsymbol{\eta}^{\boldsymbol{\alpha}},
\end{equation}
with $\boldsymbol{\Lambda}=\mathrm{diag}(\lambda_1,\dots,\lambda_m)$. Matching coefficients in the invariance equation yields linear homological equations for $\{\mathbf{w}_{\boldsymbol{\alpha}},\mathbf{r}_{\boldsymbol{\alpha}}\}$, solved order-by-order to a chosen truncation order $p$ (denoted $O(p)$ throughout).

For the one-dimensional reductions used in the SIS examples below, the reduced coordinate $\boldsymbol{\eta}$ reduces to a scalar coordinate $\eta$. In that case, the general expansion above becomes
\[
\mathbf{W}(\eta)=\mathbf{u}\eta+\sum_{k=2}^{p}\mathbf{w}_k\eta^k,
\qquad
R(\eta)=\lambda\eta+\sum_{k=2}^{p}r_k\eta^k.
\]
where $\mathbf{u}$ is the leading right eigenvector, $\mathbf{w}_k\in\mathbb{R}^N$ are lifting-map coefficients, and $r_k\in\mathbb{R}$ are reduced-dynamics coefficients. Appendix~\ref{sec:s1-methods} gives the corresponding coefficient-level homological equations for this one-dimensional case.

In all the cases used here, the dominant eigenvalue is real and sufficiently separated from the remainder of the spectrum, so a one-dimensional reduction captures the asymptotic slow dynamics accurately. This does not preclude transient improvements from higher-dimensional reductions. In particular, when the full initial condition is chosen off the reduced manifold, a two-dimensional reduction can provide a more accurate early-time approximation before both reductions approach the same long-time slow dynamics; see Sec.~S1 of the SM.

A key advantage of SSM reduction is node-resolved reconstruction. Here, the \emph{lifting map} $\mathbf{W}$ is the map from the low-dimensional reduced coordinate back to the original network state space: if the intrinsic coordinate $\eta(t)$ evolves according to $\dot{\eta}=R(\eta)$ on the SSM, then the corresponding full state is recovered as $\mathbf{x}(t)=\mathbf{W}(\eta(t))$. In this way, the reduced dynamics are solved in a low-dimensional latent variable, while the lifting map reconstructs every nodal trajectory and hence derived global observables such as the mean prevalence $\langle I\rangle=N^{-1}\sum_i x_i$. In practice, the computation of higher-order coefficients and the assembly of $(\mathbf{W},\mathbf{R})$ are automated using \texttt{SSMTool} (v2.6)~\cite{ssmtool26}.

To enlarge validity beyond the local Taylor radius, we globalize the reduced dynamics using Pad\'e-type rational approximants (gSSM)~\cite{kaszas2025globalizing}. Specifically, we replace the truncated Taylor polynomial of the reduced vector field by a rational approximation whose series matches up to a prescribed order. Unless stated otherwise, we use a diagonal-type Pad\'e approximant of type $[8/7]$ constructed from the order-15 Taylor expansion of the reduced vector field; implementation details are provided in Appendix~\ref{sec:s2}. Next, we compare SSM-reduced model performance on recovering mean and node-level dynamics (transient as well as steady-state) in various SIS networks.

\section{Results}

We simulate the SIS dynamics
\begin{equation}
\dot{x}_i=-\gamma x_i+\beta\sum_{j=1}^N A_{ij}(1-x_i)\,x_j,
\end{equation}
on networks with $N=200$, $\gamma=1.0$, $\beta=0.5$, and initialization on the dominant SSM coordinate at $\eta_0=0.01$. Here, the dominant SSM coordinate denotes the scalar intrinsic coordinate on the one-dimensional SSM tangent to the leading eigenvector $v_1$ of the linearized system, so that an aligned initial condition in the full state space is taken along the corresponding tangent direction. The value $\eta_0=0.01$ is a small prescribed amplitude chosen to initialize the dynamics near the selected equilibrium within the local regime of validity of the reduction, while keeping the setup consistent across all benchmark networks.
To verify robustness to initial conditions, we also tested aligned, orthogonal, and near-aligned states in the physical coordinates (see \hyperref[sec:s9]{Sec.~S1 of SM~\cite{sup_ssm}}).  At the same time, additional off-manifold tests showed that a two-dimensional reduction can yield a more accurate transient prediction than a one-dimensional reduction, particularly at node level and more clearly in less homogeneous networks. For the initialization tests reported in \hyperref[sec:s9]{Sec.~S1} of the SM, the trajectories considered there converged after a brief fast transient to the same effective slow evolution, which is why the one-dimensional SSM reduction accurately captures the long-term behavior in those cases. Unless otherwise noted, we use the same network instance for each synthetic topology—Erd\H{o}s–R\'enyi (ER), Small-World (SW), and Scale-Free (SF) models representing, respectively, homogeneous, shortcut rich - locally clustered, and heterogeneous connectivity—and the same empirical contact networks (Hospital, Workplace, Rural) across all analyses. Full construction details and network statistics are provided in the \hyperref[sec:s3]{Sec.~S2} of SM~\cite{sup_ssm}. Error metrics (trajectory mean squared error, steady-state error $\epsilon_i^{\mathrm{ss}}$) and the evaluation workflow are detailed in the \hyperref[sec:s5]{Sec.~S3} of SM~\cite{sup_ssm}. To quantify reduced-model accuracy, we use two primary measures. First, for the macroscopic observable $\langle I\rangle(t)$, we define the mean-squared error
\[
\mathrm{MSE}_{\langle I\rangle}^{(p)}=
\frac{1}{M}\sum_{m=1}^{M}
\Big(\widehat{\langle I\rangle}^{(p)}(t_m)-\langle I\rangle_{\mathrm{ref}}(t_m)\Big)^2,
\]
where $t_m$ denotes the common comparison grid. Second, at the node level we quantify the steady-state discrepancy by
\[
\epsilon_{i}^{ss,(p)}=\big|\hat{x}_{i}^{(p)}(T)-x_{i}^{\mathrm{ref}}(T)\big|.
\]
These metrics are used throughout to assess macroscopic and node-resolved agreement; additional diagnostics are provided in Sec.~S3 of the SM.

\begin{figure*}[t]
  \centering
  \includegraphics[width=\textwidth]{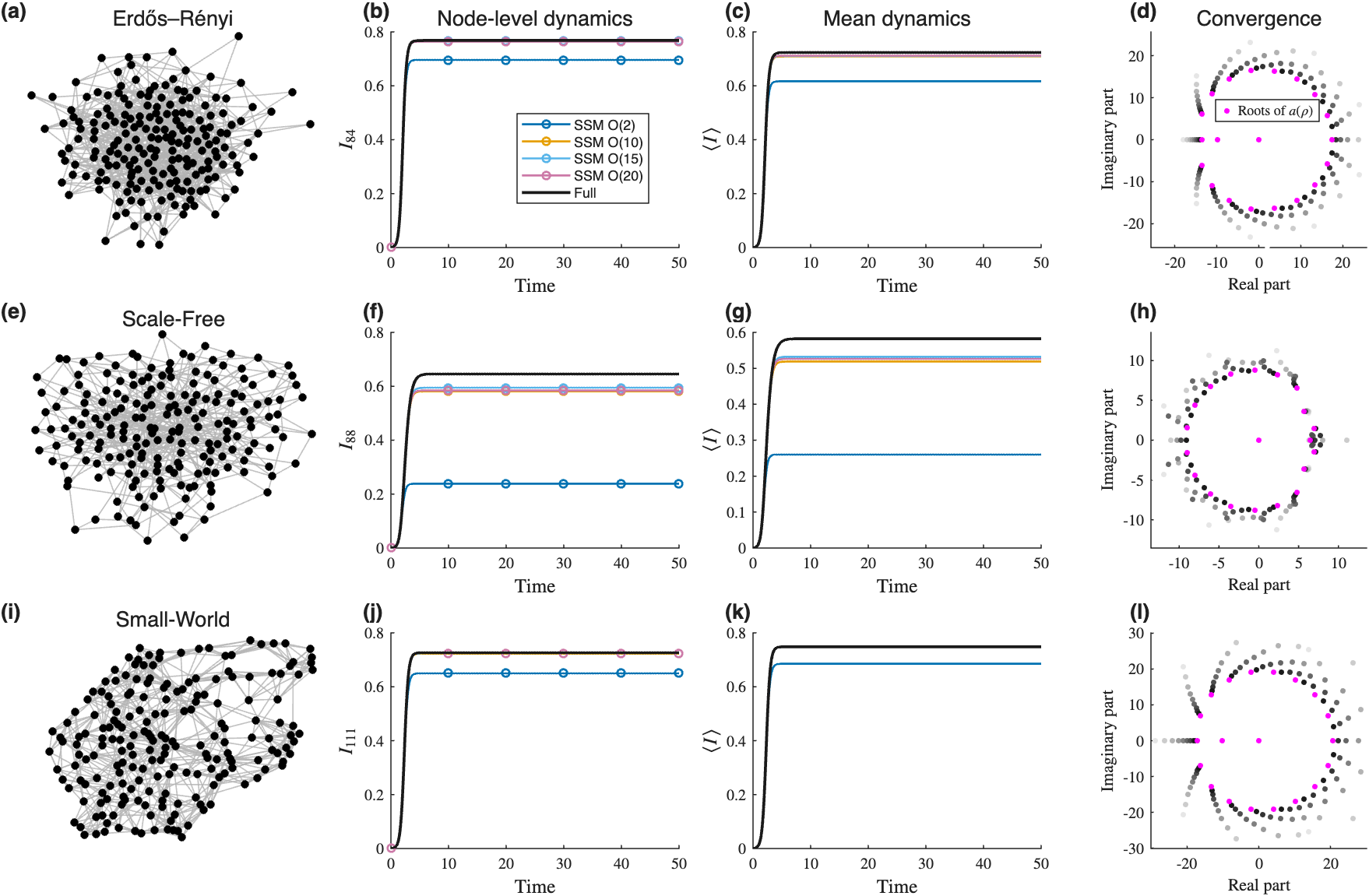}
 \caption{SIS dynamics and comparison of full and reduced trajectories Erd\H{o}s--R\'enyi (ER), Scale-Free (SF), and Small-World (SW) ($N{=}200$, $\mu{=}1.0$, $\beta{=}0.5$, $\eta {=}0.01$).
  (a,e,i) Representative network realizations. 
  (b,f,j) Node-level trajectories: $O(2)$ (blue) underestimates both transient and steady behavior; $O(10)$ (orange) is \textit{nearly indistinguishable} from the full system (black) on ER/SW; SF requires $O(15)$–$O(20)$.
  (c,g,k) Mean prevalence $\langle I\rangle$ shows the same ordering of accuracy.
  (d,h,l) \textit{Taylor convergence plot:} magenta points are zeros of the truncated radial drift $a_A(\rho)$; the positive-real zero marks the predicted nonzero steady amplitude. The dashed gray circle has radius $R_P$ computed from the outer half of the highest-order root moduli serving as an empirical Taylor radius. }
  \label{fig:sis_panels}
  \end{figure*}

\begin{figure}
  \centering
  % Prefer vector PDF; fallback to PNG if needed
    \includegraphics[width=0.8\linewidth]{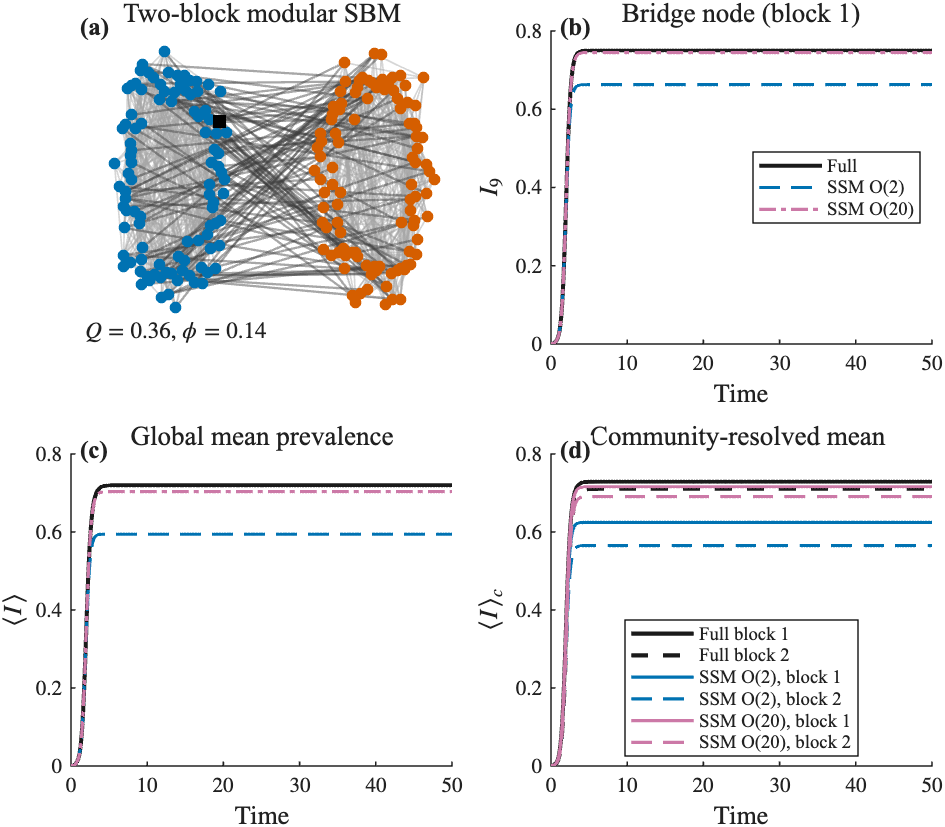}
\caption{Modular-bottleneck benchmark using a two-block stochastic block model with weak inter-community coupling. 
(a) Network colored by community; the square marks a representative bridge node. 
(b) Infection trajectory at the bridge node: $O(2)$ underestimates, $O(20)$ tracks the full system. 
(c) Global mean prevalence $\langle I\rangle$: low-order truncation is biased, higher-order matches well. 
(d) Community-level prevalence $\langle I\rangle_c$: $O(2)$ underestimates block means, $O(20)$ recovers them closely. 
Modular bottlenecks make low-order reductions less accurate, especially for bridge and community observables.}
  \label{fig:sbm_modular}
\end{figure}

In Fig.~\ref{fig:sis_panels}, we compare the full system (black) with SSM reductions at orders $O(2)$ (blue), $O(10)$ (orange), $O(15)$ (pink), and $O(20)$ (gray), reporting accuracy for both a representative node trajectory and the mean prevalence $\langle I\rangle=N^{-1}\sum_i x_i$.
Here, \(O(p)\) denotes the truncation order of the SSM Taylor expansion, with an \(O(p)\)–SSM reduction referring to the model truncated at order \(p\). For node-level time series, the ER network shows that $O(2)$-reduced model underestimates both the transient rise and the \textit{asymptotic endemic equilibrium (steady state)}, whereas  the $O(10)$–SSM reduction converges to the full trajectory and higher-order results are \textit{nearly indistinguishable} (Fig.~\ref{fig:sis_panels}b). This low-order bias has a simple interpretation: the $O(2)$ model is only a local approximation of the SSM dynamics near the reference equilibrium, so it retains only the leading nonlinear curvature of both the reduced drift and the lifting map. In the SIS system, where the infection term $A_{ij}(1-x_i)x_j$ introduces amplitude-dependent saturation, the neglected higher-order terms become important once the trajectory moves away from the immediate neighborhood of the equilibrium. As a result, the $O(2)$ truncation tends to predict both a slower transient rise and a reduced nonzero steady amplitude.
For the SF network, at low orders (e.g., $O(2)$), the SSM-reduced model exhibits the largest underestimation relative to the full dynamics; this decreases substantially at $O(10)$–$O(20)$(Fig.~\ref{fig:sis_panels}f). In the SW network, $O(10)$-approximation reproduces the full trajectory with sufficient accuracy and any further increase in order brings negligible change (Fig.~\ref{fig:sis_panels}j).

Similar accuracy trends are observed in the mean curves: $O(10)$-approximations are sufficiently accurate for ER and SW networks (Fig.~\ref{fig:sis_panels}c and Fig.~\ref{fig:sis_panels}k), while similar accuracy is achieved at $O(15)$--$O(20)$ in the case of SF network (Fig.~\ref{fig:sis_panels}g). ER and SW networks are relatively homogeneous: most nodes have comparable numbers of links (narrow degree distribution), and the leading eigenvector is broadly distributed across nodes, a property known as delocalization. In contrast, in heterogeneous networks such as SF, the leading eigenvector becomes localized on high-degree hubs such that only a few nodes have dominant weights. This localization has a direct dynamical consequence: when activity concentrates on a few hub nodes, the effective nonlinear coupling is enhanced and the local Taylor expansion of the reduced dynamics shows slower convergence. As a result, higher-order terms  of the expansion are needed to accurately represent such nonlinear interactions present in the full network. Thus, in ER and SW networks, the reduced coordinate behaves like a population average and low to moderate expansion orders already suffice for recovering the full system dynamics. In SF networks, however, higher-order Taylor expansions (or gSSM) are required for accuracy because localization on hubs effectively amplifies nonlinear effects and shrinks the radius of convergence of the SSM’s Taylor expansions. In our experience, the pattern of errors across node degrees is intuitive: at low truncation order, the reduced model captures the behavior of highly connected nodes more accurately, while discrepancies arise mainly among low-degree (peripheral) nodes. In SF networks, small residual errors can persist around hub nodes even at higher orders, but these remain within acceptable numerical tolerance (see \hyperref[sec:s4]{Sec.~S3} of the SM~\cite{sup_ssm} for detailed diagnostics).

%To assess the validity of \textit{Taylor-only} SSM model predictions, we perform a truncated polynomial roots analysis: at each truncation order $p\le 20$ of the SSM, we consider the single-mode reduced dynamics $\dot{\rho} = a(\rho)$ and plot the complex zeros of $a(\rho)$ (Fig.~\ref{fig:sis_panels}d,h,j ). 

To assess the validity of Taylor-only SSM model predictions, we perform a truncated polynomial root analysis. At each truncation order $p \le 20$, we consider the single-mode reduced dynamics $\dot{\rho}=a(\rho)$ and plot the complex zeros of the truncated polynomial $a(\rho)$ (Fig.~\ref{fig:sis_panels}d,h,j ). Here, a \emph{non-spurious} (bona fide) root means a zero that persists under increasing truncation order and remains well inside the estimated domain of convergence of the Taylor expansion, so that it can be interpreted as a genuine fixed point of the reduced dynamics. By contrast, a \emph{spurious} root is a truncation-induced zero, typically appearing near the boundary of the convergence domain and shifting substantially with $p$, and is therefore not regarded as a reliable physical prediction. As $p$ increases, the spurious roots of $a(\rho)$ cluster around the domain of convergence of the function $a(\rho)$ and any roots inside this domain are bona fide fixed points \cite{Ponsioen2019NonlinDyn}. Across all topologies, a single nontrivial transverse zero $\rho_{1}$ persists within the estimated convergence domain of $a(\rho)$, signaling a robust nonzero endemic level on the reduced coordinate. In ER and SW networks, a positive-real zero persists well within the root cloud as $p$ increases (Fig.~\ref{fig:sis_panels}d,h), consistent with rapid agreement at $O(10)$--$O(15)$ in both the transients and the steady state. In the SF network, however, a non-spurious root persists near the boundary of the convergence domain  (Fig.~\ref{fig:sis_panels}l), explaining the slower error decay and the utility of $O(15)$--$O(20)$ terms. In cases that our domain of convergence of Taylor expansions is not sufficiently large, a Pad\'{e} approximant (gSSM) still captures the full trajectory and resolves the high-prevalence regime, as demonstrated in Fig.~\ref{fig:beta_sweeps_all}.

\subsection{Targeted modular-bottleneck benchmark}

To isolate the effect of modular bottlenecks more directly, we supplement the ER, SW, and SF ensembles with a two-block stochastic block model of matched mean degree and weak inter-community coupling. This provides a controlled modular test in which transport between communities is confined to a relatively small set of bridge links. Construction details are given in \hyperref[sec:s3]{Sec.~S2} of the SM.

Figure~\ref{fig:sbm_modular} shows that the reduction remains accurate in this setting, although low-order truncations are more sensitive in the presence of bottlenecks. At the bridge node in Fig.~\ref{fig:sbm_modular}(b), the $O(2)$ approximation underestimates the full-system response, while the $O(20)$ reduction follows the full trajectory closely. The same pattern appears in the global mean prevalence in Fig.~\ref{fig:sbm_modular}(c): the low-order truncation still captures the epidemic level qualitatively, but with a visible bias that is much smaller at higher order.

The effect of modular structure is most evident in the community-level dynamics in Fig.~\ref{fig:sbm_modular}(d). The $O(2)$ reduction underestimates the mean prevalence in both communities, which indicates that low-order truncation does not fully resolve the balance between intra-community growth and bridge-mediated transfer. The $O(20)$ reduction, by contrast, reproduces the block-level means closely and remains consistent with the full-system trajectory throughout. These results do not suggest a qualitative failure of the SSM-based reduction on modular networks. They instead identify a topology-sensitive regime in which low-order truncations are less accurate, especially for bridge-node and community-resolved observables, whereas higher-order reductions recover both node-level and aggregate behavior reliably. Additional topology-aware diagnostics for this benchmark are reported in \hyperref[sec:s3]{Sec.~S7} of the SM.

\subsection{One-parameter sweeps across synthetic and real networks}

\begin{figure*}[t]
  \centering
  % Prefer vector PDF; fallback to PNG if needed
    \includegraphics[width=\textwidth]{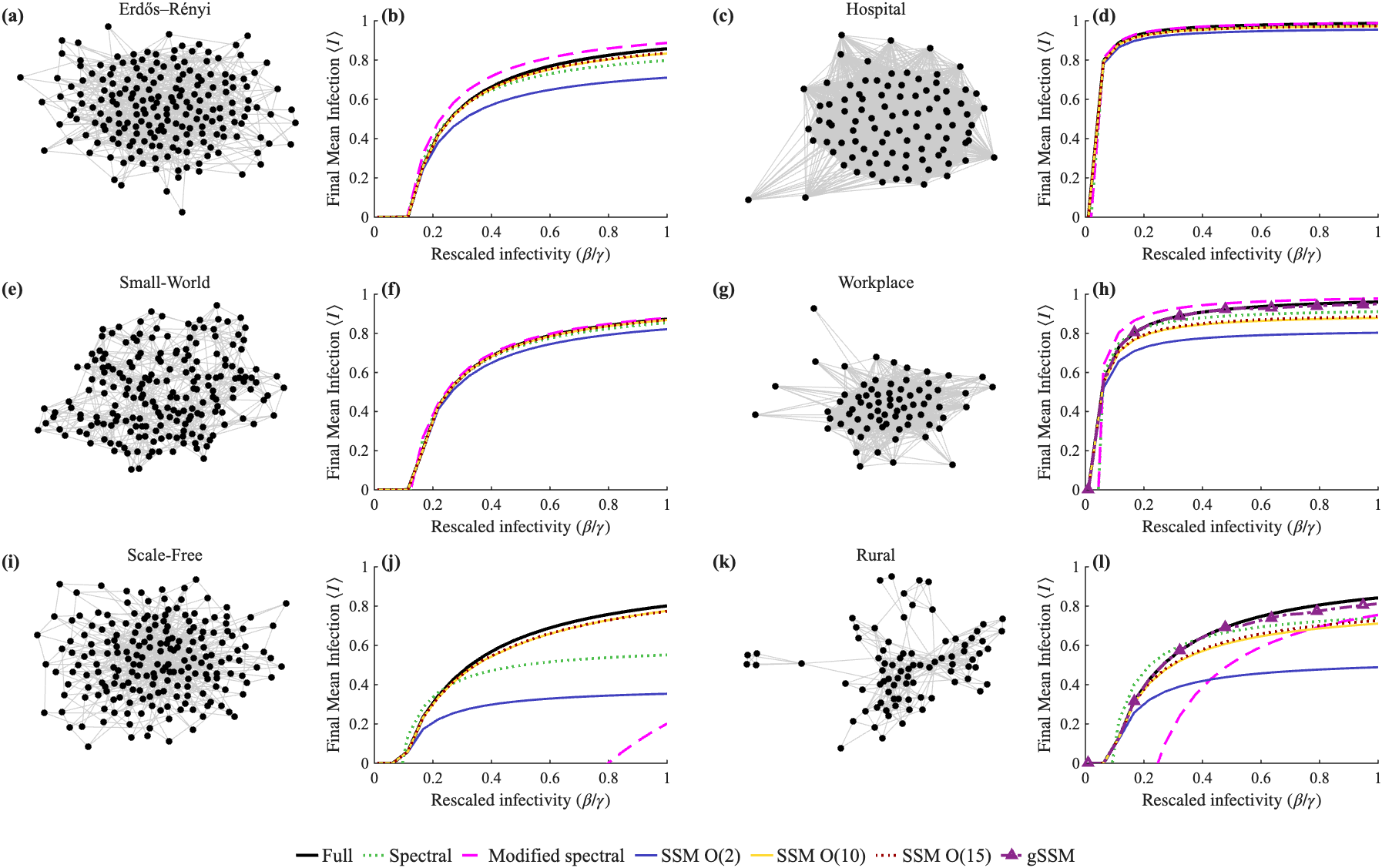}
  \caption{One-parameter sweeps of final mean infection $\langle I\rangle$ versus $\beta/\gamma$. Synthetic networks: ER, SW, SF (e.g., panels \textit{b,f,j}); empirical networks: Hospital, Workplace, Rural (e.g., panels \textit{d,h,l}). Black: full model; green dotted: spectral; magenta dashed: modified spectral; blue/orange/red: SSM $O(2)$/$O(10)$/$O(15)$; purple markers: gSSM where included. Methods for spectral baselines are in the Sec.~S4 of the SM~\cite{sup_ssm}.}
  \label{fig:beta_sweeps_all}
\end{figure*}

We now assess how SSM reduction captures a macroscopic observable and the tipping onset with respect to some control parameter. We vary $\beta/\gamma$ as our control parameter and record the final mean infection $\langle I\rangle$ across three synthetic topologies and three empirical contact networks discussed below. In Fig.~\ref{fig:beta_sweeps_all}, we compare the full system results (black) with spectral reduction (green, dotted); modified spectral reduction (magenta, dashed)~\cite{gao2016,laurence2019,masuda2022}; and SSM reduction at different orders: $O(2)$ (blue), $O(10)$ (orange), and $O(15)$ (red). Implementation details for the spectral baselines are provided in the Sec.~4 of SM~\cite{sup_ssm}.

\textit{Synthetic networks.} For the ER network (Fig.~\ref{fig:beta_sweeps_all}b), all curves stay near zero for small values of $\beta/\gamma$ and start increasing in a similar range ($\beta/\gamma\lesssim 0.2$). Beyond this range, the $O(2)$ SSM reduction underestimates the steady state for high values of $\beta/\gamma$, whereas the $O(10)$--$O(15)$ approximations accurately reproduce the full system, consistent with the nearly uniform connectivity of ER and SW networks, where nodes have similar degrees and the normalized dominant eigenvector $u$ is delocalized, meaning that its $\ell^2$ mass is distributed across many nodes rather than concentrated on a small subset. Equivalently, its inverse participation ratio $\mathrm{IPR}(u)=\sum_{i=1}^N u_i^4$ (for $\|u\|_2=1$) remains comparatively small. For the SW network, the onset and subsequent growth are approximated well by all reductions (Fig.~\ref{fig:beta_sweeps_all}f); higher-order SSM terms are required when the network is more heterogeneous because degree variation enhances nonlinear coupling among nodes. In SW networks this heterogeneity is moderate, and the leading eigenvector remains broadly spread across nodes, so higher-order SSM approximations reproduce the asymptotic mean infection with high accuracy. In contrast, on the SF network (Fig.~\ref{fig:beta_sweeps_all}j), the spectral and the modified spectral reductions provide incorrect estimates for the tipping onset and the steady state; SSM-based estimates agree with the initial near-zero regime and correctly capture the tipping onset for $\beta/\gamma\lesssim 0.15$. While $O(2)$-SSM predicts an incorrect bias in the steady states for $\beta/\gamma$-values beyond the tipping onset, this bias is eliminated upon increasing approximation order. Indeed, $O(10)$--$O(15)$ SSM reductions recover the final mean, indicating that higher-order nonlinear corrections are required under hub localization, consistent with the stronger nonlinear coupling described above.

We also evaluate SSM performance on three empirical contact networks from the \textit{SocioPatterns} collaboration, Hospital, Workplace, and Rural,  srepresenting healthcare, office, and community interaction settings~\cite{sp:hospital,sp:workplace,sp:rural}. Full dataset descriptions and preprocessing details are provided in \hyperref[sec:s3]{Sec.~S1} of the SM~\cite{sup_ssm}.  In the Hospital network (Fig.~\ref{fig:beta_sweeps_all}d), dense connectivity leads to \textit{close agreement} among all reduction methods over the range $\beta/\gamma \in (0,1)$, except for very low-order SSM. In the Workplace network (Fig.~\ref{fig:beta_sweeps_all}h), the spectral reduction underestimates the final mean infection for small values of $\beta/\gamma$ and inaccurately predicts the onset, while the modified spectral variant occasionally overshoots. SSM-based reductions converge closely to the full system for intermediate values of $\beta/\gamma$, but underestimate the outcome at larger values. The Rural network (Fig.~\ref{fig:beta_sweeps_all}l) exhibits the greatest variation across methods: spectral reduction misestimates the steady state over $\beta/\gamma \in (0,0.5)$, and the modified spectral reduction remains inaccurate across the entire range of the control parameter. SSM reductions accurately capture the initial regime for $\beta/\gamma \in (0,0.2)$ but underestimate the steady state at larger values, even at higher orders of approximation. As the Rural network displays localization, this slow convergence aligns with our earlier observations on synthetic networks regarding the limited accuracy of truncated Taylor expansions under localization. We employ the globalized SSM extension (gSSM) where necessary (purple markers), and observe good agreement with the full system across the entire range of the control parameter.

\begin{figure}
  \centering
  % Prefer vector PDF; fallback to PNG if needed
    \includegraphics[width=0.8\linewidth]{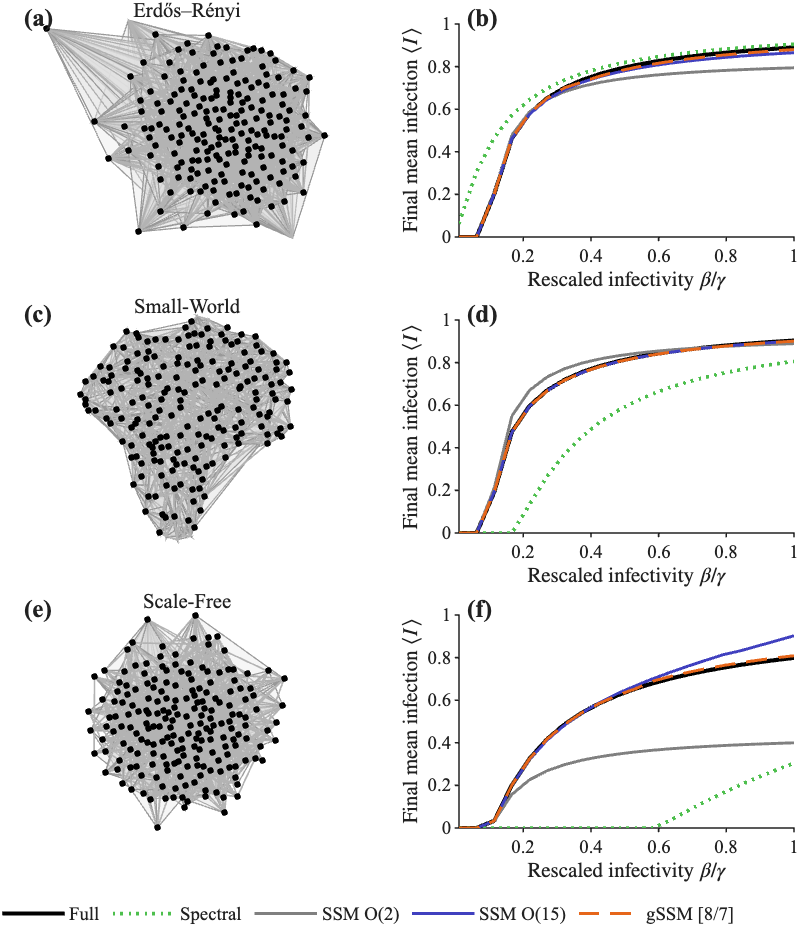}
  \caption{Final mean infection $\langle I\rangle$ versus $\beta/\gamma$ for HOI SIS with triadic interactions on ER, SW, and SF networks. Curves: full (black), spectral (green, dotted), SSM $O(2)$ (gray), SSM $O(15)$ (blue), and gSSM (orange, dashed) where included. Spectral/closure implementation details are provided in the Sec.~S5 of SM~\cite{sup_ssm}.}
  \label{fig:hoi_sis_sweep}
\end{figure}

\subsection{One-parameter sweep for HOI SIS with triadic interactions}
    
Many real systems transmit influence in groups rather than pairs. For such systems, a standard modeling approach is to lift a graph $A$ to a simplicial complex by closing cliques (e.g., adding a 2-simplex for each triangle) and allowing transmission to depend on group participation. This framework is well established and can induce discontinuous (first-order) transitions and bistability that do not appear in purely pairwise models~\cite{iacopini2019simplicial,battiston2020beyond,battiston2021higher}. We adopt this construction (triangle closure to obtain $T_{ijk}$, fixed HOI strength $\eta$) and extend the SIS dynamics as:
\begin{equation}
\label{eq:hoi-sis}
\dot{x}_i \;=\; -\gamma x_i
\;+\; \beta\sum_{j} A_{ij}(1-x_i)\,x_j
\;+\; \eta \sum_{j<k} T_{ijk}\,(1-x_i)\,x_jx_k,
\end{equation}
where $A$ is the adjacency matrix, $T_{ijk}$ encodes 2-simplices (triangles), and the parameter $\eta$ is held fixed while varying $\beta/\gamma$ as the control parameter. We use the simplicial-closure construction, where every triangle in $A$ is lifted to a 2-simplex, following standard practice in higher-order contagion
models~\cite{iacopini2019simplicial,battiston2020beyond,battiston2021higher}. Additional simplicial-closure construction and parameter details are given in \hyperref[sec:s6]{Sec.~S5} of the SM~\cite{sup_ssm}.

On the ER network (Fig.~\ref{fig:hoi_sis_sweep}b), the full HOI system and the SSM results agree closely for values of $\beta/\gamma\in(0,0.35)$, capturing both onset and early growth. Beyond this range, low-order SSM underestimates the steady-state, while $O(15)$ remains aligned. The spectral baseline performs reasonably at large values of $\beta/\gamma$ but misses the onset and produces an overly smooth transition. On the SW network (Fig.~\ref{fig:hoi_sis_sweep}d), the SSM results converge to steady-state throughout: $O(2)$ slightly overestimates $\langle I\rangle$ for $\beta/\gamma\in(0.15,0.5)$, whereas $O(15)$ (and gSSM where shown) reproduce both the knee, namely the high-curvature crossover region of the response curve $\langle I\rangle_{ss}(\beta/\gamma)$ marking the onset of rapid prevalence growth, and the asymptotic mean. The SF case (Fig.~\ref{fig:hoi_sis_sweep}f) is the most demanding: SSM reductions at all orders (and gSSM reduction) are in \textit{close agreement} with the full model near onset $\beta/\gamma\in(0,0.15)$, but $O(2)$-SSM later drifts to a false steady state; in the intermediate range $\beta/\gamma\in(0.2,0.6)$, $O(15)$-SSM and gSSM reproduce the full system curve, and for larger $\beta/\gamma$ the $O(15)$-SSM overshoots the steady-state level, whereas gSSM remains accurate. 

These trends reflect two mechanisms: (i) triadic reinforcement makes the response $\langle I\rangle_{ss}(\beta/\gamma)$ steeper and more abrupt near the transition region, which can support discontinuities and bistability; and (ii) eigenvector localization around hubs in heterogeneous graphs reduces the effectiveness of aggregate spectral projections, conditions under which higher-order SSM or gSSM reductions are advantageous~\cite{iacopini2019simplicial,battiston2020beyond,battiston2021higher}.

\subsection{Generalization beyond SIS models}
Beyond SIS dynamics, our equation-driven SSM workflow applies to generalized Lotka--Volterra, gene-regulatory, and logistic--diffusion network models across ER, SW, and SF topologies. In all cases, the SSM-reduced models similarly reproduce node-level time series and macroscopic trends; full trajectories, convergence diagnostics, and comparisons against full-order simulations are provided in \hyperref[sec:s7]{Sec.~S6} of SM~\cite{sup_ssm}.

\section{Conclusion}

We developed an equation-driven reduction framework based on spectral submanifolds (SSMs) and their rational extension (gSSM) to simplify nonlinear network dynamics while preserving both node-level and system-level fidelity. Applied to synthetic and empirical contact networks, this approach consistently advanced beyond classical spectral and mean-field surrogates. On Erd\H{o}s–R\'enyi and small-world topologies, moderate-order \(O(p)\)–SSM reductions (e.g., $O(10)$–$O(15)$) sufficed to reproduce full trajectories and steady-state levels with high accuracy. In contrast, scale-free networks revealed the limitations of eigenvector-based reductions relative to SSMs:  $O(2)$-SSMs already captured the epidemic threshold (tipping point) correctly, while higher-order \(O(p)\)–SSM reductions or gSSM recovered the nonlinear rise and high-prevalence steady state that spectral baselines misestimated under hub localization. 

A key novelty of our approach is its ability to make node-resolved prediction: the reduced coordinate, together with its lifting map, reconstructs every individual nodal trajectory alongside global observables--a capability absent in traditional spectral approaches. Also, throughout all experiments, the SSM/gSSM consistently yielded a one-dimensional reduced-order model, despite the high-dimensionality of the underlying network dynamics. This highlights both the robustness and interpretability of the proposed framework.

Extending the analysis to higher-order contagion, the triadic SIS model displayed sharper rises in prevalence and, in some regimes, discontinuous shifts in the endemic state. Here as well, SSM-based reductions consistently identified the epidemic onset tipping point, with even $O(2)$-approximations accurately identifying the tipping point despite structural heterogeneity. Quantitative agreement in the post-onset regime improved systematically with increasing polynomial order, and gSSM restored accuracy when polynomial truncations saturated for high values of the control parameter. These findings underscore SSM as a robust local predictor of tipping points and gSSM as an effective globalization tool, particularly in heterogeneous or higher-order settings. Beyond epidemic processes, the same workflow generalized seamlessly to other complex networks such as the generalized Lotka–Volterra, gene-regulation dynamics, and logistic–diffusion dynamics, where reduced models successfully reproduced both node-level and macroscopic behaviors (see Sec.~S6 of the SM~\cite{sup_ssm} for details). 

Overall, our results establish SSM/gSSM reduction as a rigorous and broadly applicable method for nonlinear networked systems. By combining node-level reconstructions with reliable tipping-point prediction, it provides a practical bridge between mechanistic models and interpretable reduced dynamics. Immediate extensions include applying the framework to time-delayed network dynamics, where interactions depend on past states~\cite{szaksz2025spectral}. Such systems have been analyzed using invariant-manifold theory for delay differential equations, providing a natural setting to extend SSM-based reductions. Longer-term directions involve incorporating temporal variability through nonautonomous formulations~\cite{li2022ii,breunung2018}, integrating uncertainty quantification via data-driven manifold inference~\cite{cenedese2022nc,liu2024chaos}, and scaling the methodology to high-dimensional empirical datasets. These avenues highlight the promise of SSM-based reductions as general tools for forecasting, intervention, and resilience analysis in complex biological, social, and engineered systems.

\subsection*{Data availability Statement}
All codes and datasets required to reproduce the results presented in this study are available at 
\href{https://drive.google.com/drive/folders/1hgUTc1x8NLxmBR-HqDLiuXflGw3PgYB6?usp=sharing}{Google Drive Repository}. 
The repository contains MATLAB \texttt{.mlx} dashboards and scripts for all models discussed in the paper, including SIS, GLV, GRN, and logistic–diffusion systems. 
Each folder includes labeled files and documentation to facilitate reproducibility of the simulations.
\subsection*{Author contributions}
M.L. and K.B. designed and carried out the research. K.B drafted the manuscript and produced the graphical illustrations. All authors contributed to the software development, and the review and editing of the paper.

\subsection*{Acknowledgements}
M.L. acknowledges financial support from the National Natural Science Foundation of China (Nos. 12302014 and 12572010). The authors also thank the anonymous reviewer for their careful reading and constructive suggestions, which substantially improved the clarity and presentation of the manuscript.
\subsection*{Competing interests}
The authors declare no competing interests.

\appendix

\section{Network-specific coefficient derivations and implementation details}
\label{sec:s1-methods}

For completeness, we collect in this Appendix the network-specific derivations of the SSM reduced dynamics used in the main text. A concise description of the general SSM/gSSM methodology (invariance equation, parameterization method, and Pad\'e globalization) is given in Sec.~\ref{sec:ssm_method_main}. Here we focus on the explicit coefficient-level derivation for the network model in Sec.~\ref{sec:s1-derivation}.

\vspace{0.5cm}

\subsection{Derivation of the SSM reduced dynamics for the network model}
\label{sec:s1-derivation}

Consider a networked dynamical system on $N$ nodes with state variables $\{x_i(t)\}_{i=1}^N$, governed by
\begin{equation}
\label{eq:S2-full}
  \dot{x}_i \;=\; F(x_i) \,+\, \sum_{j=1}^N A_{ij}\,G(x_i,x_j), 
  \qquad i=1,\dots,N,
\end{equation}
where $\mathbf{A}=[A_{ij}]$ is the adjacency matrix for an undirected graph; treatment for directed graphs is analogous.

We assume that the reduction is constructed about a selected equilibrium $\mathbf{x}^\ast$ of the full system. After shifting coordinates so that this equilibrium is mapped to the origin, we write the shifted dynamics again in the form of Eq.~\eqref{eq:S2-full}. In the SIS-type setting considered here, this yields $F(0)=0$ and $G(0,0)=0$, and hence the Taylor expansions below start at first order. In the more general case where the local and coupling terms do not vanish separately at the equilibrium, the same derivation applies after expanding the shifted full vector field about the translated origin.

\begin{align}
  F(x) &= \sum_{k\ge 1} F^{(k)} x^k, \nonumber\\
  G(x,y) &= G^{(1,0)}x + G^{(0,1)}y \nonumber\\
  &\quad + \sum_{k\ge1,\;\ell\ge1} G^{(k,\ell)} x^k y^\ell.
\end{align}
Collecting terms gives the vector form
\begin{equation}
\label{eq:S2-compact}
  \dot{\mathbf{x}} \;=\; \hat{\mathbf{A}}\,\mathbf{x}
  \;+\; \sum_{k\ge 2} F^{(k)}\,\mathbf{x}^{k}
  \;+\; \sum_{k\ge 1,\;\ell\ge 1} G^{(k,\ell)}\;\mathbf{x}^{k} * \big(\mathbf{A}\,\mathbf{x}^{\ell}\big),
\end{equation}
where $(\mathbf{x}^k)_i=x_i^k$, $*$ is the Hadamard (element-wise) product, and
\begin{equation*}
\begin{aligned}
  \hat{\mathbf{A}} \;=\;&\; F^{(1)}\mathbf{I} \;+\; G^{(1,0)}\mathbf{D} \;+\; G^{(0,1)}\mathbf{A},\\
  \mathbf{D}=\;&\;\mathrm{diag}(d_1,\dots,d_N),\; d_i=\sum_j A_{ij}.
\end{aligned}
\end{equation*}
Let $(\lambda,\mathbf{u})$ be the leading eigenpair of $\hat{\mathbf{A}}$, i.e., $\lambda$ is the eigenvalue with the largest real part. For the undirected networks considered in this derivation, we normalize the eigenvector as $\mathbf{u}^{\top}\mathbf{u}=1$ and use the graph-style gauge $\mathbf{u}^{\top}\mathbf{w}_k=0$ for all higher-order lifting coefficients $k\ge2$. Under generic nonresonance with the remaining spectrum \cite{haller2016}, we obtain the existence and uniqueness of a one-dimensional SSM tangent to $\mathrm{span}\{\mathbf{u}\}$ at the origin.

We parametrize this one-dimensional SSM using the same notation as in the main text. Thus, the scalar intrinsic coordinate is denoted by $\eta\in\mathbb{R}$, the lifting map is $\mathbf{W}(\eta)$, and the autonomous reduced dynamics are denoted by $R(\eta)$. In the present one-dimensional setting, the general SSM representation $\dot{\eta}=R(\eta)$ and $\mathbf{x}=\mathbf{W}(\eta)$ takes the Taylor form
\begin{equation}
\label{eq:S2-embed-reduced}
\begin{aligned}
  \mathbf{W}(\eta) 
  &= \mathbf{u}\,\eta + \sum_{k\ge 2} \mathbf{w}_k\,\eta^k,\\
  R(\eta) 
  &= \lambda\,\eta + \sum_{k\ge 2} r_k\,\eta^k.
\end{aligned}
\end{equation}
Here, $\mathbf{w}_k\in\mathbb{R}^N$ are the coefficients of the lifting map and $r_k\in\mathbb{R}$ are the coefficients of the scalar reduced dynamics. This notation is the one-dimensional specialization of the expansion used in Sec.~\ref{sec:ssm_method_main}.

Substituting Eq.~\eqref{eq:S2-embed-reduced} into Eq.~\eqref{eq:S2-full}, or equivalently into Eq.~\eqref{eq:S2-compact}, gives the invariance equation
\begin{equation*}
\begin{aligned}
  \mathrm{D}\mathbf{W}(\eta)\,R(\eta) \;=\;&\; \hat{\mathbf{A}}\,\mathbf{W}(\eta)
  \;+\; \sum_{k\ge 2} F^{(k)}\,\mathbf{W}(\eta)^{k} \\
  &\;+\; \sum_{k\ge 1,\;\ell\ge 1} G^{(k,\ell)}\;\mathbf{W}(\eta)^{k} * \big(\mathbf{A}\,\mathbf{W}(\eta)^{\ell}\big),
\end{aligned}
\end{equation*}
which can be solved recursively for increasing powers of $\eta$ to determine the unknown coefficients $\{\mathbf{w}_k\}$ and $\{r_k\}$ in an automated fashion \cite{ssmtool26}. We provide the explicit solution up to cubic order below.

\paragraph{Order $\eta^1$.}
This yields $\hat{\mathbf{A}}\mathbf{u}=\lambda\mathbf{u}$ by construction.

\paragraph{Order $\eta^2$.}
Collecting all $\eta^2$ terms gives
\begin{equation}
\label{eq:S2-order2-main}
\begin{aligned}
  r_2\,\mathbf{u} \;+\; 2\lambda\,\mathbf{w}_2
  \;=\;&\; \hat{\mathbf{A}}\,\mathbf{w}_2
  \;+\; F^{(2)}\,\mathbf{u}^{2} \\
  &\;+\; G^{(1,1)}\,\mathbf{u} * \big(\mathbf{A}\mathbf{u}\big).
\end{aligned}
\end{equation}
Projecting onto $\mathbf{u}^\top$ fixes the reduced coefficient
\begin{equation}
\label{eq:S2-c2}
\begin{aligned}
  r_2 \;=\;&\; F^{(2)}\,\mathbf{u}^\top \mathbf{u}^{2}
  \;+\; G^{(1,1)}\,\mathbf{u}^\top \big(\mathbf{u} * (\mathbf{A}\mathbf{u})\big),
\end{aligned}
\end{equation}
after which the second-order lifting coefficient $\mathbf{w}_2$ follows from
\begin{equation}
\label{eq:S2-x2}
\begin{aligned}
  \mathbf{w}_2 \;=\;&\; \big(\hat{\mathbf{A}}-2\lambda\,\mathbf{I}\big)^{-1}
  \left[\, r_2\,\mathbf{u} \;-\; F^{(2)}\,\mathbf{u}^{2} \right.\\
  &\left.\qquad\qquad -\; G^{(1,1)}\,\mathbf{u} * (\mathbf{A}\mathbf{u}) \,\right].
\end{aligned}
\end{equation}

\paragraph{Order $\eta^3$.}
Collecting $\eta^3$ terms yields
\begin{equation}
\label{eq:S2-order3-main}
\begin{split}
  r_3\,\mathbf{u} \;+\; 3\lambda\,\mathbf{w}_3 \;+\; 2\,\mathbf{w}_2\,r_2
  \;=\;&\; \hat{\mathbf{A}}\,\mathbf{w}_3 \\
  &\;+\; 2F^{(2)}\,\mathbf{u}*\mathbf{w}_2 \;+\; F^{(3)}\,\mathbf{u}^{3} \\
  &\;+\; G^{(1,1)}\!\Big(\mathbf{u}*(\mathbf{A}\mathbf{w}_2)
  +\mathbf{w}_2*(\mathbf{A}\mathbf{u})\Big) \\
  &\;+\; G^{(1,2)}\,\mathbf{u}*(\mathbf{A}\mathbf{u}^{2}) \\
  &\;+\; G^{(2,1)}\,\mathbf{u}^{2}*(\mathbf{A}\mathbf{u}) .
\end{split}
\end{equation}
Projecting onto $\mathbf{u}^\top$ gives
\begin{equation}
\label{eq:S2-c3}
\begin{split}
  r_3 \;=\;&\; \mathbf{u}^\top\!\Big[-2\,\mathbf{w}_2\,r_2
  \;+\; 2F^{(2)}\,\mathbf{u}*\mathbf{w}_2
  \;+\; F^{(3)}\,\mathbf{u}^{3} \\
  &\qquad\qquad\;\;+\; G^{(1,1)}\!\Big(\mathbf{u}*(\mathbf{A}\mathbf{w}_2)
  +\mathbf{w}_2*(\mathbf{A}\mathbf{u})\Big)\Big] \\
  &\;+\; \mathbf{u}^\top\!\Big[G^{(1,2)}\,\mathbf{u}*(\mathbf{A}\mathbf{u}^{2})
  \;+\; G^{(2,1)}\,\mathbf{u}^{2}*(\mathbf{A}\mathbf{u})\Big] .
\end{split}
\end{equation}
and then
\begin{equation}
\label{eq:S2-x3}
\begin{split}
  \mathbf{w}_3 \;=\;&\; \big(\hat{\mathbf{A}}-3\lambda\,\mathbf{I}\big)^{-1}\!
  \Big[
    r_3\,\mathbf{u}
    + 2\,\mathbf{w}_2\,r_2
    - 2F^{(2)}\,\mathbf{u}*\mathbf{w}_2 \\
    &\qquad\quad
    - F^{(3)}\,\mathbf{u}^{3}
    - G^{(1,1)}\!\Big(\mathbf{u}*(\mathbf{A}\mathbf{w}_2)
    +\mathbf{w}_2*(\mathbf{A}\mathbf{u})\Big) \\
    &\qquad\quad
    - G^{(1,2)}\,\mathbf{u}*(\mathbf{A}\mathbf{u}^{2})
    - G^{(2,1)}\,\mathbf{u}^{2}*(\mathbf{A}\mathbf{u})
  \Big] .
\end{split}
\end{equation}

Proceeding inductively, the $\eta^k$ balance yields a linear homological problem for the unknown lifting coefficient $\mathbf{w}_k$ and reduced coefficient $r_k$. The lower-order coefficients $\{\mathbf{w}_j\}_{j<k}$ and $\{r_j\}_{j<k}$ determine the known forcing terms at order $k$; projection onto the master direction fixes $r_k$, and the remaining component determines $\mathbf{w}_k$ on the complementary subspace.
In practice one truncates~\eqref{eq:S2-embed-reduced} at order $p$, obtaining
\[
\begin{aligned}
  R(\eta) 
  &= \lambda \eta + r_2 \eta^2 + \cdots + r_p \eta^p,\\
  \mathbf{W}(\eta) 
  &= \mathbf{u}\eta + \mathbf{w}_2 \eta^2 + \cdots + \mathbf{w}_p \eta^p.
\end{aligned}
\]
Quadratic truncation already yields a closed-form prediction for $\eta(t)$ and the nontrivial fixed level $\eta_\infty=-\lambda/r_2$ when $\lambda>0$ and $r_2\neq 0$; higher orders improve quantitative accuracy and extend the local validity.
Throughout this work, we denote by $p$ the truncation order of the Taylor expansion used in the SSM construction, e.g., $O(2)$, $O(5)$, $O(10)$, $O(15)$, or $O(20)$, and refer to each case as the corresponding $O(p)$--SSM reduction.

All coefficient solves above are linear and sparse once $\hat{\mathbf{A}}$ and the coefficients of nonlinearities are assembled. We rely on a graph-style parameterization (no re-centering) and compute $\{\mathbf{w}_k\}$ and $\{r_k\}$ up to order $p$ via order-by-order homological solves. For automating these computations, we have employed the MATLAB-based package \textsc{SSMTool} \cite{ssmtool26} that enables scalable SSM computations in physical coordinates with minimal eigenvectors, and provides diagnostics for spectral gaps and internal resonances.

\section{Globalization of invariant manifolds via Padé approximation}
\label{sec:s2}

This Appendix provides implementation-level details for the gSSM construction summarized in Sec.~\ref{sec:ssm_method_main}. A classical strategy for analytic continuation of a function near a specific point is its Padé approximant, which replaces a truncated Taylor series by a rational function capable of representing singularities that limit the convergence of pure polynomials. In our setting, the reduced dynamics on an SSM, computed upto any desired order, are extended by constructing a rational approximation of the form \cite{kaszas2025globalizing}
 
\begin{equation}
\widehat{F}(z) \;=\; 
\frac{\sum_{|\alpha|\le N} a_\alpha z^\alpha}{\,1 + \sum_{|\beta|\le M} b_\beta z^\beta}, 
\qquad b_0=1,\quad z\in\mathbb{R}^\ell,
\end{equation}
where the numerator and the denominator are multivariate polynomials of chosen degrees $N$ and $M$. The coefficients are fixed so that the series of $\widehat{F}$ matches the Taylor expansion of the reduced dynamics up to order $N+M$. This construction applies to both univariate and multivariate cases. 

Diagonal Padé approximants $[M/M]$, in which numerator and denominator have the same degree, are particularly effective and are closely related to continued–fraction representations in the univariate case~\cite{wall1948analytic,baker1996pade,cuyt2008handbook}. For meromorphic functions with unknown pole structure, classical convergence results (e.g., de Montessus–type theorems and modern refinements) ensure that appropriate Padé sequences converge almost everywhere on compact sets, with exceptions corresponding to the zero sets of the denominator~\cite{demontessus1902,baker1996pade,brezinski1991pade}. Analogous constructions and guarantees extend to multivariate and vector settings via rational/vector Padé frameworks~\cite{gravesmorris1979vector,baker1996pade,cuyt2008handbook}. We adopt this theory in our context and use the data–driven globalization strategy in~\cite{kaszas2025globalizing}.

We use this procedure to obtain globalized SSMs (gSSMs): the local Taylor coefficients from the invariance equation are preserved near the origin, while the rational form extends the validity of the reduced dynamics far beyond the Taylor radius. Throughout this work, unless noted otherwise, we adopt diagonal Padé approximants of type $[8/7]$ constructed from the order-15 Taylor expansion of the reduced vector field. This choice has proven effective in balancing accuracy and robustness across all network models studied.

%==========================================================================
% SUPPLEMENTAL MATERIAL
%==========================================================================
\newpage
\onecolumngrid

% Reset counters and use S-numbering for supplemental material
\setcounter{section}{0}
\setcounter{subsection}{0}
\setcounter{equation}{0}
\setcounter{figure}{0}
\setcounter{table}{0}
\renewcommand{\thesection}{S\arabic{section}}
\renewcommand{\thesubsection}{S\arabic{section}.\arabic{subsection}}
\renewcommand{\theequation}{S\arabic{equation}}
\renewcommand{\thefigure}{S\arabic{figure}}
\renewcommand{\thetable}{S\arabic{table}}

\begin{center}
{\large\textbf{Supplemental Material for ``Nonlinear Model Reduction of Complex Networks via Spectral Submanifolds''}}\\[0.5em]
Kaviya Bhaskaran, Shobhit Jain, Mingwu Li
\end{center}

\section{Effect of initialization: aligned, orthogonal, and near-aligned comparisons}
\label{sec:s9}
\begin{figure}[!htpb]
    \centering
    \includegraphics[width=\linewidth]{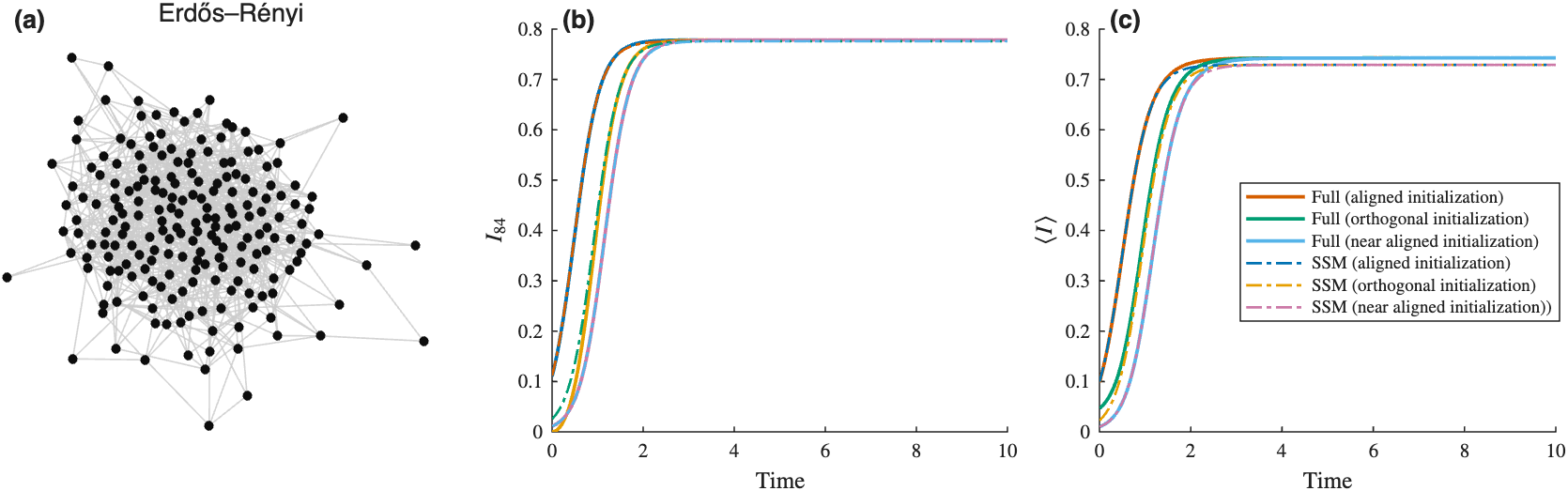}
    \caption{\textbf{Effect of initialization on the reduced dynamics.}
    (a) Erd\H{o}s--R\'enyi network with $N=200$ nodes and throughout all three simulations, the $\eta_0$ was initialized as $1.5$.
    (b) Representative node trajectory $I_{84}(t)$ for the full system launched
    from the aligned state (solid black), compared with reduced $O(20)$ SSM
    trajectories initialized on the manifold (solid blue), from an off-manifold
    projection (dashed orange), and from a small-shift projection (dash--dotted
    magenta). (c) Mean prevalence $\langle I \rangle (t)$ under the same color
    scheme. The reduced trajectories reproduce the slow dynamics accurately
    after the initial contraction, demonstrating the SSM’s robustness to
    initialization choices while clarifying that the early fast transient is
    inherently absent from the reduced model. }
    \label{fig:sis_init_ic}
\end{figure}

In Fig.~\ref{fig:sis_init_ic}, we examine how different initialization strategies
influence the transient and long-term dynamics on an Erd\H{o}s--R\'enyi network.
All simulations use the same model parameters and reduced order $O(20)$ as in the
main text. We prepared three representative initial conditions using the same
notation as in main text.

We therefore refer to the three cases as:
\begin{enumerate}
    \item[\textbf{(i)}] \textbf{Aligned initialization} (dominantly on the slow manifold),  
    $\mathbf{z}_0^{(\mathrm{A})} = \eta_0 \mathbf{v}_1$, where $\mathbf{v}_1$ is
    the leading right eigenvector of the linearized operator, tangent to the
    one-dimensional SSM.
\item[\textbf{(ii)}] \textbf{Orthogonal initialization} (transverse to the slow manifold), $\mathbf{z}_0^{(\mathrm{B})} = \eta_0 \mathbf{r}$, where
    $\mathbf{r} \perp \mathbf{v}_1$ is chosen to be Euclidean-orthogonal to the
    manifold tangent direction.

    \item[\textbf{(iii)}] \textbf{Near-aligned initialization} (slightly displaced from the manifold),  
    $\mathbf{z}_0^{(\mathrm{C})} = \eta_0 \mathbf{v}_1 + \varepsilon \mathbf{r}$,
    representing a state close to the slow manifold but perturbed by a small
    orthogonal component with $\varepsilon \ll \eta_0$.
\end{enumerate}

%------------------------------------------------------------

The full system was integrated only from the aligned state
$\mathbf{z}_0^{(\mathrm{A})}$, while the reduced dynamics on the
$O(20)$ SSM were launched from the three reduced coordinates corresponding to
cases (i)--(iii) above, i.e., either directly from $q_0$ or from the projections
of $\mathbf{z}_0^{(\mathrm{B})}$ and $\mathbf{z}_0^{(\mathrm{C})}$ onto the
slow coordinate. The network layout fig.~\ref{fig:sis_init_ic}(a), representative node-level
trajectory $I_{i^*}(t)$ fig.~\ref{fig:sis_init_ic}(b), and mean prevalence
$\langle I \rangle(t)$ fig.~\ref{fig:sis_init_ic}(c) are shown for comparison.

The blue curves, corresponding to the aligned initialization, show that
the full and reduced trajectories overlap from the start, indicating
that the initial state lies entirely on the slow manifold governed by
the dominant SSM coordinate~$\eta$.
The orange trajectories, corresponding to the orthogonal
initialization, exhibit a clear delay in activation: the state initially
evolves along fast stable directions before relaxing toward the slow
subspace, resulting in a later onset of infection growth.
The magenta curves, representing the near-aligned initialization,
display only a short transient deviation before converging to the same
trajectory as the aligned case, confirming that small displacements from
the manifold have minimal long-term influence.

The dash--dotted SSM trajectories in each color family start directly on
the manifold and thus do not capture the early fast contraction phase,
but they accurately reproduce the slow evolution once the transient
subsides.
After this short alignment period, all reduced and full trajectories
coincide on the same slow-manifold path.
These results demonstrate that the one-dimensional SSM reliably captures
the long-term network dynamics while clarifying that the fast-mode
relaxation observed in the full system governs only the brief early-time
approach to~$\mathcal{W}_{\mathrm{SSM}}$.

\begin{figure}[!htpb]
    \centering
    \includegraphics[width=\linewidth]{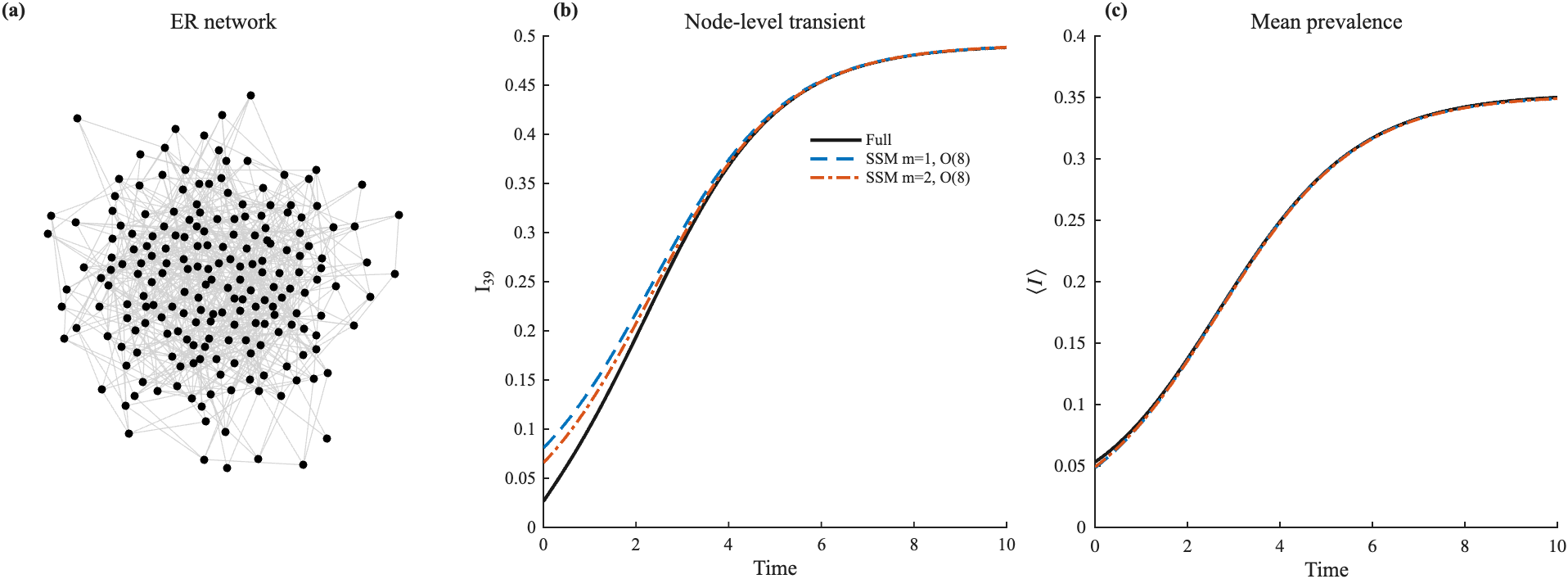}
    \caption{
    (a) Erd\H{o}s--R\'enyi network with $N=200$ nodes. The full initial condition is prescribed directly in physical coordinates and displaced from the reduced manifold by a perturbation along a third mode, while the reduced initial coordinates are obtained by projection onto the $m=1$ and $m=2$ modal subspaces.  (b) Representative node-level trajectory $I_{i}(t)$ for the full system (solid black), compared with the $m=1$ reduced trajectory (dashed blue) and the $m=2$ reduced trajectory (dash--dotted orange). The two-dimensional reduction yields a more accurate early-time approximation than the one-dimensional reduction, although the improvement is modest in the ER case.    (c) Mean prevalence $\langle I \rangle(t)$ under the same color scheme. At the macroscopic level, the difference between the $m=1$ and $m=2$ reductions is largely suppressed by averaging. In all cases, both reduced trajectories approach the same long-time slow dynamics and recover the same steady behavior as the full system.}
    \label{fig:sis_er_offmanifold_m1m2}
\end{figure}

\begin{figure}[!htpb]
    \centering
    \includegraphics[width=\linewidth]{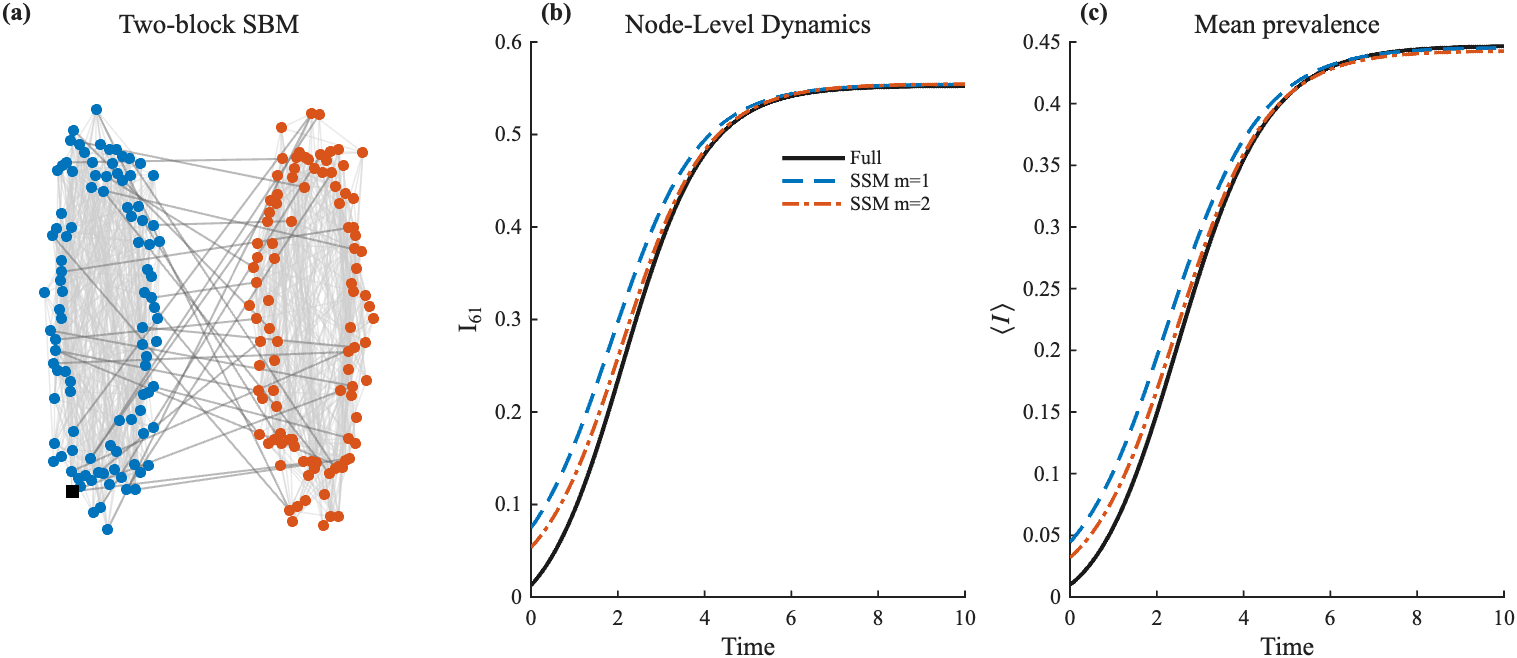}
    \caption{
    (a) Two-block SBM with $N=200$ nodes. As in Fig.~\ref{fig:sis_er_offmanifold_m1m2}, the full initial condition is prescribed directly in physical coordinates and displaced from the reduced manifold by a perturbation along a third mode, while the reduced initial coordinates are obtained by projection onto the $m=1$ and $m=2$ modal subspaces. (b) Representative node-level trajectory $I_{i}(t)$ for the full system (solid black), compared with the $m=1$ reduced trajectory (dashed blue) and the $m=2$ reduced trajectory (dash--dotted orange). In this more modular and less homogeneous setting, the transient benefit of the $m=2$ reduction is more pronounced than in the ER case. (c) Mean prevalence $\langle I \rangle(t)$ under the same color scheme. The difference between the $m=1$ and $m=2$ reductions is smaller than at node level but remains consistent with the stronger transient role of the second mode in the modular network. At longer times, both reduced trajectories converge to the same slow dynamics and recover the same steady behavior as the full system.}
    \label{fig:sis_sbm_offmanifold_m1m2}
\end{figure}

\subsection{Off-manifold initialization and transient benefit of $m=2$}

In this subsection, we examine whether increasing the reduced dimension from $m=1$ to $m=2$ improves transient prediction when the full initial condition is chosen off the reduced manifold. This directly addresses the question of whether higher-dimensional reductions provide a measurable benefit beyond the asymptotic one-dimensional slow dynamics emphasized in the main text.

For this test, we considered both an Erd\H{o}s--R\'enyi (ER) network and a two-block stochastic block model (SBM). In each case, we first constructed a two-dimensional base state from the leading two modal directions and then displaced the full initial condition off the reduced manifold by adding a perturbation along a third mode. Thus, the full-state initial condition was prescribed directly in physical coordinates and was not restricted to lie on either the one- or two-dimensional reduced manifold. The reduced initial coordinates were then obtained by projection of this same full-state initial condition onto the corresponding one-dimensional and two-dimensional modal subspaces. Consequently, the comparison tests how well the $m=1$ and $m=2$ reductions reproduce the transient evolution from the same off-manifold physical state.

Figures~S2 and~S3 summarize the resulting comparisons. Figure~S2 reports the ER case, while Fig.~S3 reports the two-block SBM case. In each figure, panel~(a) shows the network topology, panel~(b) shows a representative node-level trajectory, and panel~(c) shows the mean prevalence. The representative node was chosen so that the contribution of the second mode is appreciable, making the transient difference between the $m=1$ and $m=2$ reductions visible at node level.

The main observation is that the $m=2$ reduction yields a more accurate early-time approximation than the $m=1$ reduction. In the ER network (Fig.~S2), this improvement is relatively modest and is clearest in the node-level trajectory: the $m=2$ curve stays closer to the full system during the initial transient, whereas in the mean prevalence the difference between the two reductions is largely suppressed by averaging. In the SBM network (Fig.~S3), the transient benefit of $m=2$ is more pronounced. Here the modular structure makes the second mode more dynamically relevant, so the $m=2$ reduction provides a visibly better approximation than $m=1$ during the approach to the slow manifold.

At the same time, these tests confirm that the long-time behavior remains effectively one-dimensional for the cases considered here. Although $m=2$ improves the transient approximation, both the $m=1$ and $m=2$ reductions eventually merge with the same slow dynamics and recover the same steady behavior as the full system. Thus, the role of $m=2$ in these examples is not to change the asymptotic prediction, but to improve short-time accuracy under off-manifold initialization. This explains why the one-dimensional reductions remain sufficient for the main manuscript figures, while the additional $m=2$ results are most naturally presented as a supplementary transient test. In summary, these additional results show that higher-dimensional reductions can be beneficial when the full system is initialized away from the reduced manifold, and that this effect is most clearly visible at node level and in less homogeneous networks.

\section{Network construction, preprocessing, and numerical setup}
\label{sec:s3}

\subsection{Synthetic topologies.}
All experiments use undirected, unweighted, simple graphs on $N=200$ nodes with no self-loops. We consider three canonical ensembles:
\textit{(i) Erd\H{o}s--R\'enyi (ER):} each unordered pair is present independently with probability $p=\tfrac{2K}{N-1}$, yielding expected degree $\approx 2K$; we use $K=4$ (so $\mathbb{E}[\mathrm{deg}]\approx 8$). 
\textit{(ii) Scale-free (SF):} a target degree sequence is sampled from a truncated power law with exponent $\gamma=3$ ; a configuration-model pairing of stubs produces a simple graph, followed by a minimal connectivity fix (if multiple components occur, we connect each to the largest component by a single edge). 
\textit{(iii) Small-world (SW):} a ring lattice with $2K_{\text{neigh}}$ nearest neighbors (we use $K_{\text{neigh}}=4$ for mean degree $8$) is rewired with probability $p_{\text{rew}}=0.10$ per existing edge (Watts--Strogatz--like). 
For the targeted modular-bottleneck benchmark introduced in the revised manuscript, we additionally consider a two-block stochastic block model (SBM) with equal community sizes $N/2$ and matched expected mean degree. Writing $p_{\mathrm{out}}=\rho\,p_{\mathrm{in}}$ with $\rho<1$ for the inter- to intra-community connection ratio, we choose $p_{\mathrm{in}}$ so that $(N/2-1)p_{\mathrm{in}}+(N/2)p_{\mathrm{out}}\approx 8$, and then fix $\rho$ to obtain weak inter-community coupling and hence a clearly bottlenecked modular structure.
Adjacency matrices $A$ are symmetric $\{0,1\}$-valued.

\subsection{Empirical networks (SocioPatterns).}
We use three real contact networks from the SocioPatterns collaboration: (i) a hospital ward in Lyon, France (Dec.\ 6–10, 2010), recording face-to-face proximity among patients and health-care workers; (ii) an office building in France (Jun.\ 24–Jul.\ 3, 2013), capturing workplace contacts; and (iii) a rural village in Malawi, providing observational contact data at the household/community scale.
For each dataset, we symmetrize and binarize the temporal edges over the observation window and analyze the largest connected component.
The hospital data span 46 health-care workers and 29 patients over \(\sim\)72 hours, as documented in the released metadata.
Dataset details and access: Hospital ward, Workplace, and Rural Malawi (SocioPatterns). Temporal SocioPatterns edge lists are first aggregated into static weighted adjacency matrices by summing contact durations (or counts). These matrices are then symmetrized, binarized, and restricted to the largest connected component before performing the $\beta$–sweeps.
\footnote{SocioPatterns datasets: Hospital ward (Lyon, 2010) \cite{sp:hospital};
Workplace (France, 2013) \cite{sp:workplace};
Rural Malawi village \cite{sp:rural}.
Hospital participant counts from the CRAN mirror of the SocioPatterns documentation \cite{cran:hospital}.}

\begin{table*}[!htpb]
	\caption{Global numerical settings used in all simulations.}
	\label{tab:global_settings}
	\begin{ruledtabular}
		\begin{tabular}{lll}
			Item & Description & Value / note \\
			\colrule
			Time horizon   & Simulation time $t_f$, samples
			& $t_f \in [20,100]$; $n_{\text{steps}} = 10^{3}$--$1.5\times10^{3}$ \\
			SSM orders     & Taylor truncations
			& $O(2),\,O(5),\,O(10),\,O(15),\,O(20)$ \\
			Initialization & Along dominant eigenvector
			& $\eta_{0} \in [0.01,0.05]$ \\
			Integrator     & Solver
			& MATLAB \texttt{ode45}, default tolerances \\
			Networks       & Realizations
			& Fixed per topology, reused in all tests \\
		\end{tabular}
	\end{ruledtabular}
\end{table*}

%------------------------------------------------------------------------------
\subsection{Global numerical settings}

All models share the same numerical setup for time-integration and reduction.  
These global settings are listed in Table~\ref{tab:global_settings}.  
For consistent comparisons, the same network realizations were used across all experiments.

\section{Error metrics for SIS time series (full vs.\ SSM)}
\label{sec:s4}

\paragraph{Setup and notation.}
Let $\{t_m\}_{m=1}^{M}\subset[0,T]$ be the common time grid used for comparisons.  
The trajectory of node $i$ in the full system is denoted $x_i^{\mathrm{ref}}(t_m)$, while the reduced dynamics trajectory on an $O(p)$–SSM reduction is $\hat{x}_i^{(p)}(t_m)$, obtained by interpolation over the same grid.  
The mean prevalence is 
\[
\langle I\rangle(t)=\frac{1}{N}\sum_{i=1}^{N}x_i(t),
\]
with reduced counterpart $\widehat{\langle I\rangle}^{(p)}(t)$.  
The degree of node $i$ is $k_i=\sum_j A_{ij}$. To evaluate the accuracy of the $O(p)$–SSM reductions, we employ five complementary metrics:  

1. \textbf{Macroscopic mean-squared error (MSE): Mean error vs.\ order.}  
This measures the mean-squared difference between the reduced and full-system prevalence curves,  
\[
\mathrm{MSE}_{\langle I\rangle}^{(p)}=\frac{1}{M}\sum_{m=1}^{M}\Big(\widehat{\langle I\rangle}^{(p)}(t_m)-\langle I\rangle^{\mathrm{ref}}(t_m)\Big)^2.
\]  
$\mathrm{MSE}_{\langle I\rangle}^{(p)}$ quantifies how well the reduced dynamics captures the global infection level; decreasing values with increasing $p$ indicate systematic convergence.

2. \textbf{Degree–error relation.}  
For any node $i$, we compute the absolute steady-state error,  
\[
\epsilon_i^{\mathrm{ss},(p)}=\bigl|\hat{x}_i^{(p)}(T)-x_i^{\mathrm{ref}}(T)\bigr|,
\]  
and plot it against the node degree $k_i$. This shows whether errors are concentrated in low-degree nodes (often harder to approximate) or spread more uniformly across the network.

3. \textbf{Nodewise steady-state distribution.}  
The values for  $\epsilon_i^{\mathrm{ss},(p)}$ above are summarized across all nodes as boxplots for each order $p$.  This provides a compact view of how the distribution of steady-state errors tightens with increasing SSM order.

4. \textbf{Time-to-steady-state (TSS).}  
We define the relaxation time as the point when the trajectories remain within a fixed tolerance $\tau$ of their final steady state. For the full and reduced systems,  
\[
\mathrm{TSS}_{\mathrm{full}}=\min\Bigl\{t:\tfrac{1}{N}\sum_{i=1}^{N}|x_i^{\mathrm{ref}}(t)-x_{i,\infty}^{\mathrm{ref}}|\le\tau\Bigr\},
\]  
\[
\mathrm{TSS}_{\mathrm{red}}^{(p)}=\min\Bigl\{t:\tfrac{1}{N}\sum_{i=1}^{N}|\hat{x}_i^{(p)}(t)-\hat{x}_{i,\infty}^{(p)}|\le\tau\Bigr\}.
\]  
Comparing these times indicates how well the $O(p)$–SSM reduction reproduces the transient convergence rate of the full system.

5. \textbf{Global MAE and spatial correlation.}  
The global mean absolute error (MAE) is defined as  
\[
\mathrm{MAE}^{(p)}=\frac{1}{MN}\sum_{m=1}^{M}\sum_{i=1}^{N}\big|\hat{x}_i^{(p)}(t_m)-x_i^{\mathrm{ref}}(t_m)\big|,
\]  
while the spatial correlation compares steady-state values node-by-node as 
\[
\mathrm{Corr}^{(p)}_{\mathrm{sp}}=\mathrm{corr}\!\bigl(\hat{\mathbf{x}}^{(p)}(T),\,\mathbf{x}^{\mathrm{ref}}(T)\bigr).
\]  
Together, these capture both the overall trajectory accuracy and whether the final spatial pattern across nodes is faithfully reproduced.

These metrics collectively evaluate our reduced models over multiple scales: the global prevalence curve (MSE), node-level steady states (degree–error and distributions), transient relaxation (TSS), and overall fidelity across time and space (MAE and correlation). All metrics are computed on the same network instances, with reduced trajectories reconstructed via $W^{(p)}$ to node-level coordinates and then resampled onto the full-system time grid for consistency.

\begin{figure*}[!htpb]
  \centering
  \includegraphics[width=\linewidth]{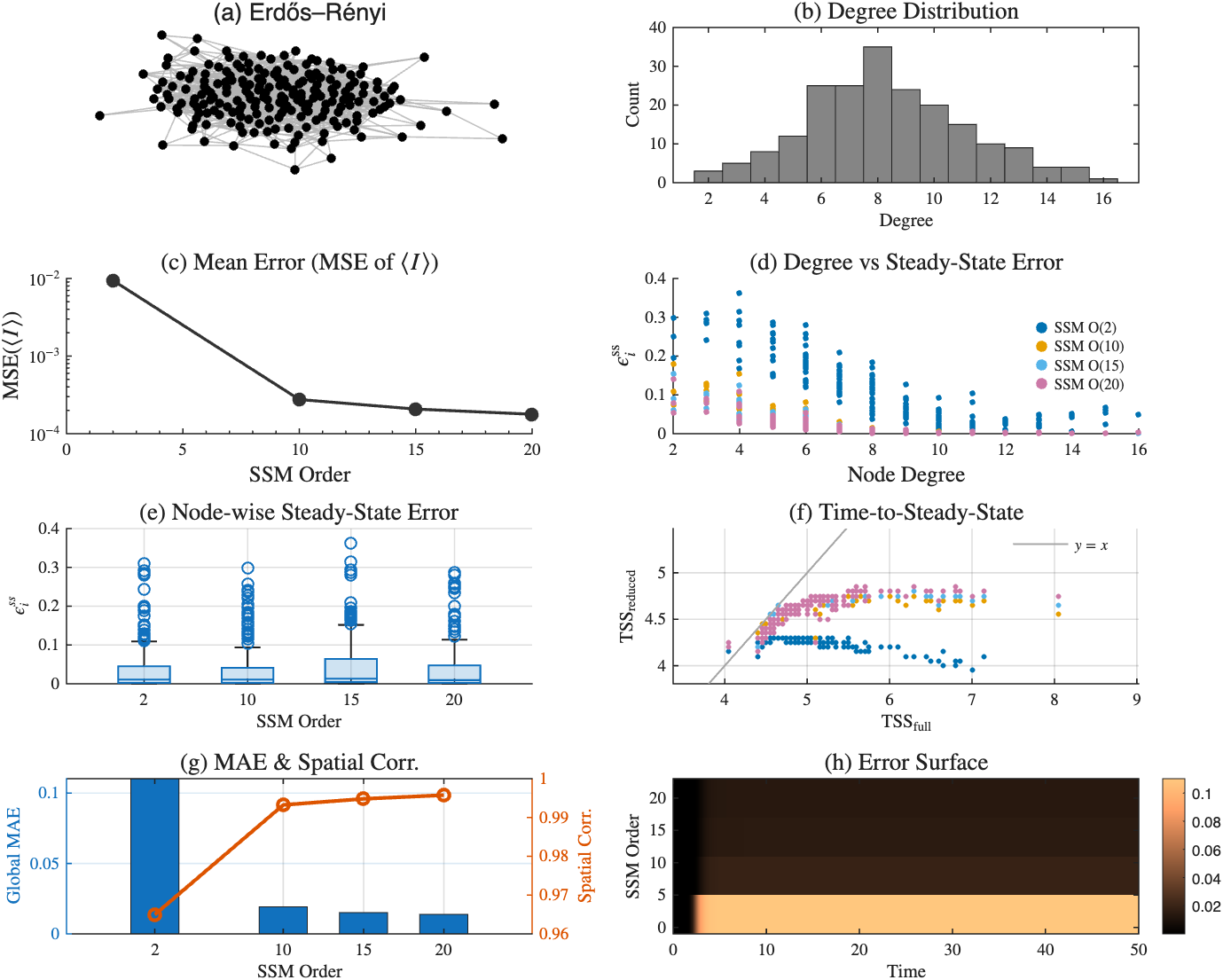}
  \caption{%
\textbf{ER error diagnostics.} 
(a) Network layout. 
(b) Degree distribution. 
(c) Mean-prevalence MSE vs.\ $O(p)$–SSM order. 
(d) Steady-state node error $\epsilon_i^{\mathrm{ss}}$ vs.\ degree $k_i$, shown as scatter points in dark blue ($O(2)$), orange ($O(10)$), blue ($O(15)$), and pink ($O(20)$). 
(e) Boxcharts of $\epsilon_i^{\mathrm{ss}}$ grouped by order. 
(f) Time-to-steady-state: $\mathrm{TSS}_{\mathrm{red}}$ vs.\ $\mathrm{TSS}_{\mathrm{full}}$, using the same color scheme as panel (d). 
(g) Global MAE (blue bars, left axis) and steady-state spatial correlation (orange line, right axis) vs.\ order. 
(h) Time–order error surface $E^{(p)}(t)$ with colormap representing residual error values.}
  \label{fig:si_er_metrics}
\end{figure*}

\begin{figure*}[!htpb]
  \centering
 \includegraphics[width=\textwidth]{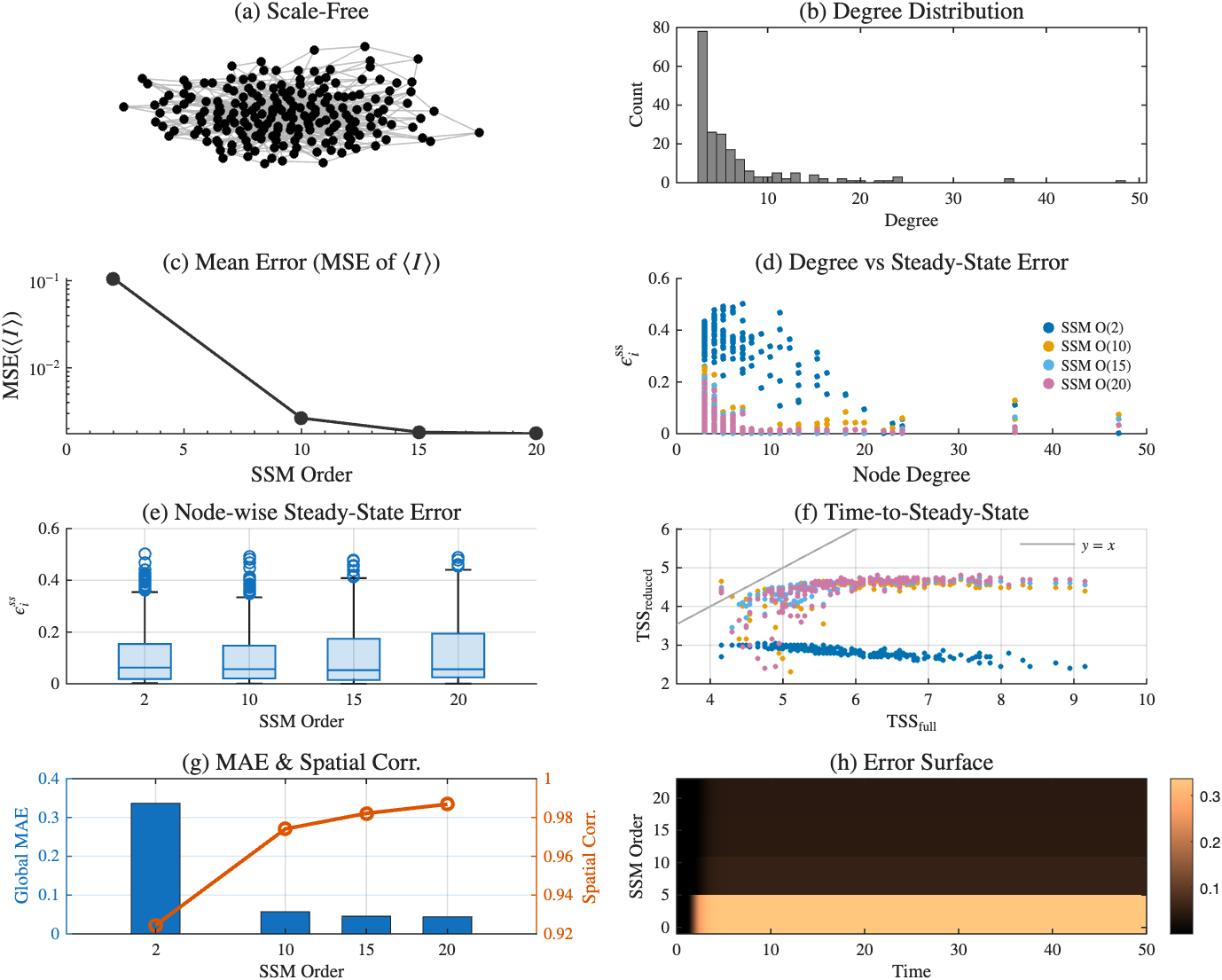}
  \caption{%
\textbf{SF error diagnostics.} 
(a) Network layout. 
(b) Degree distribution. 
(c) Mean-prevalence MSE vs.\ $O(p)$–SSM order. 
(d) Steady-state node error $\epsilon_i^{\mathrm{ss}}$ vs.\ degree $k_i$, shown as scatter points in dark blue ($O(2)$), orange ($O(10)$), blue ($O(15)$), and pink ($O(20)$). 
(e) Boxcharts of $\epsilon_i^{\mathrm{ss}}$ grouped by order. 
(f) Time-to-steady-state: $\mathrm{TSS}_{\mathrm{red}}$ vs.\ $\mathrm{TSS}_{\mathrm{full}}$, using the same color scheme as panel (d). 
(g) Global MAE (blue bars, left axis) and steady-state spatial correlation (orange line, right axis) vs.\ order. 
(h) Time–order error surface $E^{(p)}(t)$ with colormap representing residual error values.}
  \label{fig:si_sf_metrics}
\end{figure*}

\begin{figure*}[!htpb]
  \centering
\includegraphics[width=\textwidth]{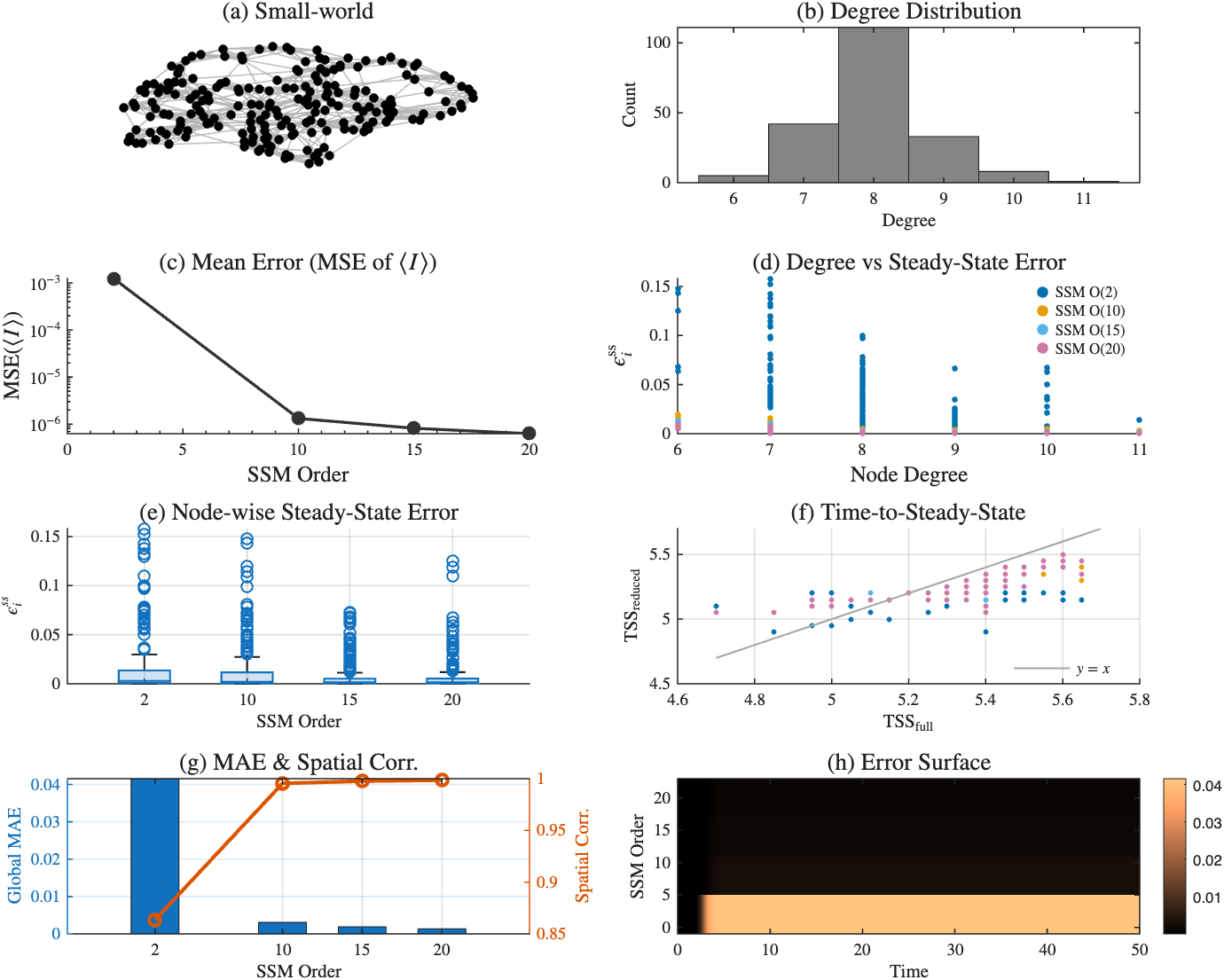}
  \caption{%
\textbf{SW error diagnostics.} 
(a) Network layout. 
(b) Degree distribution. 
(c) Mean-prevalence MSE vs.\ $O(p)$–SSM order. 
(d) Steady-state node error $\epsilon_i^{\mathrm{ss}}$ vs.\ degree $k_i$, shown as scatter points in dark blue ($O(2)$), orange ($O(10)$), blue ($O(15)$), and pink ($O(20)$). 
(e) Boxcharts of $\epsilon_i^{\mathrm{ss}}$ grouped by order. 
(f) Time-to-steady-state: $\mathrm{TSS}_{\mathrm{red}}$ vs.\ $\mathrm{TSS}_{\mathrm{full}}$, using the same color scheme as panel (d). 
(g) Global MAE (blue bars, left axis) and steady-state spatial correlation (orange line, right axis) vs.\ order. 
(h) Time–order error surface $E^{(p)}(t)$ with colormap representing residual error values.}
  \label{fig:si_sw_metrics}
\end{figure*}

\paragraph{Erd\H{o}s--R\'enyi (ER): error diagnostics.}
In the ER case, the network structure shown in Fig.~\ref{fig:si_er_metrics}a is visually homogeneous, and the degree histogram in panel~(b) confirms this by displaying a narrow, unimodal distribution centered near the mean degree. This structural regularity is reflected in the accuracy of the $O(p)$–SSM reductions. The macroscopic error in the mean prevalence (panel~c) decreases almost monotonically with increasing $p$, dropping by nearly two orders of magnitude between $O(2)$ and $O(20)$. By $O(10)$, the $O(p)$–SSM reduction already reproduces the full mean dynamics with negligible deviation, indicating that only moderate order is needed for accurate system-level predictions.  

At the node level, steady-state errors are strongly anti-correlated with degree (panel~d): low-degree nodes carry the bulk of the residual at low order, whereas high-degree nodes are approximated more faithfully from the start. This scatter decays rapidly with increasing $p$, and by $O(15)$–$O(20)$ the errors are nearly uniform across nodes. The same trend appears in the distribution of final errors across all nodes (panel~e): medians fall, interquartile ranges narrow, and outliers disappear with increasing order, demonstrating progressively tighter fidelity.  

Dynamical relaxation is also well captured at moderate orders. The time-to-steady-state comparison in panel~(f) shows that the lowest-order model ($O(2)$–SSM) relaxes slightly faster than the full system, while higher orders fall directly on the diagonal, confirming that the reduced and full models converge on the same timescale. At the global level (panel~g), the mean absolute error decreases steadily with order, while the steady-state spatial correlation approaches unity, showing that both overall levels and cross-node patterns are faithfully reproduced. Finally, the error surface in panel~(h) reveals where discrepancies persist: they are concentrated in early transients at low order and vanish almost entirely at higher orders. Together, these diagnostics demonstrate that the structural homogeneity of ER networks and the delocalization of their dominant spectral modes allow the $O(p)$–SSM reduction to rapidly converge. Moderate orders such as $O(10)$ are sufficient to recover both macroscopic observables and detailed node-level behavior with high accuracy.

\paragraph{Scale--Free (SF): error diagnostics.}
In the SF case, the network layout in Fig.~\ref{fig:si_sf_metrics}a clearly reveals hub nodes, and the heavy-tailed degree distribution in panel~(b) confirms strong structural heterogeneity. This variability directly affects model reduction. The macroscopic error in the mean prevalence (panel~c) decreases with $O(p)$–SSM order, dropping by roughly an order of magnitude by $O(10)$, but then levels off, reflecting slower convergence compared to the ER network due to the influence of hubs and peripheral nodes.  

At the node level, steady-state error versus degree (panel~d) shows a strong anti-correlation: low-degree peripheral nodes carry the largest residuals at low order, while the scatter progressively decays towards zero at higher orders. Small residuals can persist even for hubs, but they remain bounded and diminish with order. The overall distribution of steady-state errors (panel~e) narrows as the order increases, with reduced medians and spread, though the distribution remains broader than in the ER case, consistent with the heavy-tailed topology.  

Dynamical convergence is also topology-dependent. The time-to-steady-state analysis (panel~f) shows that the lowest-order $O(2)$–SSM relaxes too quickly relative to the full system, while higher orders ($O(10)$–$O(20)$) align more closely with the diagonal, correcting the relaxation time bias. At the global scale (panel~g), the mean absolute error steadily decreases with order, and the spatial correlation between full and reduced steady states rises to $\approx 0.98$–$1.00$ by mid-to-high orders, confirming that both global levels and cross-node structures are recovered. The error surface in panel~(h) localizes most discrepancies to early transients at low order, with little error persisting once the order is increased.  

Overall, the strong heterogeneity of SF networks slows the rate of Taylor-order convergence and makes low-degree nodes the most challenging to approximate. Nonetheless, by moderate-to-high orders, the $O(p)$–SSM reductions accurately reproduce both macroscopic trends and node-level patterns, offering robust fidelity even in heavy-tailed topologies.

\paragraph{Small--World (SW): error diagnostics.}
In the SW network, the layout in Fig.~\ref{fig:si_sw_metrics}a highlights clustered connectivity with a few shortcuts, while the narrow degree distribution in panel~(b) indicates only mild heterogeneity. This structure favors efficient reduction. The macroscopic error in the mean prevalence (panel~c) decreases rapidly with order: from about $10^{-3}$ at $O(2)$ to below $10^{-6}$ by $O(15)$–$O(20)$, with $O(10)$–SSM reductions already nearly indistinguishable from the full system. At the node level, steady-state error versus degree (panel~d) is small across all nodes and loses its weak degree dependence as the order increases; correspondingly, the overall distribution of errors (panel~e) becomes tighter, with lower medians and reduced spread.  

Dynamically, the time-to-steady-state analysis (panel~f) shows that $O(2)$–SSM relaxes slightly too quickly, while higher orders ($O(15)$–$O(20)$) fall close to the diagonal, reflecting accurate reproduction of relaxation times. Global error metrics reinforce this picture: the mean absolute error falls steadily with order, and the spatial correlation between reduced and full steady states (panel~g) approaches unity by mid order, confirming that both overall levels and cross-node structure are well preserved. The error surface in panel~(h) localizes residual discrepancies to early transients at low order, which vanish as the order increases. Together, these diagnostics illustrate that the moderate heterogeneity and delocalized modes of SW networks make them particularly well suited for $O(p)$–SSM reduction: accurate node-level and system-level dynamics are already achieved at moderate orders, with little to gain beyond $O(10)$.

\section{Spectral and modified spectral baselines}
\label{sec:s5}
This section summarizes the two eigenmode-based baselines used in the
$\beta/\gamma$ sweeps: a classical \textit{spectral} reduction and a \textit{modified
spectral} variant. Both project the SIS dynamics onto a one-dimensional
macroscopic variable aligned with an eigenvector of the adjacency matrix $A$ and
produce an explicit prediction for the final mean infection $\langle I\rangle$
as a function of $\beta/\gamma$. The implementations below corresponds to the  spectral reductions in the literature \cite{gao2016,laurence2019,masuda2022}.

\paragraph{Spectral. \cite{gao2016}}
Let $\lambda_1$ and $u^{(1)}$ denote the dominant eigenvalue and the
corresponding right eigenvector of $A$ (largest real part). Define the
degree vector $k=(k_1,\dots,k_N)^\top$ with $k_i=\sum_j A_{ij}$ and the
normalized weights
\[
a \;=\; \frac{u^{(1)}}{\mathbf{1}^\top u^{(1)}},\qquad
b \;=\; \frac{a\odot a}{\mathbf{1}^\top (a\odot a)} \, ,
\]
where $\odot$ denotes elementwise multiplication and $\mathbf{1}$ is the
all-ones vector. A degree-weighted factor
\[
\widehat{\beta}_s \;=\; \frac{k^\top b}{k^\top a}
\]
rescales the macroscopic projection. The resulting
spectral prediction for the final mean infection is
\begin{equation}
\widehat{\langle I\rangle}_{\mathrm{spec}}(\beta/\gamma)
\;=\; \max\!\left\{\,0,\; 1 - \frac{1}{(\beta/\gamma)\,\lambda_1\,\widehat{\beta}_s}\right\}.
\label{eq:spectral_rom}
\end{equation}
The linear onset is then $(\beta/\gamma)_c = 1/\lambda_1$; $\widehat{\beta}_s$
affects the saturation level but not the threshold.

\paragraph{Modified spectral. \cite{laurence2019,masuda2022}.
}
In heterogeneous graphs, the leading eigenvector may be localized around hubs.
To mitigate this, we scan eigenpairs $\{(\lambda_i,u^{(i)})\}$ and choose the
index $i^\star$ that minimizes a degree–eigenvector mismatch score,
\[
i^\star \;=\; \arg\min_i
\sum_{n=1}^N \bigl(k_n-\lambda_i\bigr)^2 \,\bigl(u^{(i)}_n\bigr)^2 ,
\]
which favors less localized modes. Using the selected
eigenvalue $\lambda_{i^\star}$, the modified spectral prediction is
\begin{equation}
\widehat{\langle I\rangle}_{\mathrm{mod}}(\beta/\gamma)
\;=\; \max\!\left\{\,0,\; 1 - \frac{1}{(\beta/\gamma)\,\lambda_{i^\star}}\right\}.
\label{eq:modspectral_rom}
\end{equation}
Here $(\beta/\gamma)_c = 1/\lambda_{i^\star}$, and the saturation level follows
directly from the same expression.

For each $\beta$, we evaluate \eqref{eq:spectral_rom} and
\eqref{eq:modspectral_rom} and compare them to the full SIS simulations and
SSM/gSSM reductions. On near-homogeneous topologies (e.g., ER, SW), the
dominant eigenvector is delocalized and both baselines track the sweep well.
On heavy-tailed graphs (e.g., SF), eigenvector localization shifts the effective
coupling toward hubs, causing threshold and plateau biases; the anti-localization
choice in the modified spectral baseline partly alleviates this, but SSM/gSSM
consistently provide more accurate node- and system-level predictions in our tests.

\section{Higher–order (triadic) SIS: mathematical implementation and parameters}
\label{sec:s6}
In this work, higher–order effects are implemented via \emph{simplicial (triangle) closure} of the pairwise contact network. Starting from the adjacency matrix $\mathbf{A}=[A_{ij}]$, we identify all triangles (3-cliques) $\{i,j,k\}$ in the underlying graph and encode them by the third-order indicator tensor $T_{ijk}$:
\[
T_{ijk} =
\begin{cases}
1, & \text{if } \{i,j,k\}\ \text{forms a triangle in } \mathbf{A},\\
0, & \text{otherwise},
\end{cases}
\qquad T_{ijk}=T_{ikj}=T_{jik}=\cdots
\]
(i.e., $T_{ijk}$ is symmetric in its indices). The higher–order SIS dynamics used throughout the manuscript are
\begin{equation}
\label{eq:hoi-sis}
\dot{x}_i \;=\; -\gamma\,x_i
\;+\; \beta \sum_{j=1}^N A_{ij}\,(1-x_i)\,x_j
\;+\; \eta \sum_{j<k} T_{ijk}\,(1-x_i)\,x_jx_k,
\qquad i=1,\dots,N,
\end{equation}
where $\gamma$ is the recovery rate, $\beta$ is the pairwise transmission rate, and $\eta$ controls the strength of triadic reinforcement. The restriction $j<k$ avoids double counting of unordered pairs within each triangle.

This simplicial-closure construction is a standard modeling choice for higher–order contagion and can generate discontinuous transitions and bistability that are absent in purely pairwise SIS dynamics~\cite{iacopini2019simplicial,battiston2020beyond,battiston2021higher}. Unless stated otherwise, all higher–order SIS simulations in the main text use Eq.~\eqref{eq:hoi-sis} with $T_{ijk}$ obtained from triangle closure of the corresponding $\mathbf{A}$. We hold the higher–order strength $\eta$ fixed while varying $\beta/\gamma$ as the control parameter (consistent with the manuscript). The remaining parameter choices are summarized in Table~\ref{tab:hoi_params}.

\begin{table}[!htpb]
	\caption{Parameters for the HOI SIS (triadic) model.}
	\label{tab:hoi_params}
	\begin{ruledtabular}
		\begin{tabular}{lll}
			Symbol & Meaning & Value / rule \\
			\colrule
			$\beta$     & Pairwise transmission rate       & Sweep in $\beta/\gamma$ \\
			$\gamma$    & Recovery rate                    & $1.0$ \\
			$\eta$      & Triadic infection strength       & $1.0$ \\
			$A$         & Adjacency matrix                 & Fixed per topology \\
			$T_{ijk}$   & Triangle tensor                  & Clique closure of $A$ \\
			$h_i$       & Triad load of node $i$           & $\#\{(j,k): T_{ijk} = 1\}$ \\
			$N$         & Number of nodes                  & $200$ \\
		\end{tabular}
	\end{ruledtabular}
\end{table}

\section{Other examples}
\label{sec:s7}
\subsection{Generalized Lotka--Volterra (GLV) }

We model a facilitative GLV system on a graph with adjacency matrix $A$:
\begin{equation}
\label{eq:glv_facilitation_ecology}
\begin{split}
\dot{N}_i
=\;& r\,N_i
\;+\; m \sum_{j=1}^N A_{ij}\,N_j \\
&\;+\; \alpha\,N_i \sum_{j=1}^N A_{ij}\,N_j
\;-\; c\,N_i^{2},
\qquad 1\le i\le N,
\end{split}
\end{equation}
where $r$ is the intrinsic growth rate, $m$ is a linear dispersal or immigration term from neighbors, $\alpha$ is the strength of pairwise facilitation (mutualistic gain), and $c>0$ represents intraspecific self-regulation (density dependence).  

\begin{figure*}[htbp!]
  \centering
  \includegraphics[width=\textwidth]{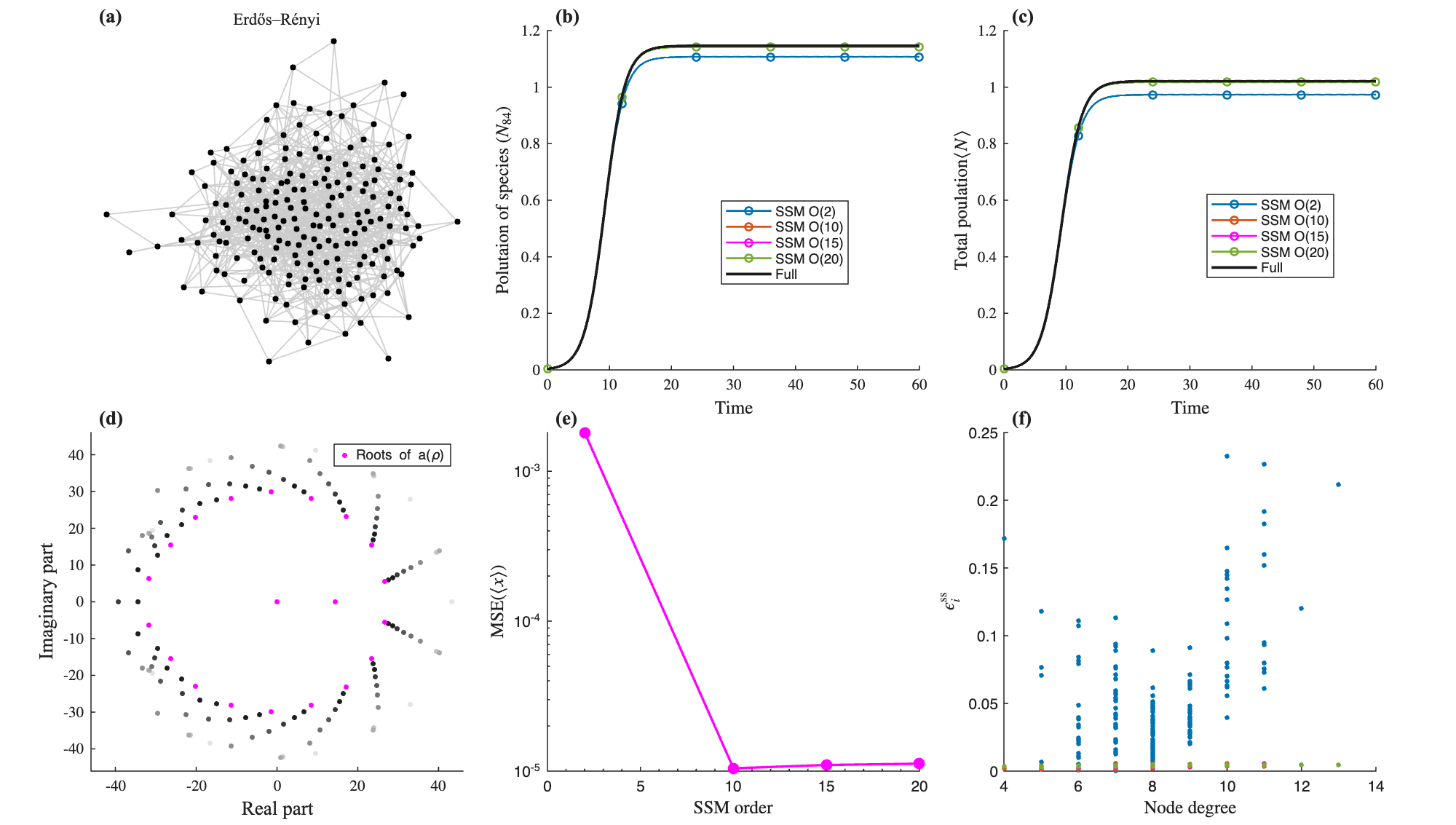}
\caption{\textbf{Generalized Lotka--Volterra dynamics on an Erd\H{o}s--R\'enyi network.} 
(a) Network layout with $N=200$ nodes. 
(b) Representative node abundance $N_{84}(t)$ for the full system (black) and $O(p)$--SSM reductions with $p\in\{2,10,15,20\}$ (colors as shown). 
(c) Mean abundance $\langle N\rangle(t)=N^{-1}\sum_i N_i$ with the same color scheme. 
(d) Root diagnostic from the scalar amplitude function $a(\rho)$: brighter shades denote higher $p$, with the highest-order roots highlighted in magenta. 
(e) Mean-squared error (MSE) of $\langle N\rangle(t)$ versus $p$, showing near-geometric decay. 
(f) Degree versus steady-state error $\epsilon_i^{\mathrm{ss}}$, with scatter points colored by $p$.  
Moderate $p$ (e.g., $O(10)$) reproduces both node-level and macroscopic dynamics with high fidelity, while $O(2)$ systematically underestimates the equilibrium.}
\label{fig:glv_er}
\end{figure*}

\begin{figure*}[htbp!]
  \centering
  \includegraphics[width=\textwidth]{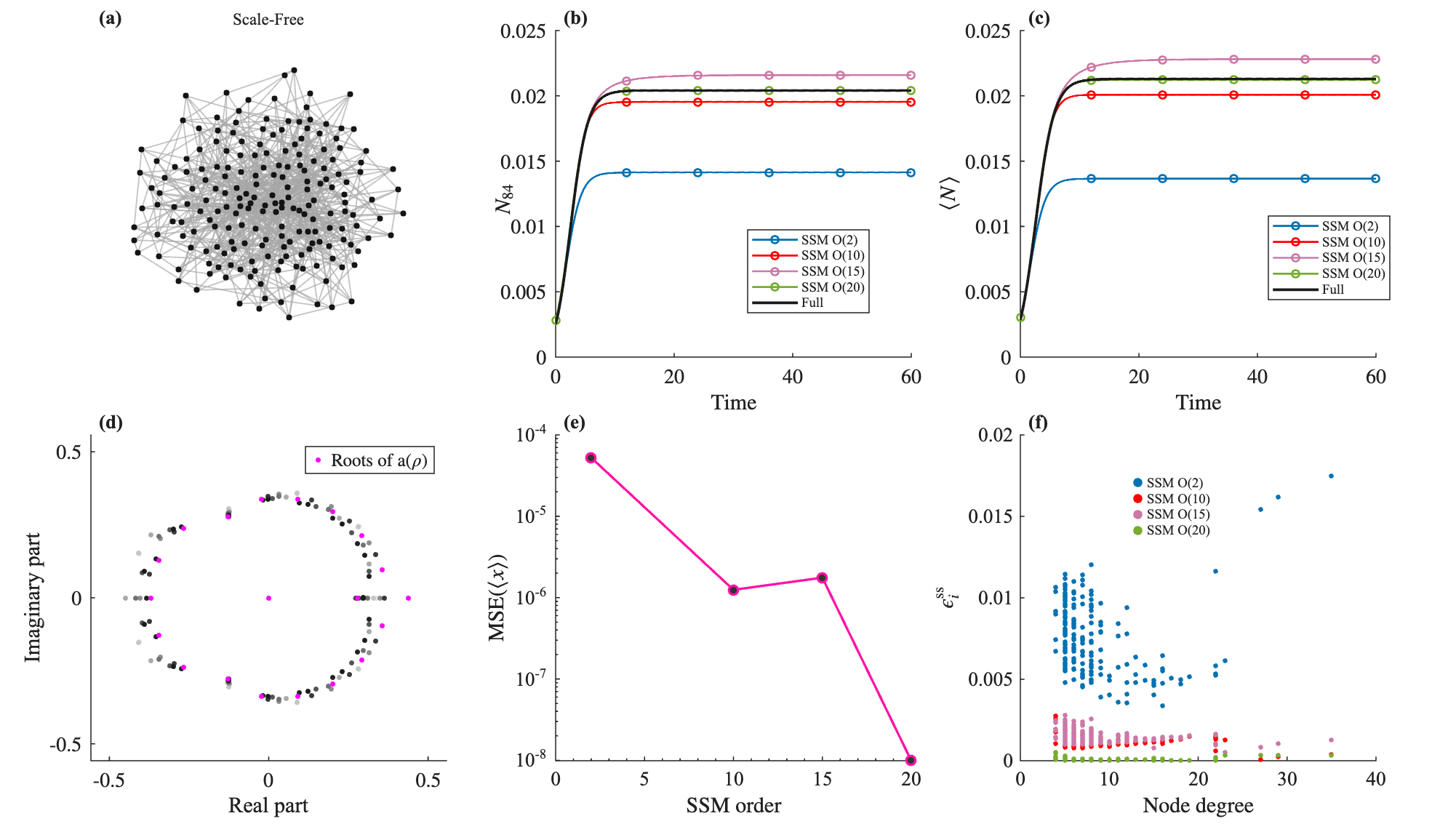}
\caption{\textbf{Generalized Lotka--Volterra dynamics on a Scale-Free network.} 
(a) Network layout with $N=200$ nodes. 
(b) Representative node abundance $N_{84}(t)$ for the full system (black) and $O(p)$--SSM reductions with $p\in\{2,10,15,20\}$. 
(c) Mean abundance $\langle N\rangle(t)=N^{-1}\sum_i N_i$. 
(d) Root diagnostic from $a(\rho)$: brighter shades denote higher $p$, with the highest-order roots highlighted in magenta. 
(e) MSE of $\langle N\rangle(t)$ versus $p$, showing slower decay than in ER/SW. 
(f) Degree versus $\epsilon_i^{\mathrm{ss}}$ colored by $p$.  
Higher orders ($O(15)$–$O(20)$) are required to fully recover both node- and system-level dynamics due to hub localization and a smaller Taylor radius.}
  \label{fig:glv_sf}
\end{figure*}

\begin{figure*}[htbp!]
  \centering
  \includegraphics[width=\textwidth]{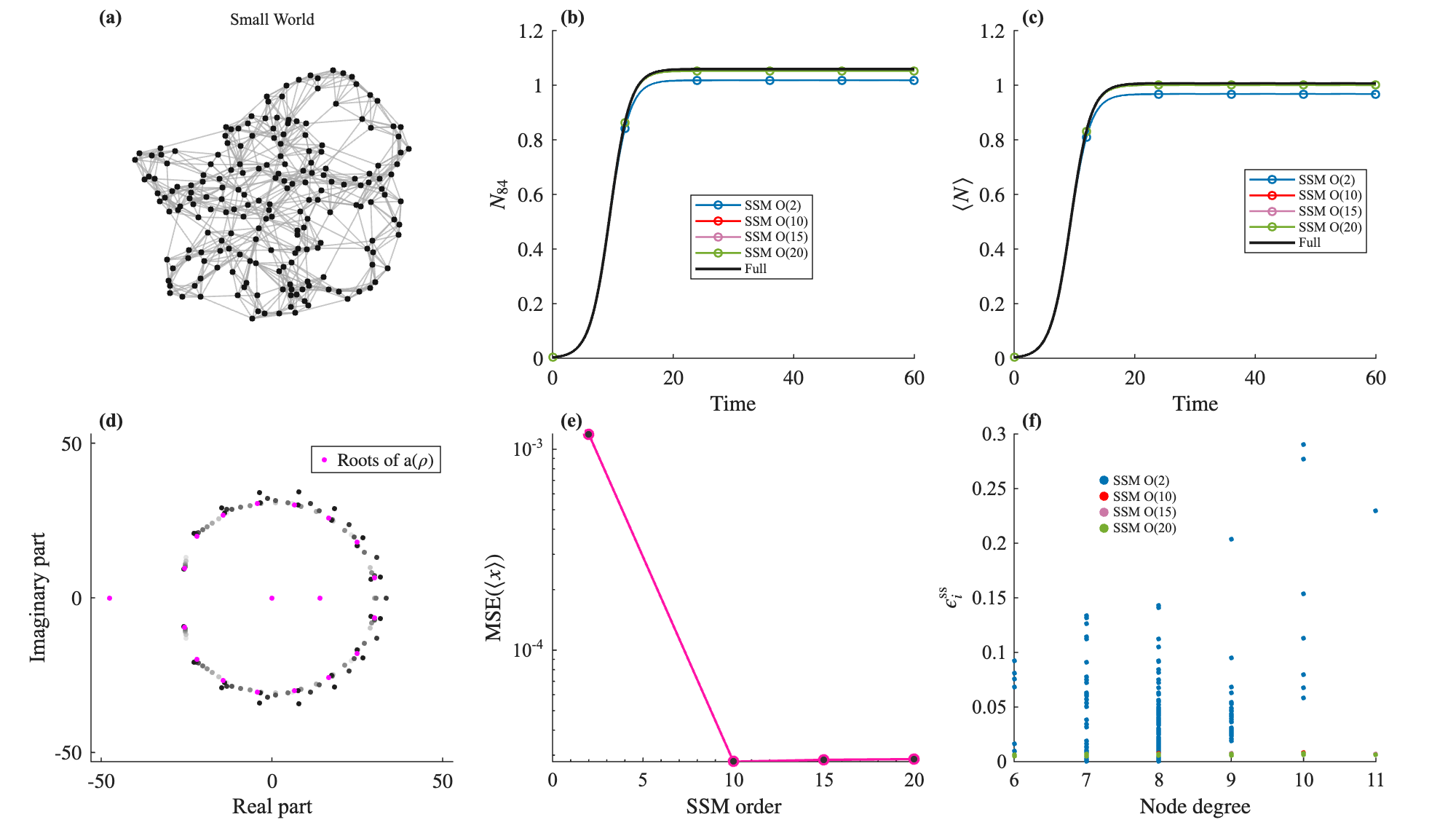}
\caption{\textbf{Generalized Lotka--Volterra dynamics on a Small-World network.} 
(a) Network layout with $N=200$ nodes. 
(b) Representative node abundance $N_{84}(t)$ for the full system (black) and $O(p)$--SSM reductions with $p\in\{2,10,15,20\}$. 
(c) Mean abundance $\langle N\rangle(t)=N^{-1}\sum_i N_i$. 
(d) Root diagnostic from $a(\rho)$: brighter shades denote higher $p$, with the highest-order roots highlighted in magenta. 
(e) MSE of $\langle N\rangle(t)$ versus $p$, showing rapid decay by $O(10)$. 
(f) Degree versus $\epsilon_i^{\mathrm{ss}}$ colored by $p$.  
By $O(10)$, both node-level and macroscopic trajectories are nearly indistinguishable from the full dynamics, with uniformly small errors across degrees.}
  \label{fig:glv_sw}
\end{figure*}

\begin{figure*}[htbp!]
  \centering
  \includegraphics[width=\textwidth]{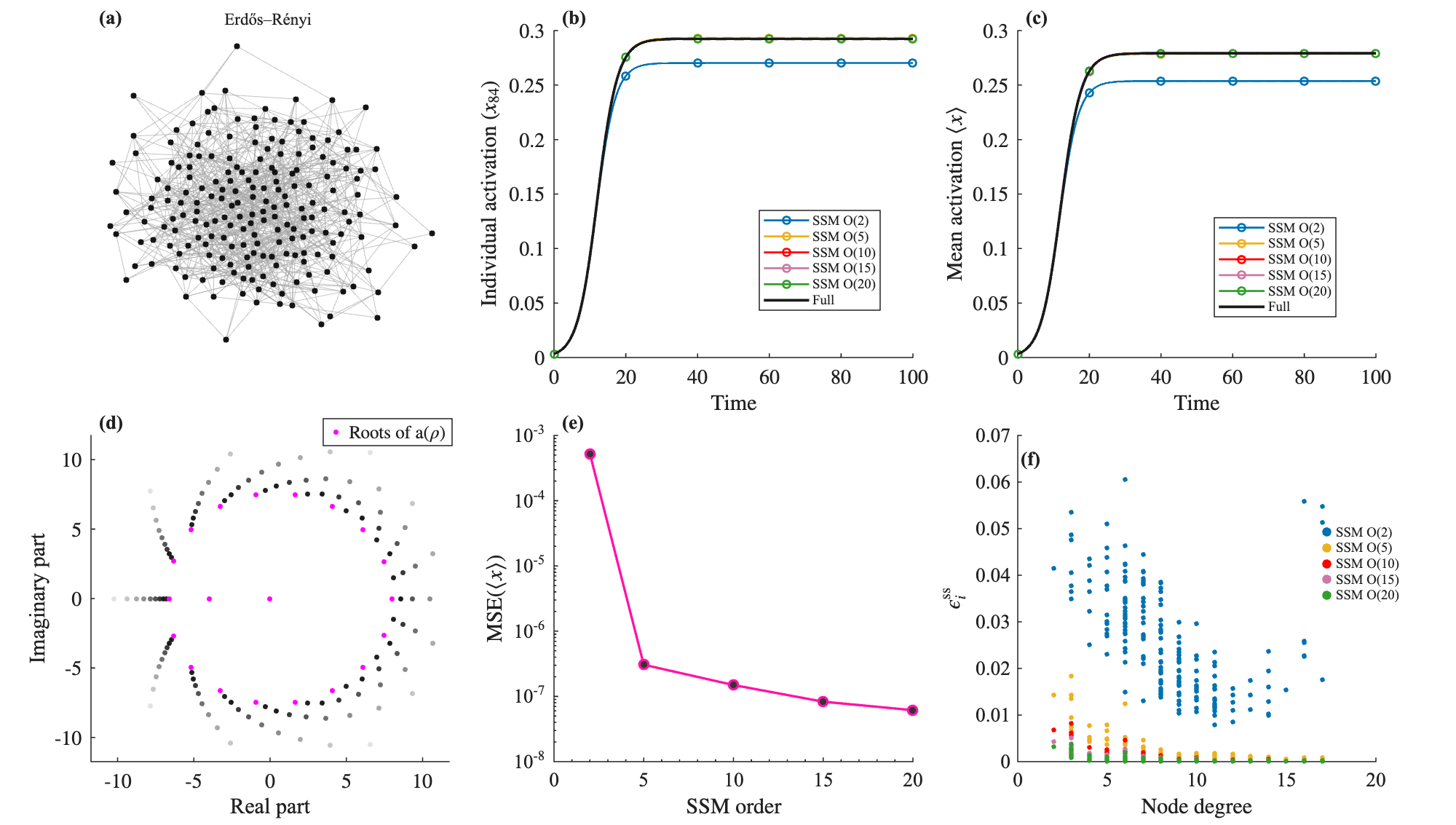}
\caption{\textbf{Gene–regulatory dynamics on an Erd\H{o}s--R\'enyi network.} 
(a) Network layout with $N=200$ nodes. 
(b) Representative node activation $x_{84}(t)$ for the full system (black) and $O(p)$--SSM reductions ($p\in\{2,5,10,15,20\}$; colors as shown). 
(c) Mean activation $\langle x\rangle(t)=N^{-1}\sum_i x_i(t)$. 
(d) Root diagnostic from $a(\rho)$: brighter shades denote higher $p$, with the highest order roots highlighted in magenta. 
(e) MSE of $\langle x\rangle(t)$ versus $p$, showing near-geometric decay. 
(f) Degree versus $\epsilon_i^{\mathrm{ss}}$ colored by $p$. 
Moderate $p$ reproduces node-level and macroscopic dynamics with high fidelity; $O(2)$ underestimates the equilibrium.}
  \label{fig:grn_er}
\end{figure*}

\begin{figure*}[htbp!]
  \centering
  \includegraphics[width=\textwidth]{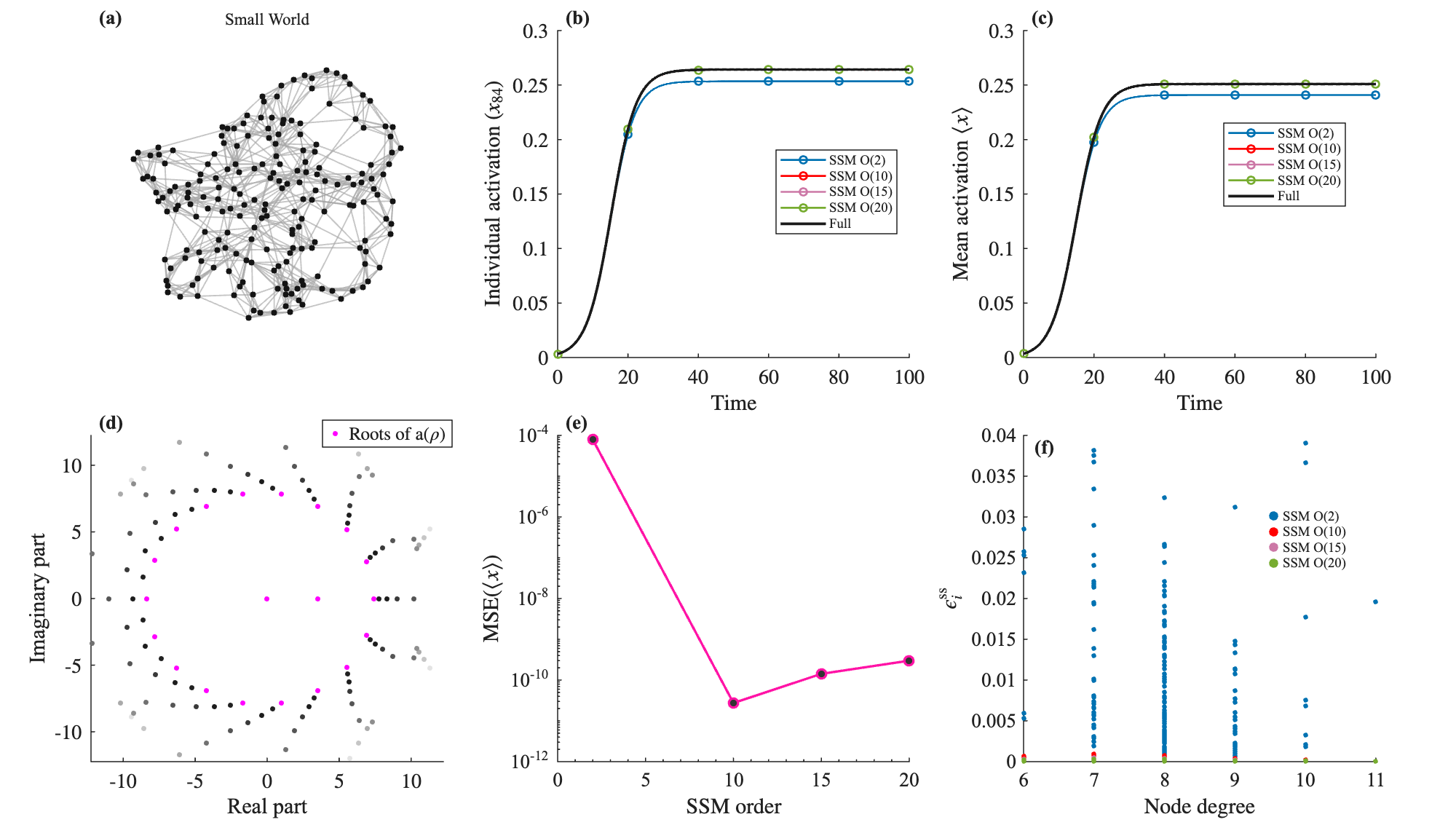}
\caption{\textbf{Gene–regulatory dynamics on a Small-World network.} 
(a) Network layout with $N=200$ nodes. 
(b) Representative node activation $x_{84}(t)$ for the full system (black) and $O(p)$--SSM reductions ($p\in\{2,10,15,20\}$). 
(c) Mean activation $\langle x\rangle(t)=N^{-1}\sum_i x_i(t)$. 
(d) Root diagnostic from $a(\rho)$: brighter shades denote higher $p$, with the highest order roots highlighted in magenta. 
(e) MSE of $\langle x\rangle(t)$ versus $p$, showing near-geometric decay. 
(f) Degree versus $\epsilon_i^{\mathrm{ss}}$ colored by $p$. 
By $O(10)$, both node- and system-level dynamics are reproduced with negligible deviation.}
  \label{fig:grn_sw}
\end{figure*}

\begin{figure*}[htbp!]
  \centering
  \includegraphics[width=\textwidth]{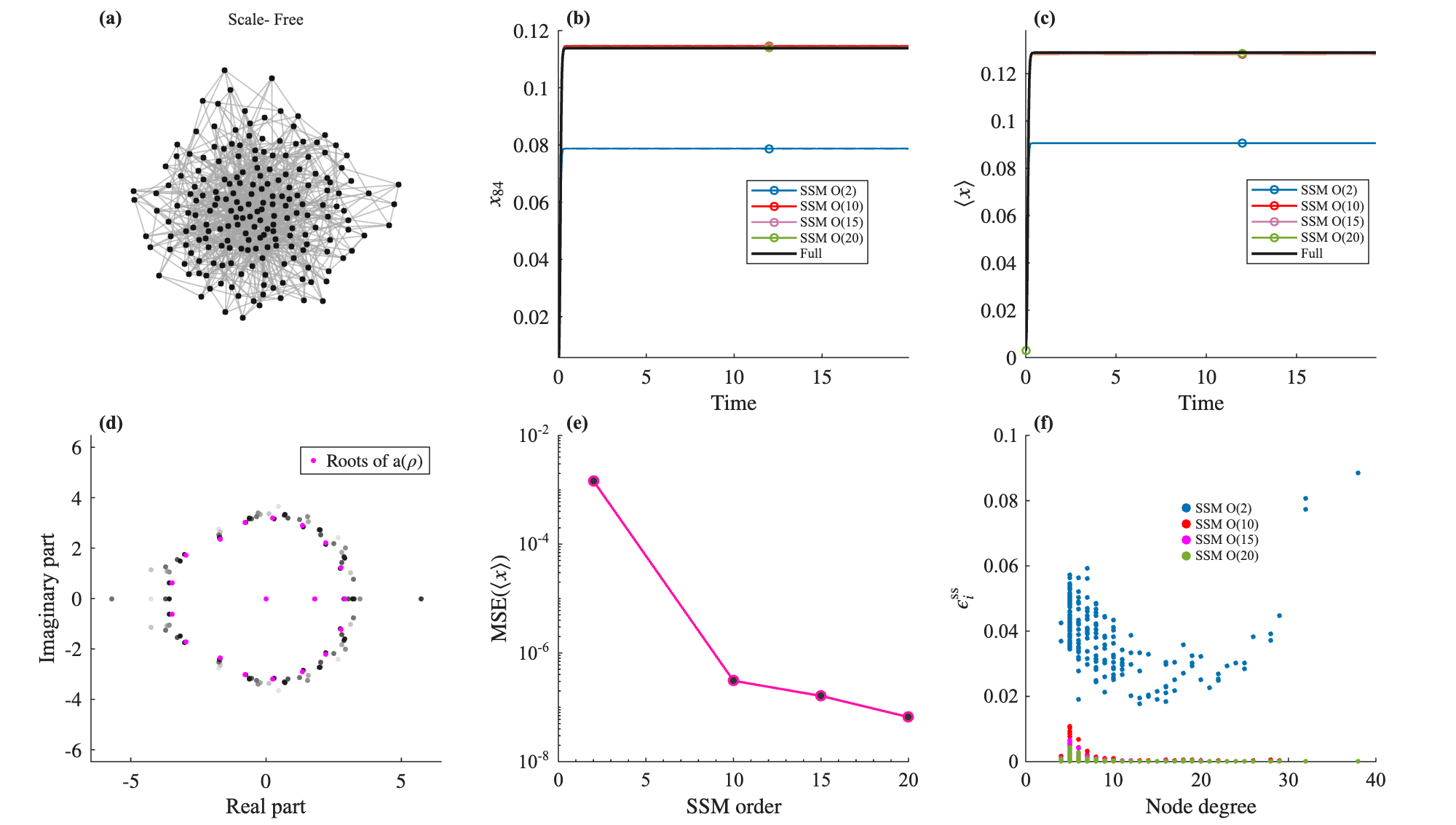}
\caption{\textbf{Gene–regulatory dynamics on a Scale-Free network.} 
(a) Network layout with $N=200$ nodes. 
(b) Representative node activation $x_{84}(t)$ for the full system (black) and $O(p)$--SSM reductions ($p\in\{2,10,15,20\}$). 
(c) Mean activation $\langle x\rangle(t)=N^{-1}\sum i\, x_i(t)$. 
(d) Root diagnostic from $a(\rho)$: brighter shades denote higher $p$, with the highest order roots highlighted in magenta. 
(e) MSE of $\langle x\rangle(t)$ versus $p$. 
(f) Degree versus $\epsilon_i^{\mathrm{ss}}$ colored by $p$. 
Higher $p$ ($15$–$20$) yields nearly indistinguishable trajectories from the full system despite strong heterogeneity.}
  \label{fig:grn_sf}
\end{figure*}

% ===================== ER: Logistic + diffusion =========================
\begin{figure*}[htbp!]
  \centering
  \includegraphics[width=\textwidth]{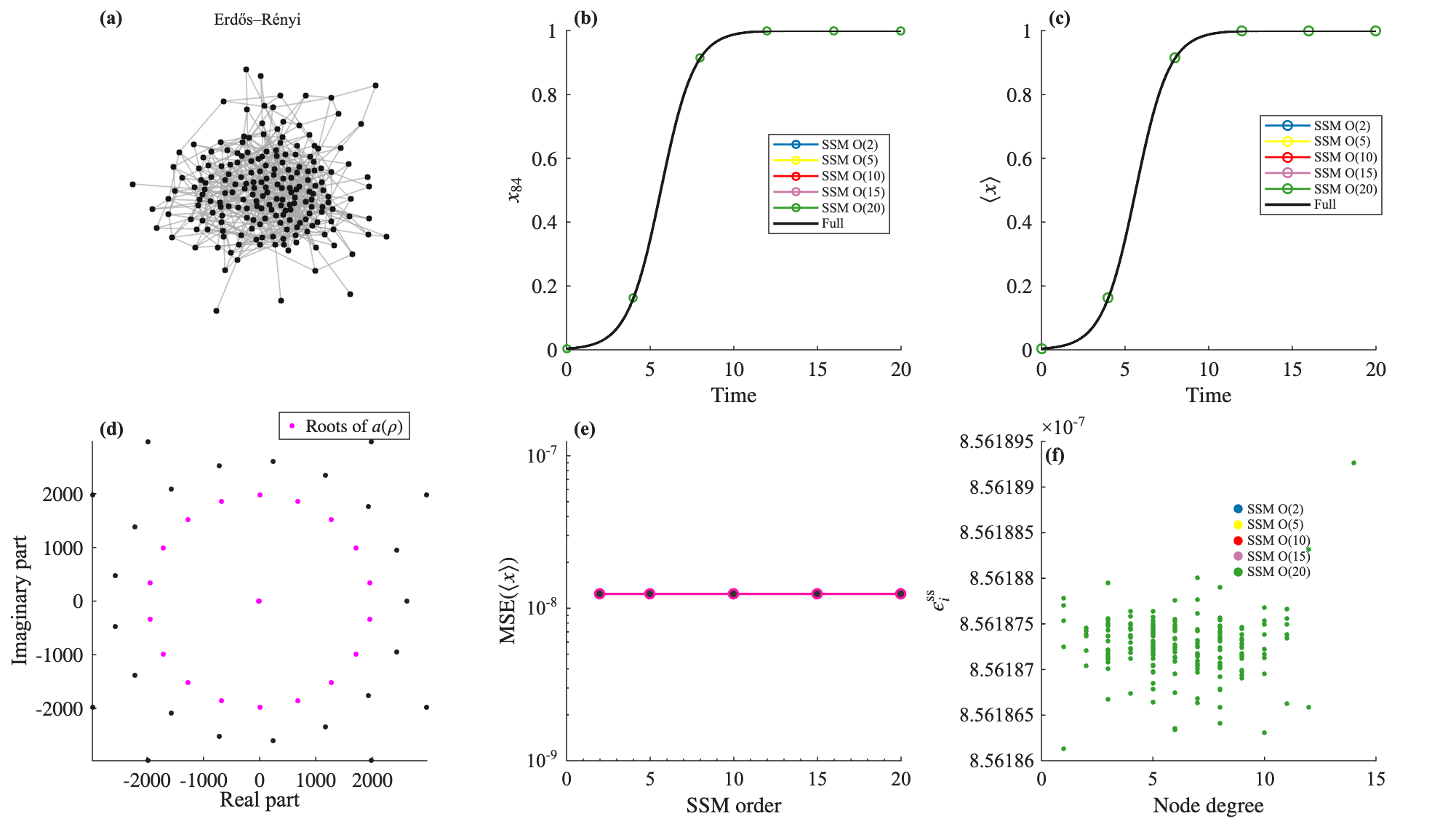}
\caption{\textbf{Logistic growth with diffusive coupling on an Erd\H{o}s--R\'enyi network.} 
(a) Network layout with $N=200$ nodes. 
(b) Representative node activation $x_{84}(t)$ for the full system (black) and $O(p)$--SSM reductions ($p\in\{2,5,10,15,20\}$; colors as shown). 
(c) Mean activation $\langle x\rangle(t)=N^{-1}\sum_i x_i(t)$. 
(d) Root diagnostic from $a(\rho)$: brighter shades denote higher $p$, with the highest-order roots highlighted in magenta. 
(e) MSE of $\langle x\rangle(t)$ remains negligible at $\sim 10^{-8}$ across all $p$. 
(f) Degree versus $\epsilon_i^{\mathrm{ss}}$ colored by $p$. 
Across all considered orders, the $O(p)$--SSM reductions reproduce the full-system trajectories with negligible deviation at both node and system scales.}
  \label{fig:lwd_er}
\end{figure*}

% ===================== SW: Logistic + diffusion =========================
\begin{figure*}[htbp!]
  \centering
  \includegraphics[width=\textwidth]{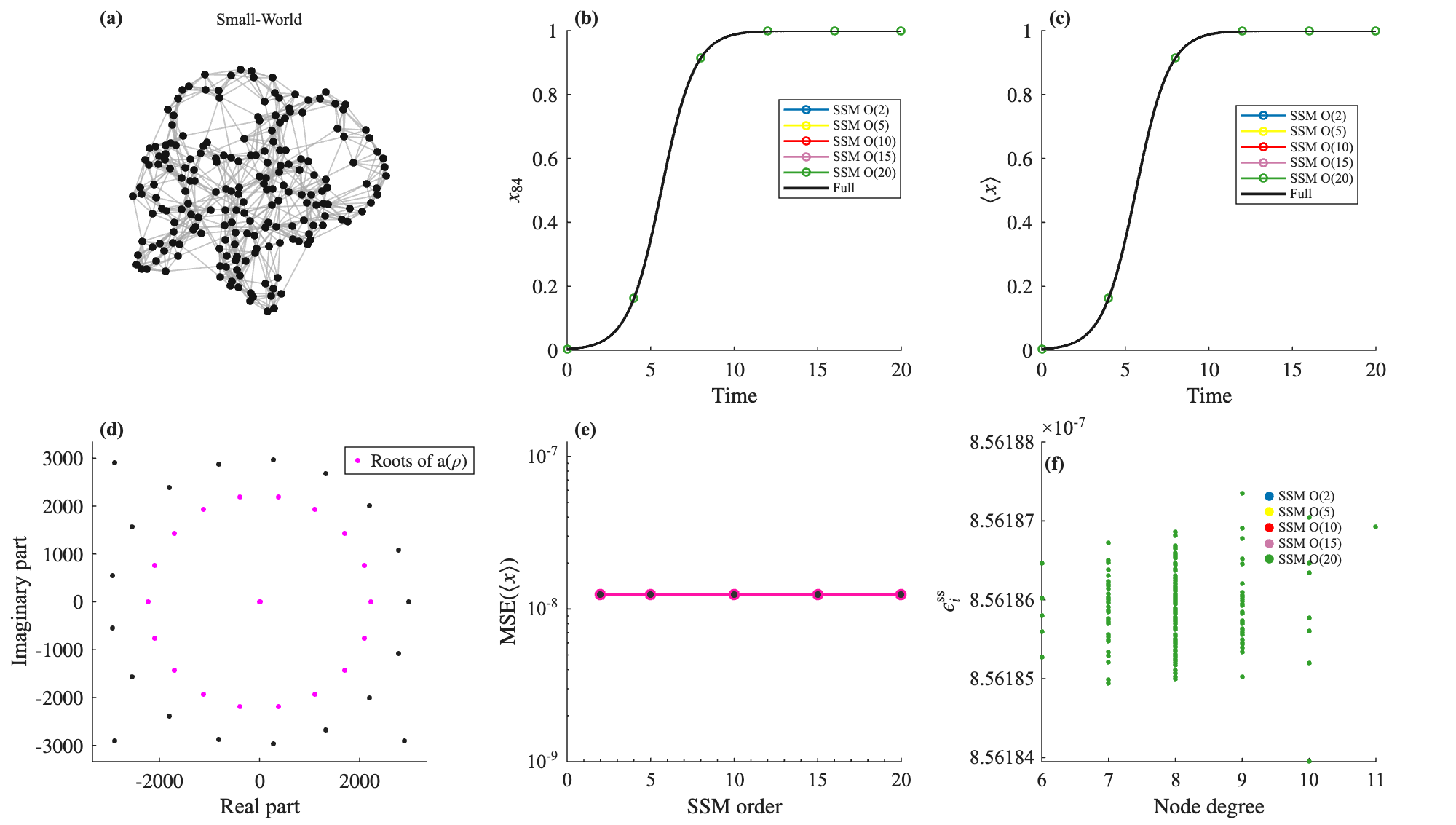}
\caption{\textbf{Logistic growth with diffusive coupling on a Small-World network.} 
(a) Network layout with $N=200$ nodes. 
(b) Representative node activation $x_{84}(t)$ for the full system (black) and $O(p)$--SSM reductions ($p\in\{2,5,10,15,20\}$). 
(c) Mean activation $\langle x\rangle(t)=N^{-1}\sum_i x_i(t)$. 
(d) Root diagnostic from $a(\rho)$: brighter shades denote higher $p$, with the highest-order roots highlighted in magenta. 
(e) MSE of $\langle x\rangle(t)$ versus $p$, flat at $\sim 10^{-8}$. 
(f) Degree versus $\epsilon_i^{\mathrm{ss}}$ colored by $p$. 
All $O(p)$--SSM reductions track the full system without discernible error.}
  \label{fig:lwd_sw}
\end{figure*}

% ===================== SF: Logistic + diffusion =========================
\begin{figure*}[htbp!]
  \centering
  \includegraphics[width=\textwidth]{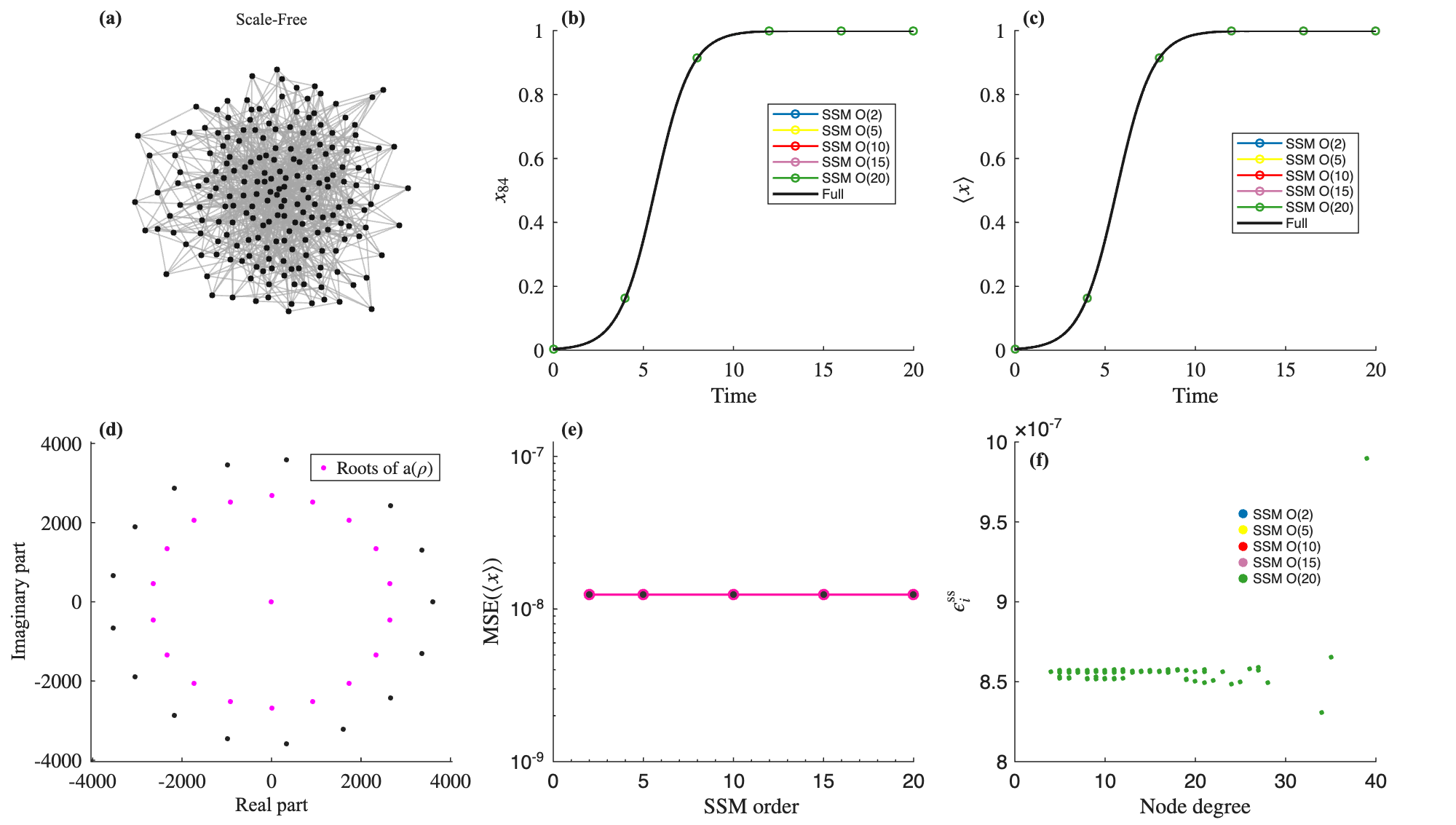}
 \caption{\textbf{Logistic growth with diffusive coupling on a Scale-Free network.} 
(a) Network layout with $N=200$ nodes. 
(b) Representative node activation $x_{84}(t)$ for the full system (black) and $O(p)$--SSM reductions ($p\in\{2,5,10,15,20\}$). 
(c) Mean activation $\langle x\rangle(t)=N^{-1}\sum_i x_i(t)$. 
(d) Root diagnostic from $a(\rho)$: brighter shades denote higher $p$, with the highest-order roots highlighted in magenta. 
(e) MSE of $\langle x\rangle(t)$ versus $p$, flat at $\sim 10^{-8}$. 
(f) Degree versus $\epsilon_i^{\mathrm{ss}}$ colored by $p$. 
Diffusion suppresses localization; even $O(2)$ yields trajectories that are nearly indistinguishable from the full system at both node and system levels.}
  \label{fig:lwd_sf}
\end{figure*}

Unless noted otherwise, we follow a common protocol across topologies: $N=200$ nodes, $O(p)$--SSM reductions with $p\in\{2,10,15,20\}$, initialization of the reduced coordinate at $q_0=0.01$ along the dominant eigenvector, and a final simulation horizon $t_f=60$. We compare full trajectories with $O(p)$--SSM reductions at both the node level and the mean abundance $\langle N\rangle = N^{-1}\sum_i N_i$. The parameters used are summarized in Table~\ref{tab:glv_params}.

\begin{table}[!htpb]
	\caption{GLV parameters used across experiments.}
	\label{tab:glv_params}
	\begin{ruledtabular}
		\begin{tabular}{lll}
			Symbol & Meaning & Value used \\
			\hline
			$r$      & intrinsic growth rate          & $0.20$ \\
			$m$      & linear coupling (dispersal)    & $0.05$ \\
			$\alpha$ & pairwise facilitation strength & $0.05$ \\
			$c$      & self--limitation (saturation)  & $1.00$ \\
			$A$      & adjacency                      & fixed per topology \\
			$N$      & number of nodes                & $200$ \\
		\end{tabular}
	\end{ruledtabular}
\end{table}

\paragraph{Results across topologies.} 
In the Erd\H{o}s–R\'enyi network (Fig.~\ref{fig:glv_er}a–f), the representative node trajectory in panel~(b) and the mean abundance in panel~(c) reveal that $O(2)$ captures the qualitative rise and saturation but settles below the true equilibrium. At $O(10)$, the reduced dynamics are nearly indistinguishable from the full system, and $O(15)$–$O(20)$ bring no visible change. The root diagnostic in panel~(d) supports this rapid convergence: the highest-order zeros lie well inside a wide annulus, consistent with the near–geometric decay of the mean error in panel~(e). The steady-state error in panel~(f) concentrates among low-degree nodes at low order but collapses toward zero as $p$ increases.  

The scale-free network (Fig.~\ref{fig:glv_sf}a–f) presents a more stringent test. Panels~(b,c) show visible bias at $O(2)$ and $O(10)$, while $O(15)$–$O(20)$ recover the correct levels. The roots in panel~(d) cluster nearer the outer boundary, indicating a reduced analyticity margin for Taylor expansions and explaining slower convergence. Correspondingly, panel~(e) shows more gradual MSE decay, though still improving by several orders of magnitude by $O(20)$. Degree–error scatter in panel~(f) confirms that low-degree nodes are hardest at low order, whereas hubs align well by mid order.  

For the Small-World network (Fig.~\ref{fig:glv_sw}a–f), convergence is fast: $O(10)$ suffices to match both node-level and mean trajectories; higher orders add only marginal refinements. The roots in panel~(d) show a stable positive-real zero well within a broad annulus, aligning with the sharp MSE drop from $O(2)$ to $O(10)$ in panel~(e). By $O(15)$, degree–error scatter in panel~(f) is uniformly small across all degrees.  

Overall, ER and SW networks are accurately represented with moderate $p$, while SF networks benefit from higher $p$ to resolve localized dynamics. Root maps anticipate these convergence patterns, and degree–error plots highlight where residual inaccuracies may persist.

\subsection{Gene–regulatory (activation–repression) dynamics}
\label{sec:grn}

\paragraph{Equation.}
We consider a simple gene–regulatory network model capturing activation and repression effects between connected nodes. Each node represents a gene whose activation level $x_i(t)$ evolves according to the balance of decay, activation, and inhibition processes described by
\begin{equation}
\label{eq:si_grn}
\begin{split}
\dot{x}_i
=\;& -\mu\,x_i \;+\; \gamma \sum_{j=1}^N A_{ij} x_j
\;-\; b\,x_i^{2} \\
&\;-\; k_{\mathrm{rep}} \sum_{j=1}^N A_{ij} x_i x_j ,
\qquad i=1,\dots,N .
\end{split}
\end{equation}
The parameters and their values used in the simulations are summarized in Table~\ref{tab:grn_params}.
% Preamble (if not already):
% \documentclass[aps,prx,reprint]{revtex4-2}

\begin{table}[!htpb]
	\caption{GRN activation--repression parameters.}
	\label{tab:grn_params}
	\begin{ruledtabular}
		\begin{tabular}{lll}
			Symbol & Meaning & Value used \\
			\hline
			$\mu$              & Degradation / decay              & $0.20$  \\
			$\gamma$           & Linear activation gain           & $0.060$ \\
			$b$                & Local self--saturation           & $0.80$  \\
			$k_{\mathrm{rep}}$ & Pairwise repression strength     & $0.040$ \\
			$A$                & Adjacency                        & Fixed per topology \\
			$N$                & Number of nodes                  & $200$   \\
		\end{tabular}
	\end{ruledtabular}
\end{table}

\paragraph{Results across topologies.}
For the Erd\H{o}s–R\'enyi (ER) network in Fig.~\ref{fig:grn_er}, the representative node trajectory in panel~(b) and the mean activation in panel~(c) illustrate that $O(2)$ captures the qualitative trend but settles at a slightly lower steady level than the full system. By $O(10)$, the reduced dynamics are effectively indistinguishable from the reference trajectory, and $O(15)$–$O(20)$ bring no visible change, confirming rapid convergence. The roots map in panel~(d) shows a clear and persistent positive–real zero of the amplitude function located comfortably inside a broad annulus of complex roots, aligning with the near–geometric decay of the mean–prevalence MSE in panel~(e). The degree–error plot in panel~(f) confirms that residuals at low order are largely confined to low–degree nodes; by $O(10)$, errors are small and uniformly distributed.

The Small–World (SW) case in Fig.~\ref{fig:grn_sw} shows a similar pattern but with even faster convergence. Panels~(b,c) demonstrate that $O(10)$ suffices to replicate both node–level and mean trajectories, with higher orders adding only negligible refinements. The root map in panel~(d) shows a wide analyticity margin surrounding the persistent positive–real zero, consistent with the sharp drop in $\mathrm{MSE}(\langle x\rangle)$ between $O(2)$ and $O(10)$ in panel~(e). The degree–error scatter in panel~(f) is tightly banded by $O(10)$, indicating nearly uniform accuracy across degrees.

The Scale–Free (SF) topology in Fig.~\ref{fig:grn_sf} presents a sharper challenge. Panels~(b,c) show that at $O(2)$ the reduced trajectories significantly underestimate both node–level and mean steady states. Increasing to $O(10)$ reduces this bias, and by $O(15)$–$O(20)$ the $O(p)$--SSM curves are nearly indistinguishable from the full dynamics, though convergence is visibly slower than in ER or SW. The root map in panel~(d) explains this behavior: dominant zeros of the amplitude function cluster closer to the outer boundary, indicating a smaller effective Taylor radius and slower MSE decay in panel~(e). The degree–error scatter in panel~(f) shows residuals concentrated among low–degree nodes at low order; as $p$ increases, nodewise errors diminish substantially across the degree spectrum.

Taken together, ER and SW networks are accurately captured by mid–order $O(p)$--SSM ($p\approx10$), while SF networks benefit from higher $p$ ($15$–$20$) to resolve localized dynamics. In all cases, the persistent positive–real zero of the amplitude function serves as a compact reliability check for the Taylor–only reduction.

\subsection{Logistic growth with diffusive coupling}

We study logistic growth with diffusion on a graph with adjacency matrix $A$ and Laplacian 
$L=\mathrm{diag}(\deg)-A$, where $\deg$ denotes the node-degree vector whose $i$th entry is 
$k_i=\sum_j A_{ij}$. The governing dynamics are
\begin{equation}
  \dot{x}_i
  \;=\;
  r\,x_i\!\left(1-\frac{x_i}{K}\right)
  \;+\; D\sum_{j=1}^N L_{ij}\,x_j,
  \qquad 1\le i\le N ,
  \label{eq:logistic_diffusive_model}
\end{equation}
where $x_i(t)\in[0,K]$ is the state at node $i$. We construct reduced models on a single-mode SSM at orders $O(2)$, $O(5)$, $O(10)$, $O(15)$, and $O(20)$, and perform time integration initialized on the SSM.

% PRX/REVTeX (two-column spanning)
\begin{table}[!htpb]
	\caption{Logistic--diffusion parameters.}
	\label{tab:logdiff_params}
	\begin{ruledtabular}
		\begin{tabular}{lll}
			Symbol & Meaning & Value used \\
			\colrule
			$r$ & Logistic growth rate  & $1.0$ \\
			$K$ & Carrying capacity     & $1.0$ \\
			$D$ & Diffusion coefficient & $0.05$ \\
			$A$ & Adjacency             & Fixed per topology \\
			$N$ & Number of nodes       & $200$ \\
		\end{tabular}
	\end{ruledtabular}
\end{table}

\paragraph{Results across topologies.}
For the Erd\H{o}s--R\'enyi case (Fig.~\ref{fig:lwd_er}a–f), the representative node trajectory in panel~(b) and the mean abundance in panel~(c) are visually indistinguishable between the full system and all $O(p)$--SSM reductions, even at $O(2)$. The root diagnostic in panel~(d) shows a large domain of analyticity with a stable positive-real zero persisting well within the domain. Consistent with this observation, the MSE of the macroscopic observable (panel~e) remains negligible at $\sim 10^{-8}$ for all orders. Degree-resolved errors (panel~f) are similarly negligible, confirming uniform accuracy across all nodes.

The Small-World case (Fig.~\ref{fig:lwd_sw}a–f) exhibits the same behavior. Node-level trajectories (panel~b) and the mean dynamics (panel~c) are nearly identical for the full and reduced systems. The roots in panel~(d) again form a wide, regular annulus; the macroscopic error curve (panel~e) stays at the numerical floor; and steady-state errors (panel~f) vanish across degrees.

For the heterogeneous Scale-Free network (Fig.~\ref{fig:lwd_sf}a–f), diffusion suppresses localization effects. Panels~(b,c) show that full and reduced trajectories remain nearly indistinguishable across all $p$. The root map in panel~(d) indicates a comfortable analytic buffer; the mean error in panel~(e) remains flat at the numerical floor; and degree-wise errors (panel~f) are near machine precision.

Together, these results show that diffusion homogenizes the dynamics across the network, collapsing them onto a single smooth mode. Consequently, even the lowest-order $O(2)$–SSM suffices to reproduce both transient and steady-state dynamics with full fidelity, regardless of topology. 
For the logistic–diffusion model, we verified analytically that all higher–order SSM coefficients vanish (\(c_k=0\) for \(k\ge3\)), yielding an exact quadratic reduced dynamics 
\(\dot{q}=rq - (r/K)\,(\sum_i u_i^3)q^2\).
Consequently, the \(O(2)\)–SSM reduction reproduces the full system exactly, consistent with the numerical results reported above.

\begin{figure*}[!htpb]
    \centering
    \includegraphics[width=\textwidth]{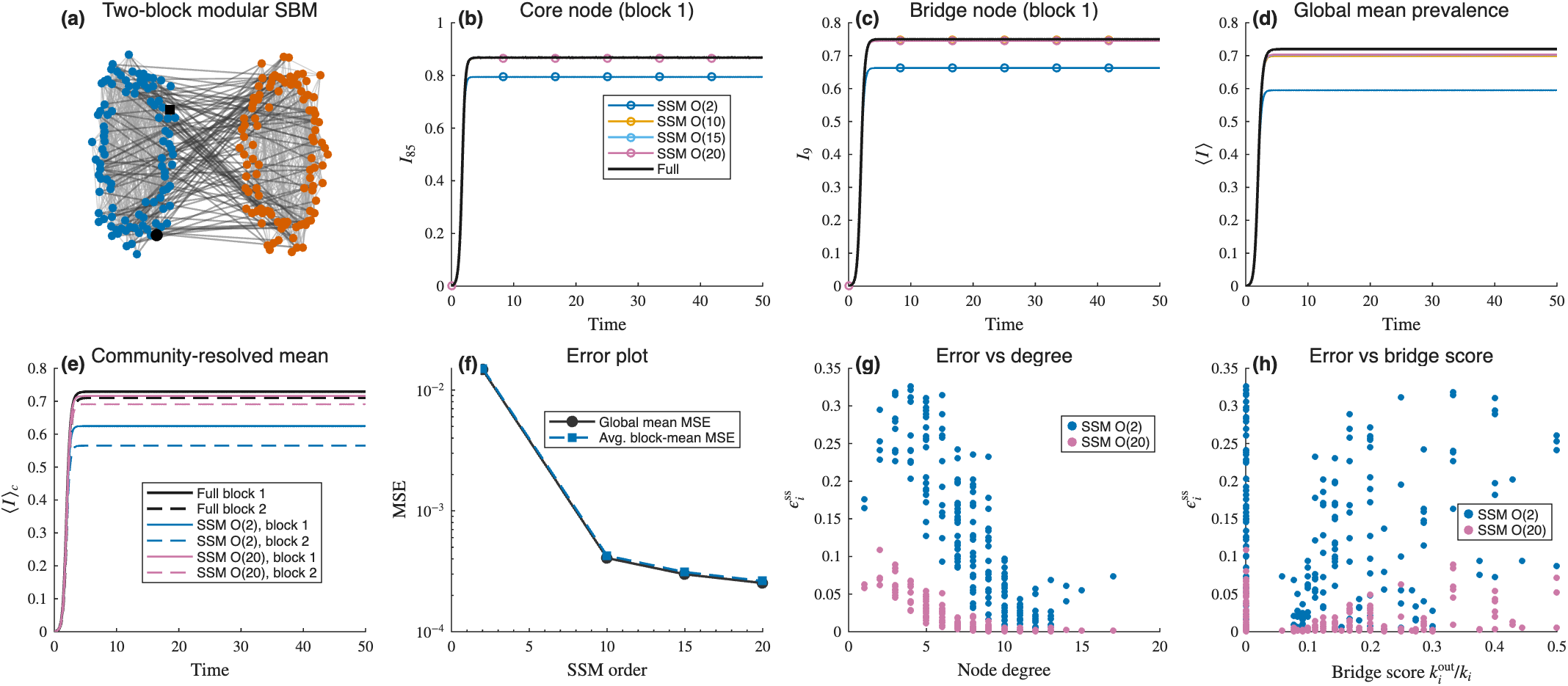}
    \caption{
   \textbf{Full topology-aware diagnostics for the modular stochastic-block-model benchmark.}
    (a) Two-block modular SBM realization, with nodes colored by community; the marked circle denotes a representative bridge node and the marked square denotes a representative core node.
    (b) Representative core-node trajectory $I_{85}(t)$ for the full system and $O(p)$--SSM reductions with $p\in\{2,10,15,20\}$.
    (c) Representative bridge-node trajectory $I_{9}(t)$ under the same comparison.
    (d) Global mean prevalence $\langle I\rangle(t)$.
    (e) Community-resolved mean prevalence $\langle I\rangle_c(t)$ in the two blocks.
    (f) Mean-squared error of the global mean and the average block mean versus SSM order.
    (g) Steady-state nodewise error $\epsilon_i^{\mathrm{ss}}$ versus degree.
    (h) Steady-state nodewise error $\epsilon_i^{\mathrm{ss}}$ versus bridge score $k_i^{\mathrm{out}}/k_i$.
    The benchmark shows that modular bottlenecks do not induce a qualitative breakdown of the reduction, but they do make low-order truncations more sensitive, especially in bridge-mediated and community-resolved observables; increasing the order from $O(2)$ to $O(20)$ strongly suppresses these errors.
    }
    \label{fig:sm_sbm_modular_full}
\end{figure*}

\section{Modular stochastic-block-model benchmark}
\label{sec:sm_sbm_modular}

To examine modular structure in a controlled setting, we supplement the ER, SW, and SF ensembles with a two-block stochastic block model (SBM) with matched mean degree and weak inter-community coupling. The purpose of this benchmark is to isolate modular bottlenecks, which are not directly prescribed by the ER, SW, and SF constructions. The network is built under the same conventions as in Sec.~S2.1: undirected, unweighted, simple graphs on $N=200$ nodes, with equal-size communities and no self-loops. We match the expected mean degree to the main synthetic benchmarks and choose the ratio of inter-community to intra-community connection probabilities so as to obtain a clearly bottlenecked modular structure. Unless noted otherwise, the SIS dynamics, initialization protocol, time integration, and $O(p)$--SSM truncation orders are the same as in the main-text SIS experiments, and the error metrics follow Sec.~S3.

Figure~\ref{fig:sm_sbm_modular_full} shows the full set of diagnostics for this modular benchmark and complements the condensed figure in the main text. Panel (a) displays a representative two-block realization with dense intra-community connectivity and a relatively sparse set of inter-community links. Inter-community transport is therefore constrained to a small set of bridge nodes, making this a more direct test of modular bottlenecks than the SW ensemble.

Panels (b) and (c) show the node-level dynamics. The reduction remains accurate in this setting, but the effect of truncation order is more visible than in the better-connected cases. For the representative core node in panel (b), the $O(2)$ reduction underestimates the endemic level, while the higher-order reductions, particularly $O(15)$ and $O(20)$, track the full trajectory closely. Panel (c) shows the same pattern for a representative bridge node in a form that is more directly tied to the modular structure: the $O(2)$ approximation again underestimates the response, whereas the $O(20)$ reduction nearly overlaps with the full-system trajectory. The modular bottleneck therefore does not lead to a qualitative breakdown at the node level, but it does make low-order truncations more sensitive.

This dependence on order also appears in the aggregate observables. In panel (d), the global mean prevalence is captured qualitatively even at low order, but the $O(2)$ reduction remains visibly biased. Increasing the order reduces this bias substantially, and the $O(20)$ reduction reproduces the full mean prevalence closely. Panel (e) makes the modular effect more explicit through the community-resolved mean dynamics. There, the $O(2)$ reduction underestimates the mean prevalence in both communities, which indicates that low-order truncation does not fully represent the balance between intra-community growth and bridge-mediated transfer. The $O(20)$ reduction, by contrast, reproduces the block-level means closely throughout the trajectory.

Panel (f) quantifies the improvement with order. The MSE of the global mean and the average block mean both drop sharply as the truncation order increases from $O(2)$ to $O(10)$, followed by a smaller but still clear improvement from $O(10)$ to $O(20)$. The modular benchmark therefore does not indicate a breakdown of the SSM framework. It instead identifies a topology-sensitive regime in which higher-order truncation noticeably improves both macroscopic and community-level accuracy.

Panels (g) and (h) show where the remaining steady-state errors are concentrated. In panel (g), the $O(2)$ errors are spread broadly across the network, with especially visible deviations among low- and intermediate-degree nodes, whereas the $O(20)$ errors are much smaller across nearly the full degree range. Panel (h) reports the same comparison against the bridge score $k_i^{\mathrm{out}}/k_i$, which measures the fraction of edges leaving a node’s own community. The $O(2)$ reduction shows appreciable errors across a wide range of bridge scores, while the $O(20)$ reduction suppresses these discrepancies substantially. These diagnostics indicate that modular bottlenecks chiefly expose the limitations of low-order truncations, especially near community interfaces, whereas higher-order reductions recover both node-level and aggregate behavior reliably.

This benchmark complements the ER, SW, and SF examples by isolating a structural effect that is only implicit in those ensembles, namely the role of modular bottlenecks in shaping reduction accuracy. The conclusion is not that modularity causes a qualitative failure of the SSM-based reduction. Rather, modularity increases the sensitivity of bridge-node and community-resolved observables to truncation order, which makes topology-aware diagnostics particularly useful in this setting.

%==========================================================================
\clearpage  % flush all pending floats before references
\bibliography{sample}

%apsrev4-2.bst 2019-01-14 (MD) hand-edited version of apsrev4-1.bst
%Control: key (0)
%Control: author (8) initials jnrlst
%Control: editor formatted (1) identically to author
%Control: production of article title (0) allowed
%Control: page (0) single
%Control: year (1) truncated
%Control: production of eprint (0) enabled
\begin{thebibliography}{45}%
\makeatletter
\providecommand \@ifxundefined [1]{%
 \@ifx{#1\undefined}
}%
\providecommand \@ifnum [1]{%
 \ifnum #1\expandafter \@firstoftwo
 \else \expandafter \@secondoftwo
 \fi
}%
\providecommand \@ifx [1]{%
 \ifx #1\expandafter \@firstoftwo
 \else \expandafter \@secondoftwo
 \fi
}%
\providecommand \natexlab [1]{#1}%
\providecommand \enquote  [1]{``#1''}%
\providecommand \bibnamefont  [1]{#1}%
\providecommand \bibfnamefont [1]{#1}%
\providecommand \citenamefont [1]{#1}%
\providecommand \href@noop [0]{\@secondoftwo}%
\providecommand \href [0]{\begingroup \@sanitize@url \@href}%
\providecommand \@href[1]{\@@startlink{#1}\@@href}%
\providecommand \@@href[1]{\endgroup#1\@@endlink}%
\providecommand \@sanitize@url [0]{\catcode `\\12\catcode `\$12\catcode
  `\&12\catcode `\#12\catcode `\^12\catcode `\_12\catcode `\%12\relax}%
\providecommand \@@startlink[1]{}%
\providecommand \@@endlink[0]{}%
\providecommand \url  [0]{\begingroup\@sanitize@url \@url }%
\providecommand \@url [1]{\endgroup\@href {#1}{\urlprefix }}%
\providecommand \urlprefix  [0]{URL }%
\providecommand \Eprint [0]{\href }%
\providecommand \doibase [0]{https://doi.org/}%
\providecommand \selectlanguage [0]{\@gobble}%
\providecommand \bibinfo  [0]{\@secondoftwo}%
\providecommand \bibfield  [0]{\@secondoftwo}%
\providecommand \translation [1]{[#1]}%
\providecommand \BibitemOpen [0]{}%
\providecommand \bibitemStop [0]{}%
\providecommand \bibitemNoStop [0]{.\EOS\space}%
\providecommand \EOS [0]{\spacefactor3000\relax}%
\providecommand \BibitemShut  [1]{\csname bibitem#1\endcsname}%
\let\auto@bib@innerbib\@empty
%</preamble>
\bibitem [{\citenamefont {Pascual}\ and\ \citenamefont
  {Dunne}(2006)}]{pascual2006}%
  \BibitemOpen
  \bibinfo {editor} {\bibfnamefont {M.}~\bibnamefont {Pascual}}\ and\ \bibinfo
  {editor} {\bibfnamefont {J.~A.}\ \bibnamefont {Dunne}},\ eds.,\ \href@noop {}
  {\emph {\bibinfo {title} {Ecological Networks: Linking Structure to Dynamics
  in Food Webs}}},\ Santa Fe Institute Studies on the Sciences of Complexity\
  (\bibinfo  {publisher} {Oxford University Press},\ \bibinfo {address}
  {Oxford},\ \bibinfo {year} {2006})\BibitemShut {NoStop}%
\bibitem [{\citenamefont {Anderson}\ and\ \citenamefont
  {May}(1991)}]{anderson1991}%
  \BibitemOpen
  \bibfield  {author} {\bibinfo {author} {\bibfnamefont {R.~M.}\ \bibnamefont
  {Anderson}}\ and\ \bibinfo {author} {\bibfnamefont {R.~M.}\ \bibnamefont
  {May}},\ }\href@noop {} {\emph {\bibinfo {title} {Infectious Diseases of
  Humans: Dynamics and Control}}},\ Oxford Science Publications\ (\bibinfo
  {publisher} {Oxford University Press},\ \bibinfo {address} {Oxford},\
  \bibinfo {year} {1991})\BibitemShut {NoStop}%
\bibitem [{\citenamefont {Dobson}\ \emph {et~al.}(2007)\citenamefont {Dobson},
  \citenamefont {Carreras}, \citenamefont {Lynch},\ and\ \citenamefont
  {Newman}}]{dobson2007}%
  \BibitemOpen
  \bibfield  {author} {\bibinfo {author} {\bibfnamefont {I.}~\bibnamefont
  {Dobson}}, \bibinfo {author} {\bibfnamefont {B.~A.}\ \bibnamefont
  {Carreras}}, \bibinfo {author} {\bibfnamefont {V.~E.}\ \bibnamefont
  {Lynch}},\ and\ \bibinfo {author} {\bibfnamefont {D.~E.}\ \bibnamefont
  {Newman}},\ }\bibfield  {title} {\bibinfo {title} {Complex systems analysis
  of series of blackouts: Cascading failure, critical points, and
  self-organization},\ }\href {https://doi.org/10.1063/1.2737822} {\bibfield
  {journal} {\bibinfo  {journal} {Chaos}\ }\textbf {\bibinfo {volume} {17}},\
  \bibinfo {pages} {026103} (\bibinfo {year} {2007})}\BibitemShut {NoStop}%
\bibitem [{\citenamefont {Rumelhart}\ \emph {et~al.}(1986)\citenamefont
  {Rumelhart}, \citenamefont {Hinton},\ and\ \citenamefont
  {Williams}}]{rumelhart1986}%
  \BibitemOpen
  \bibfield  {author} {\bibinfo {author} {\bibfnamefont {D.~E.}\ \bibnamefont
  {Rumelhart}}, \bibinfo {author} {\bibfnamefont {G.~E.}\ \bibnamefont
  {Hinton}},\ and\ \bibinfo {author} {\bibfnamefont {R.~J.}\ \bibnamefont
  {Williams}},\ }\bibfield  {title} {\bibinfo {title} {Learning representations
  by back-propagating errors},\ }\href {https://doi.org/10.1038/323533a0}
  {\bibfield  {journal} {\bibinfo  {journal} {Nature}\ }\textbf {\bibinfo
  {volume} {323}},\ \bibinfo {pages} {533} (\bibinfo {year}
  {1986})}\BibitemShut {NoStop}%
\bibitem [{\citenamefont {Gao}\ \emph {et~al.}(2016)\citenamefont {Gao},
  \citenamefont {Barzel},\ and\ \citenamefont {Barab{\'a}si}}]{gao2016}%
  \BibitemOpen
  \bibfield  {author} {\bibinfo {author} {\bibfnamefont {J.}~\bibnamefont
  {Gao}}, \bibinfo {author} {\bibfnamefont {B.}~\bibnamefont {Barzel}},\ and\
  \bibinfo {author} {\bibfnamefont {A.-L.}\ \bibnamefont {Barab{\'a}si}},\
  }\bibfield  {title} {\bibinfo {title} {Universal resilience patterns in
  complex networks},\ }\href {https://doi.org/10.1038/nature16948} {\bibfield
  {journal} {\bibinfo  {journal} {Nature}\ }\textbf {\bibinfo {volume} {530}},\
  \bibinfo {pages} {307} (\bibinfo {year} {2016})}\BibitemShut {NoStop}%
\bibitem [{\citenamefont {Landi}\ \emph {et~al.}(2018)\citenamefont {Landi},
  \citenamefont {Minoarivelo}, \citenamefont {Br{\"a}nnstr{\"o}m},
  \citenamefont {Hui},\ and\ \citenamefont {Dieckmann}}]{landi2018}%
  \BibitemOpen
  \bibfield  {author} {\bibinfo {author} {\bibfnamefont {P.}~\bibnamefont
  {Landi}}, \bibinfo {author} {\bibfnamefont {H.~O.}\ \bibnamefont
  {Minoarivelo}}, \bibinfo {author} {\bibfnamefont {{\AA}.}~\bibnamefont
  {Br{\"a}nnstr{\"o}m}}, \bibinfo {author} {\bibfnamefont {C.}~\bibnamefont
  {Hui}},\ and\ \bibinfo {author} {\bibfnamefont {U.}~\bibnamefont
  {Dieckmann}},\ }\bibfield  {title} {\bibinfo {title} {Complexity and
  stability of ecological networks: a review of the theory},\ }\href
  {https://doi.org/10.1007/s10144-018-0628-3} {\bibfield  {journal} {\bibinfo
  {journal} {Population Ecology}\ }\textbf {\bibinfo {volume} {60}},\ \bibinfo
  {pages} {319} (\bibinfo {year} {2018})}\BibitemShut {NoStop}%
\bibitem [{\citenamefont {Jiang}\ \emph {et~al.}(2018)\citenamefont {Jiang},
  \citenamefont {Huang}, \citenamefont {Seager}, \citenamefont {Hastings},\
  and\ \citenamefont {Lai}}]{jiang2018}%
  \BibitemOpen
  \bibfield  {author} {\bibinfo {author} {\bibfnamefont {J.}~\bibnamefont
  {Jiang}}, \bibinfo {author} {\bibfnamefont {Z.-G.}\ \bibnamefont {Huang}},
  \bibinfo {author} {\bibfnamefont {T.~P.}\ \bibnamefont {Seager}}, \bibinfo
  {author} {\bibfnamefont {A.}~\bibnamefont {Hastings}},\ and\ \bibinfo
  {author} {\bibfnamefont {Y.-C.}\ \bibnamefont {Lai}},\ }\bibfield  {title}
  {\bibinfo {title} {Predicting tipping points in mutualistic networks through
  dimension reduction},\ }\href {https://doi.org/10.1073/pnas.1714958115}
  {\bibfield  {journal} {\bibinfo  {journal} {Proceedings of the National
  Academy of Sciences}\ }\textbf {\bibinfo {volume} {115}},\ \bibinfo {pages}
  {E639} (\bibinfo {year} {2018})}\BibitemShut {NoStop}%
\bibitem [{\citenamefont {Laurence}\ \emph {et~al.}(2019)\citenamefont
  {Laurence}, \citenamefont {Doyon}, \citenamefont {Dub{\'e}},\ and\
  \citenamefont {Desrosiers}}]{laurence2019}%
  \BibitemOpen
  \bibfield  {author} {\bibinfo {author} {\bibfnamefont {E.}~\bibnamefont
  {Laurence}}, \bibinfo {author} {\bibfnamefont {N.}~\bibnamefont {Doyon}},
  \bibinfo {author} {\bibfnamefont {L.~J.}\ \bibnamefont {Dub{\'e}}},\ and\
  \bibinfo {author} {\bibfnamefont {P.}~\bibnamefont {Desrosiers}},\ }\bibfield
   {title} {\bibinfo {title} {Spectral dimension reduction of complex dynamical
  networks},\ }\href {https://doi.org/10.1103/PhysRevX.9.011042} {\bibfield
  {journal} {\bibinfo  {journal} {Physical Review X}\ }\textbf {\bibinfo
  {volume} {9}},\ \bibinfo {pages} {011042} (\bibinfo {year}
  {2019})}\BibitemShut {NoStop}%
\bibitem [{\citenamefont {Masuda}\ and\ \citenamefont
  {Kundu}(2022)}]{masuda2022}%
  \BibitemOpen
  \bibfield  {author} {\bibinfo {author} {\bibfnamefont {N.}~\bibnamefont
  {Masuda}}\ and\ \bibinfo {author} {\bibfnamefont {P.}~\bibnamefont {Kundu}},\
  }\bibfield  {title} {\bibinfo {title} {Dimension reduction of dynamical
  systems on networks with leading and nonleading eigenvectors of adjacency
  matrices},\ }\href {https://doi.org/10.1103/PhysRevResearch.4.023257}
  {\bibfield  {journal} {\bibinfo  {journal} {Physical Review Research}\
  }\textbf {\bibinfo {volume} {4}},\ \bibinfo {pages} {023257} (\bibinfo {year}
  {2022})}\BibitemShut {NoStop}%
\bibitem [{\citenamefont {Vegu{\'e}}\ \emph {et~al.}(2023)\citenamefont
  {Vegu{\'e}}, \citenamefont {Thibeault}, \citenamefont {Desrosiers},\ and\
  \citenamefont {Allard}}]{vegue2023}%
  \BibitemOpen
  \bibfield  {author} {\bibinfo {author} {\bibfnamefont {M.}~\bibnamefont
  {Vegu{\'e}}}, \bibinfo {author} {\bibfnamefont {V.}~\bibnamefont
  {Thibeault}}, \bibinfo {author} {\bibfnamefont {P.}~\bibnamefont
  {Desrosiers}},\ and\ \bibinfo {author} {\bibfnamefont {A.}~\bibnamefont
  {Allard}},\ }\bibfield  {title} {\bibinfo {title} {Dimension reduction of
  dynamics on modular and heterogeneous directed networks},\ }\href
  {https://doi.org/10.1093/pnasnexus/pgad150} {\bibfield  {journal} {\bibinfo
  {journal} {PNAS Nexus}\ }\textbf {\bibinfo {volume} {2}},\ \bibinfo {pages}
  {pgad150} (\bibinfo {year} {2023})}\BibitemShut {NoStop}%
\bibitem [{\citenamefont {Tu}\ \emph {et~al.}(2021)\citenamefont {Tu},
  \citenamefont {D'Odorico},\ and\ \citenamefont {Suweis}}]{tu2021}%
  \BibitemOpen
  \bibfield  {author} {\bibinfo {author} {\bibfnamefont {C.}~\bibnamefont
  {Tu}}, \bibinfo {author} {\bibfnamefont {P.}~\bibnamefont {D'Odorico}},\ and\
  \bibinfo {author} {\bibfnamefont {S.}~\bibnamefont {Suweis}},\ }\bibfield
  {title} {\bibinfo {title} {Dimensionality reduction of complex dynamical
  systems},\ }\href {https://doi.org/10.1016/j.isci.2020.101912} {\bibfield
  {journal} {\bibinfo  {journal} {iScience}\ }\textbf {\bibinfo {volume}
  {24}},\ \bibinfo {pages} {101912} (\bibinfo {year} {2021})}\BibitemShut
  {NoStop}%
\bibitem [{\citenamefont {Tu}\ \emph {et~al.}(2023{\natexlab{a}})\citenamefont
  {Tu}, \citenamefont {Wu}, \citenamefont {Luo}, \citenamefont {Jiang},\ and\
  \citenamefont {Pan}}]{tu2022discrete}%
  \BibitemOpen
  \bibfield  {author} {\bibinfo {author} {\bibfnamefont {C.}~\bibnamefont
  {Tu}}, \bibinfo {author} {\bibfnamefont {Y.}~\bibnamefont {Wu}}, \bibinfo
  {author} {\bibfnamefont {J.}~\bibnamefont {Luo}}, \bibinfo {author}
  {\bibfnamefont {Y.}~\bibnamefont {Jiang}},\ and\ \bibinfo {author}
  {\bibfnamefont {X.}~\bibnamefont {Pan}},\ }\bibfield  {title} {\bibinfo
  {title} {Dimensionality reduction in discrete-time dynamical systems},\
  }\href@noop {} {\bibfield  {journal} {\bibinfo  {journal} {Communications in
  Nonlinear Science and Numerical Simulation}\ }\textbf {\bibinfo {volume}
  {123}},\ \bibinfo {pages} {107268} (\bibinfo {year}
  {2023}{\natexlab{a}})}\BibitemShut {NoStop}%
\bibitem [{\citenamefont {Tu}\ \emph {et~al.}(2023{\natexlab{b}})\citenamefont
  {Tu}, \citenamefont {Luo}, \citenamefont {Fan},\ and\ \citenamefont
  {Pan}}]{tu2023stochastic}%
  \BibitemOpen
  \bibfield  {author} {\bibinfo {author} {\bibfnamefont {C.}~\bibnamefont
  {Tu}}, \bibinfo {author} {\bibfnamefont {J.}~\bibnamefont {Luo}}, \bibinfo
  {author} {\bibfnamefont {Y.}~\bibnamefont {Fan}},\ and\ \bibinfo {author}
  {\bibfnamefont {X.}~\bibnamefont {Pan}},\ }\bibfield  {title} {\bibinfo
  {title} {Dimensionality reduction in stochastic complex dynamical networks},\
  }\href@noop {} {\bibfield  {journal} {\bibinfo  {journal} {Chaos, Solitons \&
  Fractals}\ }\textbf {\bibinfo {volume} {175}},\ \bibinfo {pages} {114034}
  (\bibinfo {year} {2023}{\natexlab{b}})}\BibitemShut {NoStop}%
\bibitem [{\citenamefont {Thibeault}\ \emph {et~al.}(2020)\citenamefont
  {Thibeault}, \citenamefont {St-Onge}, \citenamefont {Dub{\'e}},\ and\
  \citenamefont {Desrosiers}}]{thibeault2020}%
  \BibitemOpen
  \bibfield  {author} {\bibinfo {author} {\bibfnamefont {V.}~\bibnamefont
  {Thibeault}}, \bibinfo {author} {\bibfnamefont {G.}~\bibnamefont {St-Onge}},
  \bibinfo {author} {\bibfnamefont {L.~J.}\ \bibnamefont {Dub{\'e}}},\ and\
  \bibinfo {author} {\bibfnamefont {P.}~\bibnamefont {Desrosiers}},\ }\bibfield
   {title} {\bibinfo {title} {Threefold way to the dimension reduction of
  dynamics on networks: An application to synchronization},\ }\href
  {https://doi.org/10.1103/PhysRevResearch.2.043215} {\bibfield  {journal}
  {\bibinfo  {journal} {Physical Review Research}\ }\textbf {\bibinfo {volume}
  {2}},\ \bibinfo {pages} {043215} (\bibinfo {year} {2020})}\BibitemShut
  {NoStop}%
\bibitem [{\citenamefont {Wu}\ \emph {et~al.}(2023)\citenamefont {Wu},
  \citenamefont {Duan},\ and\ \citenamefont {Xiao}}]{wu2023entropy}%
  \BibitemOpen
  \bibfield  {author} {\bibinfo {author} {\bibfnamefont {C.}~\bibnamefont
  {Wu}}, \bibinfo {author} {\bibfnamefont {D.}~\bibnamefont {Duan}},\ and\
  \bibinfo {author} {\bibfnamefont {R.}~\bibnamefont {Xiao}},\ }\bibfield
  {title} {\bibinfo {title} {A novel dimension reduction method with
  information entropy to evaluate network resilience},\ }\href
  {https://doi.org/10.1016/j.physa.2023.128727} {\bibfield  {journal} {\bibinfo
   {journal} {Physica A: Statistical Mechanics and its Applications}\ }\textbf
  {\bibinfo {volume} {620}},\ \bibinfo {pages} {128727} (\bibinfo {year}
  {2023})}\BibitemShut {NoStop}%
\bibitem [{\citenamefont {Mohammadi}\ \emph {et~al.}(2020)\citenamefont
  {Mohammadi}, \citenamefont {Amini},\ and\ \citenamefont
  {Arabnia}}]{mohammadi2019}%
  \BibitemOpen
  \bibfield  {author} {\bibinfo {author} {\bibfnamefont {F.~G.}\ \bibnamefont
  {Mohammadi}}, \bibinfo {author} {\bibfnamefont {M.~H.}\ \bibnamefont
  {Amini}},\ and\ \bibinfo {author} {\bibfnamefont {H.~R.}\ \bibnamefont
  {Arabnia}},\ }\bibfield  {title} {\bibinfo {title} {Applications of
  nature-inspired algorithms for dimension reduction: enabling efficient data
  analytics},\ }in\ \href@noop {} {\emph {\bibinfo {booktitle} {Optimization,
  learning, and control for interdependent complex networks}}}\ (\bibinfo
  {publisher} {Springer},\ \bibinfo {year} {2020})\ pp.\ \bibinfo {pages}
  {67--84}\BibitemShut {NoStop}%
\bibitem [{\citenamefont {Prasse}\ and\ \citenamefont
  {Mieghem}(2022)}]{prasse2022pnas}%
  \BibitemOpen
  \bibfield  {author} {\bibinfo {author} {\bibfnamefont {B.}~\bibnamefont
  {Prasse}}\ and\ \bibinfo {author} {\bibfnamefont {P.~V.}\ \bibnamefont
  {Mieghem}},\ }\bibfield  {title} {\bibinfo {title} {Predicting network
  dynamics without requiring the knowledge of the interaction graph},\ }\href
  {https://doi.org/10.1073/pnas.2205517119} {\bibfield  {journal} {\bibinfo
  {journal} {Proceedings of the National Academy of Sciences}\ }\textbf
  {\bibinfo {volume} {119}},\ \bibinfo {pages} {e2205517119} (\bibinfo {year}
  {2022})}\BibitemShut {NoStop}%
\bibitem [{\citenamefont {Prasse}\ \emph {et~al.}(2021)\citenamefont {Prasse},
  \citenamefont {Devriendt},\ and\ \citenamefont {Mieghem}}]{prasse2021chaos}%
  \BibitemOpen
  \bibfield  {author} {\bibinfo {author} {\bibfnamefont {B.}~\bibnamefont
  {Prasse}}, \bibinfo {author} {\bibfnamefont {K.}~\bibnamefont {Devriendt}},\
  and\ \bibinfo {author} {\bibfnamefont {P.~V.}\ \bibnamefont {Mieghem}},\
  }\bibfield  {title} {\bibinfo {title} {Clustering for epidemics on networks:
  A geometric approach},\ }\href {https://doi.org/10.1063/5.0048779} {\bibfield
   {journal} {\bibinfo  {journal} {Chaos}\ }\textbf {\bibinfo {volume} {31}},\
  \bibinfo {pages} {063115} (\bibinfo {year} {2021})}\BibitemShut {NoStop}%
\bibitem [{\citenamefont {Ding}\ \emph {et~al.}(2024)\citenamefont {Ding},
  \citenamefont {Huang}, \citenamefont {Magdon-Ismail},\ and\ \citenamefont
  {Gao}}]{ding2024arxiv}%
  \BibitemOpen
  \bibfield  {author} {\bibinfo {author} {\bibfnamefont {Y.}~\bibnamefont
  {Ding}}, \bibinfo {author} {\bibfnamefont {Z.}~\bibnamefont {Huang}},
  \bibinfo {author} {\bibfnamefont {M.}~\bibnamefont {Magdon-Ismail}},\ and\
  \bibinfo {author} {\bibfnamefont {J.}~\bibnamefont {Gao}},\ }\bibfield
  {title} {\bibinfo {title} {Predicting time series of networked dynamical
  systems without knowing topology},\ }\href@noop {} {\bibfield  {journal}
  {\bibinfo  {journal} {arXiv preprint arXiv:2412.18734}\ } (\bibinfo {year}
  {2024})}\BibitemShut {NoStop}%
\bibitem [{\citenamefont {Thibeault}\ \emph {et~al.}(2024)\citenamefont
  {Thibeault}, \citenamefont {Allard},\ and\ \citenamefont
  {Desrosiers}}]{thibeault2024natphys}%
  \BibitemOpen
  \bibfield  {author} {\bibinfo {author} {\bibfnamefont {V.}~\bibnamefont
  {Thibeault}}, \bibinfo {author} {\bibfnamefont {A.}~\bibnamefont {Allard}},\
  and\ \bibinfo {author} {\bibfnamefont {P.}~\bibnamefont {Desrosiers}},\
  }\bibfield  {title} {\bibinfo {title} {The low-rank hypothesis of complex
  systems},\ }\href {https://doi.org/10.1038/s41567-023-02344-7} {\bibfield
  {journal} {\bibinfo  {journal} {Nature Physics}\ }\textbf {\bibinfo {volume}
  {20}},\ \bibinfo {pages} {294} (\bibinfo {year} {2024})}\BibitemShut
  {NoStop}%
\bibitem [{\citenamefont {Kasz{\'a}s}\ and\ \citenamefont
  {Haller}(2025)}]{kaszas2025globalizing}%
  \BibitemOpen
  \bibfield  {author} {\bibinfo {author} {\bibfnamefont {B.}~\bibnamefont
  {Kasz{\'a}s}}\ and\ \bibinfo {author} {\bibfnamefont {G.}~\bibnamefont
  {Haller}},\ }\bibfield  {title} {\bibinfo {title} {Globalizing manifold-based
  reduced models for equations and data},\ }\href
  {https://doi.org/10.1038/s41467-025-61252-9} {\bibfield  {journal} {\bibinfo
  {journal} {Nature Communications}\ }\textbf {\bibinfo {volume} {16}},\
  \bibinfo {pages} {61252} (\bibinfo {year} {2025})}\BibitemShut {NoStop}%
\bibitem [{\citenamefont {Haller}\ and\ \citenamefont
  {Ponsioen}(2016)}]{haller2016}%
  \BibitemOpen
  \bibfield  {author} {\bibinfo {author} {\bibfnamefont {G.}~\bibnamefont
  {Haller}}\ and\ \bibinfo {author} {\bibfnamefont {S.}~\bibnamefont
  {Ponsioen}},\ }\bibfield  {title} {\bibinfo {title} {Nonlinear normal modes
  and spectral submanifolds: existence, uniqueness and use in model
  reduction},\ }\href {https://doi.org/10.1007/s11071-016-2974-z} {\bibfield
  {journal} {\bibinfo  {journal} {Nonlinear Dynamics}\ }\textbf {\bibinfo
  {volume} {86}},\ \bibinfo {pages} {1493} (\bibinfo {year}
  {2016})}\BibitemShut {NoStop}%
\bibitem [{\citenamefont {Jain}\ and\ \citenamefont {Haller}(2022)}]{jain2022}%
  \BibitemOpen
  \bibfield  {author} {\bibinfo {author} {\bibfnamefont {S.}~\bibnamefont
  {Jain}}\ and\ \bibinfo {author} {\bibfnamefont {G.}~\bibnamefont {Haller}},\
  }\bibfield  {title} {\bibinfo {title} {How to compute invariant manifolds and
  their reduced dynamics in high-dimensional finite element models},\ }\href
  {https://doi.org/10.1007/s11071-021-06957-4} {\bibfield  {journal} {\bibinfo
  {journal} {Nonlinear Dynamics}\ }\textbf {\bibinfo {volume} {107}},\ \bibinfo
  {pages} {1417} (\bibinfo {year} {2022})}\BibitemShut {NoStop}%
\bibitem [{\citenamefont {Breunung}\ and\ \citenamefont
  {Haller}(2018)}]{breunung2018}%
  \BibitemOpen
  \bibfield  {author} {\bibinfo {author} {\bibfnamefont {T.}~\bibnamefont
  {Breunung}}\ and\ \bibinfo {author} {\bibfnamefont {G.}~\bibnamefont
  {Haller}},\ }\bibfield  {title} {\bibinfo {title} {Explicit backbone curves
  from spectral submanifolds of forced-damped nonlinear mechanical systems},\
  }\href {https://doi.org/10.1098/rspa.2018.0083} {\bibfield  {journal}
  {\bibinfo  {journal} {Proceedings of the Royal Society A}\ }\textbf {\bibinfo
  {volume} {474}},\ \bibinfo {pages} {20180083} (\bibinfo {year}
  {2018})}\BibitemShut {NoStop}%
\bibitem [{\citenamefont {Ponsioen}\ \emph {et~al.}(2020)\citenamefont
  {Ponsioen}, \citenamefont {Renson}, \citenamefont {{van der Veen}},\ and\
  \citenamefont {Haller}}]{ponsioen2020}%
  \BibitemOpen
  \bibfield  {author} {\bibinfo {author} {\bibfnamefont {S.}~\bibnamefont
  {Ponsioen}}, \bibinfo {author} {\bibfnamefont {L.}~\bibnamefont {Renson}},
  \bibinfo {author} {\bibfnamefont {G.~W.~H.}\ \bibnamefont {{van der Veen}}},\
  and\ \bibinfo {author} {\bibfnamefont {G.}~\bibnamefont {Haller}},\
  }\bibfield  {title} {\bibinfo {title} {Model reduction to spectral
  submanifolds and forced response curve computation},\ }\href
  {https://doi.org/10.1016/j.jsv.2020.115640} {\bibfield  {journal} {\bibinfo
  {journal} {Journal of Sound and Vibration}\ }\textbf {\bibinfo {volume}
  {488}},\ \bibinfo {pages} {115640} (\bibinfo {year} {2020})}\BibitemShut
  {NoStop}%
\bibitem [{\citenamefont {Li}\ \emph {et~al.}(2022)\citenamefont {Li},
  \citenamefont {Jain},\ and\ \citenamefont {Haller}}]{li2022i}%
  \BibitemOpen
  \bibfield  {author} {\bibinfo {author} {\bibfnamefont {M.}~\bibnamefont
  {Li}}, \bibinfo {author} {\bibfnamefont {S.}~\bibnamefont {Jain}},\ and\
  \bibinfo {author} {\bibfnamefont {G.}~\bibnamefont {Haller}},\ }\bibfield
  {title} {\bibinfo {title} {Nonlinear analysis of forced mechanical systems
  with internal resonance using spectral submanifolds, part i: Periodic
  response and forced response curve},\ }\href
  {https://doi.org/10.1007/s11071-022-07714-x} {\bibfield  {journal} {\bibinfo
  {journal} {Nonlinear Dynamics}\ }\textbf {\bibinfo {volume} {110}},\ \bibinfo
  {pages} {1005} (\bibinfo {year} {2022})}\BibitemShut {NoStop}%
\bibitem [{\citenamefont {Li}\ and\ \citenamefont {Haller}(2022)}]{li2022ii}%
  \BibitemOpen
  \bibfield  {author} {\bibinfo {author} {\bibfnamefont {M.}~\bibnamefont
  {Li}}\ and\ \bibinfo {author} {\bibfnamefont {G.}~\bibnamefont {Haller}},\
  }\bibfield  {title} {\bibinfo {title} {Nonlinear analysis of forced
  mechanical systems with internal resonance using spectral submanifolds, part
  ii: Bifurcation and quasi-periodic response},\ }\href
  {https://doi.org/10.1007/s11071-022-07715-w} {\bibfield  {journal} {\bibinfo
  {journal} {Nonlinear Dynamics}\ }\textbf {\bibinfo {volume} {110}},\ \bibinfo
  {pages} {1045} (\bibinfo {year} {2022})}\BibitemShut {NoStop}%
\bibitem [{\citenamefont {Cenedese}\ \emph {et~al.}(2022)\citenamefont
  {Cenedese}, \citenamefont {Ax{\aa}s}, \citenamefont {B{\"a}uerlein},
  \citenamefont {Avila},\ and\ \citenamefont {Haller}}]{cenedese2022nc}%
  \BibitemOpen
  \bibfield  {author} {\bibinfo {author} {\bibfnamefont {M.}~\bibnamefont
  {Cenedese}}, \bibinfo {author} {\bibfnamefont {J.}~\bibnamefont {Ax{\aa}s}},
  \bibinfo {author} {\bibfnamefont {B.}~\bibnamefont {B{\"a}uerlein}}, \bibinfo
  {author} {\bibfnamefont {K.}~\bibnamefont {Avila}},\ and\ \bibinfo {author}
  {\bibfnamefont {G.}~\bibnamefont {Haller}},\ }\bibfield  {title} {\bibinfo
  {title} {Data-driven modeling and prediction of non-linearizable dynamics via
  spectral submanifolds},\ }\href {https://doi.org/10.1038/s41467-022-28518-y}
  {\bibfield  {journal} {\bibinfo  {journal} {Nature Communications}\ }\textbf
  {\bibinfo {volume} {13}},\ \bibinfo {pages} {872} (\bibinfo {year}
  {2022})}\BibitemShut {NoStop}%
\bibitem [{\citenamefont {Liu}\ \emph {et~al.}(2024)\citenamefont {Liu},
  \citenamefont {Ax{\aa}s},\ and\ \citenamefont {Haller}}]{liu2024chaos}%
  \BibitemOpen
  \bibfield  {author} {\bibinfo {author} {\bibfnamefont {A.}~\bibnamefont
  {Liu}}, \bibinfo {author} {\bibfnamefont {J.}~\bibnamefont {Ax{\aa}s}},\ and\
  \bibinfo {author} {\bibfnamefont {G.}~\bibnamefont {Haller}},\ }\bibfield
  {title} {\bibinfo {title} {Data-driven modeling and forecasting of chaotic
  dynamics on inertial manifolds constructed as spectral submanifolds},\ }\href
  {https://doi.org/10.1063/5.0183754} {\bibfield  {journal} {\bibinfo
  {journal} {Chaos}\ }\textbf {\bibinfo {volume} {34}},\ \bibinfo {pages}
  {033140} (\bibinfo {year} {2024})}\BibitemShut {NoStop}%
\bibitem [{\citenamefont {Bhaskaran}\ \emph {et~al.}(2025)\citenamefont
  {Bhaskaran}, \citenamefont {Jain},\ and\ \citenamefont {Li}}]{sup_ssm}%
  \BibitemOpen
  \bibfield  {author} {\bibinfo {author} {\bibfnamefont {K.}~\bibnamefont
  {Bhaskaran}}, \bibinfo {author} {\bibfnamefont {S.}~\bibnamefont {Jain}},\
  and\ \bibinfo {author} {\bibfnamefont {M.}~\bibnamefont {Li}},\ }\href@noop
  {} {\bibinfo {title} {Supplemental material for ``nonlinear spectral model
  reduction of networked systems''}} (\bibinfo {year} {2025}),\ \bibinfo {note}
  {contains extended derivations, implementation details, and additional
  figures.}\BibitemShut {Stop}%
\bibitem [{\citenamefont {Jain}\ \emph {et~al.}(2024)\citenamefont {Jain},
  \citenamefont {Li}, \citenamefont {Thurnher},\ and\ \citenamefont
  {Haller}}]{ssmtool26}%
  \BibitemOpen
  \bibfield  {author} {\bibinfo {author} {\bibfnamefont {S.}~\bibnamefont
  {Jain}}, \bibinfo {author} {\bibfnamefont {M.}~\bibnamefont {Li}}, \bibinfo
  {author} {\bibfnamefont {T.}~\bibnamefont {Thurnher}},\ and\ \bibinfo
  {author} {\bibfnamefont {G.}~\bibnamefont {Haller}},\ }\href
  {https://doi.org/10.5281/zenodo.13909795} {\bibinfo {title} {Ssmtool 2.6:
  Computation of invariant manifolds in high-dimensional mechanics problems
  (v2.6)}},\ \bibinfo {howpublished} {Zenodo} (\bibinfo {year} {2024}),\
  \bibinfo {note} {software}\BibitemShut {NoStop}%
\bibitem [{\citenamefont {Ponsioen}\ \emph {et~al.}(2019)\citenamefont
  {Ponsioen}, \citenamefont {Pedergnana},\ and\ \citenamefont
  {Haller}}]{Ponsioen2019NonlinDyn}%
  \BibitemOpen
  \bibfield  {author} {\bibinfo {author} {\bibfnamefont {S.}~\bibnamefont
  {Ponsioen}}, \bibinfo {author} {\bibfnamefont {T.}~\bibnamefont
  {Pedergnana}},\ and\ \bibinfo {author} {\bibfnamefont {G.}~\bibnamefont
  {Haller}},\ }\bibfield  {title} {\bibinfo {title} {Analytic prediction of
  isolated forced response curves from spectral submanifolds},\ }\href
  {https://doi.org/10.1007/s11071-019-05023-4} {\bibfield  {journal} {\bibinfo
  {journal} {Nonlinear Dynamics}\ }\textbf {\bibinfo {volume} {98}},\ \bibinfo
  {pages} {2755} (\bibinfo {year} {2019})}\BibitemShut {NoStop}%
\bibitem [{\citenamefont {{SocioPatterns Collaboration}}(2010)}]{sp:hospital}%
  \BibitemOpen
  \bibfield  {author} {\bibinfo {author} {\bibnamefont {{SocioPatterns
  Collaboration}}},\ }\href@noop {} {\bibinfo {title} {Hospital ward contact
  network, lyon, 2010}},\ \bibinfo {howpublished}
  {\url{https://www.sociopatterns.org/datasets/hospital-ward-dynamic-contact-network/}}
  (\bibinfo {year} {2010}),\ \bibinfo {note} {accessed 2025}\BibitemShut
  {NoStop}%
\bibitem [{\citenamefont {{SocioPatterns Collaboration}}(2013)}]{sp:workplace}%
  \BibitemOpen
  \bibfield  {author} {\bibinfo {author} {\bibnamefont {{SocioPatterns
  Collaboration}}},\ }\href@noop {} {\bibinfo {title} {Office building contact
  network, france, 2013}},\ \bibinfo {howpublished}
  {\url{https://www.sociopatterns.org/datasets/contacts-in-a-workplace/}}
  (\bibinfo {year} {2013}),\ \bibinfo {note} {accessed 2025}\BibitemShut
  {NoStop}%
\bibitem [{\citenamefont {{SocioPatterns Collaboration}}(2014)}]{sp:rural}%
  \BibitemOpen
  \bibfield  {author} {\bibinfo {author} {\bibnamefont {{SocioPatterns
  Collaboration}}},\ }\href@noop {} {\bibinfo {title} {Rural malawi contact
  network}},\ \bibinfo {howpublished}
  {\url{https://www.sociopatterns.org/datasets/rural-malawi/}} (\bibinfo {year}
  {2014}),\ \bibinfo {note} {accessed 2025}\BibitemShut {NoStop}%
\bibitem [{\citenamefont {Iacopini}\ \emph {et~al.}(2019)\citenamefont
  {Iacopini}, \citenamefont {Petri}, \citenamefont {Barrat},\ and\
  \citenamefont {Latora}}]{iacopini2019simplicial}%
  \BibitemOpen
  \bibfield  {author} {\bibinfo {author} {\bibfnamefont {I.}~\bibnamefont
  {Iacopini}}, \bibinfo {author} {\bibfnamefont {G.}~\bibnamefont {Petri}},
  \bibinfo {author} {\bibfnamefont {A.}~\bibnamefont {Barrat}},\ and\ \bibinfo
  {author} {\bibfnamefont {V.}~\bibnamefont {Latora}},\ }\bibfield  {title}
  {\bibinfo {title} {Simplicial models of social contagion},\ }\href
  {https://doi.org/10.1038/s41467-019-10431-6} {\bibfield  {journal} {\bibinfo
  {journal} {Nature Communications}\ }\textbf {\bibinfo {volume} {10}},\
  \bibinfo {pages} {2485} (\bibinfo {year} {2019})}\BibitemShut {NoStop}%
\bibitem [{\citenamefont {Battiston}\ \emph {et~al.}(2020)\citenamefont
  {Battiston}, \citenamefont {Cencetti}, \citenamefont {Iacopini},
  \citenamefont {Latora}, \citenamefont {Lucas}, \citenamefont {Patania},
  \citenamefont {Young},\ and\ \citenamefont {Petri}}]{battiston2020beyond}%
  \BibitemOpen
  \bibfield  {author} {\bibinfo {author} {\bibfnamefont {F.}~\bibnamefont
  {Battiston}}, \bibinfo {author} {\bibfnamefont {G.}~\bibnamefont {Cencetti}},
  \bibinfo {author} {\bibfnamefont {I.}~\bibnamefont {Iacopini}}, \bibinfo
  {author} {\bibfnamefont {V.}~\bibnamefont {Latora}}, \bibinfo {author}
  {\bibfnamefont {M.}~\bibnamefont {Lucas}}, \bibinfo {author} {\bibfnamefont
  {A.}~\bibnamefont {Patania}}, \bibinfo {author} {\bibfnamefont {J.-G.}\
  \bibnamefont {Young}},\ and\ \bibinfo {author} {\bibfnamefont
  {G.}~\bibnamefont {Petri}},\ }\bibfield  {title} {\bibinfo {title} {Networks
  beyond pairwise interactions: Structure and dynamics},\ }\href
  {https://doi.org/10.1016/j.physrep.2020.05.004} {\bibfield  {journal}
  {\bibinfo  {journal} {Physics Reports}\ }\textbf {\bibinfo {volume} {874}},\
  \bibinfo {pages} {1} (\bibinfo {year} {2020})}\BibitemShut {NoStop}%
\bibitem [{\citenamefont {Battiston}\ \emph {et~al.}(2021)\citenamefont
  {Battiston}, \citenamefont {Amico}, \citenamefont {Barrat}, \citenamefont
  {Bianconi}, \citenamefont {de~Arruda}, \citenamefont {Franceschiello},
  \citenamefont {Iacopini}, \citenamefont {K{\'e}fi}, \citenamefont {Latora},
  \citenamefont {Moreno}, \citenamefont {Murray}, \citenamefont {Peixoto},
  \citenamefont {Vaccarino},\ and\ \citenamefont
  {Petri}}]{battiston2021higher}%
  \BibitemOpen
  \bibfield  {author} {\bibinfo {author} {\bibfnamefont {F.}~\bibnamefont
  {Battiston}}, \bibinfo {author} {\bibfnamefont {E.}~\bibnamefont {Amico}},
  \bibinfo {author} {\bibfnamefont {A.}~\bibnamefont {Barrat}}, \bibinfo
  {author} {\bibfnamefont {G.}~\bibnamefont {Bianconi}}, \bibinfo {author}
  {\bibfnamefont {G.~F.}\ \bibnamefont {de~Arruda}}, \bibinfo {author}
  {\bibfnamefont {B.}~\bibnamefont {Franceschiello}}, \bibinfo {author}
  {\bibfnamefont {I.}~\bibnamefont {Iacopini}}, \bibinfo {author}
  {\bibfnamefont {S.}~\bibnamefont {K{\'e}fi}}, \bibinfo {author}
  {\bibfnamefont {V.}~\bibnamefont {Latora}}, \bibinfo {author} {\bibfnamefont
  {Y.}~\bibnamefont {Moreno}}, \bibinfo {author} {\bibfnamefont {M.~M.}\
  \bibnamefont {Murray}}, \bibinfo {author} {\bibfnamefont {T.~P.}\
  \bibnamefont {Peixoto}}, \bibinfo {author} {\bibfnamefont {F.}~\bibnamefont
  {Vaccarino}},\ and\ \bibinfo {author} {\bibfnamefont {G.}~\bibnamefont
  {Petri}},\ }\bibfield  {title} {\bibinfo {title} {The physics of higher-order
  interactions in complex systems},\ }\href
  {https://doi.org/10.1038/s41567-021-01371-4} {\bibfield  {journal} {\bibinfo
  {journal} {Nature Physics}\ }\textbf {\bibinfo {volume} {17}},\ \bibinfo
  {pages} {1093} (\bibinfo {year} {2021})}\BibitemShut {NoStop}%
\bibitem [{\citenamefont {Szaksz}\ \emph {et~al.}(2025)\citenamefont {Szaksz},
  \citenamefont {Orosz},\ and\ \citenamefont {Stepan}}]{szaksz2025spectral}%
  \BibitemOpen
  \bibfield  {author} {\bibinfo {author} {\bibfnamefont {B.}~\bibnamefont
  {Szaksz}}, \bibinfo {author} {\bibfnamefont {G.}~\bibnamefont {Orosz}},\ and\
  \bibinfo {author} {\bibfnamefont {G.}~\bibnamefont {Stepan}},\ }\bibfield
  {title} {\bibinfo {title} {Spectral submanifolds in time delay systems},\
  }\href@noop {} {\bibfield  {journal} {\bibinfo  {journal} {Nonlinear
  Dynamics}\ }\textbf {\bibinfo {volume} {113}},\ \bibinfo {pages} {14265}
  (\bibinfo {year} {2025})}\BibitemShut {NoStop}%
\bibitem [{\citenamefont {Wall}(1948)}]{wall1948analytic}%
  \BibitemOpen
  \bibfield  {author} {\bibinfo {author} {\bibfnamefont {H.~S.}\ \bibnamefont
  {Wall}},\ }\href@noop {} {\emph {\bibinfo {title} {Analytic Theory of
  Continued Fractions}}}\ (\bibinfo  {publisher} {D. Van Nostrand},\ \bibinfo
  {address} {New York},\ \bibinfo {year} {1948})\BibitemShut {NoStop}%
\bibitem [{\citenamefont {Baker}\ and\ \citenamefont
  {Graves-Morris}(1996)}]{baker1996pade}%
  \BibitemOpen
  \bibfield  {author} {\bibinfo {author} {\bibfnamefont {G.~A.}\ \bibnamefont
  {Baker}}\ and\ \bibinfo {author} {\bibfnamefont {P.}~\bibnamefont
  {Graves-Morris}},\ }\href@noop {} {\emph {\bibinfo {title} {Pad{\'e}
  Approximants}}},\ \bibinfo {edition} {2nd}\ ed.\ (\bibinfo  {publisher}
  {Cambridge University Press},\ \bibinfo {address} {Cambridge, UK},\ \bibinfo
  {year} {1996})\BibitemShut {NoStop}%
\bibitem [{\citenamefont {Cuyt}\ \emph {et~al.}(2008)\citenamefont {Cuyt},
  \citenamefont {Petersen}, \citenamefont {Verdonk}, \citenamefont
  {Waadeland},\ and\ \citenamefont {Jones}}]{cuyt2008handbook}%
  \BibitemOpen
  \bibfield  {author} {\bibinfo {author} {\bibfnamefont {A.}~\bibnamefont
  {Cuyt}}, \bibinfo {author} {\bibfnamefont {V.}~\bibnamefont {Petersen}},
  \bibinfo {author} {\bibfnamefont {B.}~\bibnamefont {Verdonk}}, \bibinfo
  {author} {\bibfnamefont {H.}~\bibnamefont {Waadeland}},\ and\ \bibinfo
  {author} {\bibfnamefont {W.~B.}\ \bibnamefont {Jones}},\ }\href@noop {}
  {\emph {\bibinfo {title} {Handbook of Continued Fractions for Special
  Functions}}}\ (\bibinfo  {publisher} {Springer},\ \bibinfo {address} {New
  York},\ \bibinfo {year} {2008})\BibitemShut {NoStop}%
\bibitem [{\citenamefont {de~Montessus~de Ballore}(1902)}]{demontessus1902}%
  \BibitemOpen
  \bibfield  {author} {\bibinfo {author} {\bibfnamefont {R.}~\bibnamefont
  {de~Montessus~de Ballore}},\ }\bibfield  {title} {\bibinfo {title} {Sur les
  fractions continues alg{\'e}briques},\ }\href@noop {} {\bibfield  {journal}
  {\bibinfo  {journal} {Bulletin de la Soci{\'e}t{\'e} Math{\'e}matique de
  France}\ }\textbf {\bibinfo {volume} {30}},\ \bibinfo {pages} {28} (\bibinfo
  {year} {1902})}\BibitemShut {NoStop}%
\bibitem [{\citenamefont {Brezinski}(1991)}]{brezinski1991pade}%
  \BibitemOpen
  \bibfield  {author} {\bibinfo {author} {\bibfnamefont {C.}~\bibnamefont
  {Brezinski}},\ }\href@noop {} {\emph {\bibinfo {title} {Pad{\'e}-Type
  Approximation and General Orthogonal Polynomials}}}\ (\bibinfo  {publisher}
  {Birkh{\"a}user},\ \bibinfo {address} {Basel},\ \bibinfo {year}
  {1991})\BibitemShut {NoStop}%
\bibitem [{\citenamefont {Graves-Morris}(1979)}]{gravesmorris1979vector}%
  \BibitemOpen
  \bibfield  {author} {\bibinfo {author} {\bibfnamefont {P.~R.}\ \bibnamefont
  {Graves-Morris}},\ }\href@noop {} {\emph {\bibinfo {title} {Vector and Matrix
  Pad{\'e} Approximations}}}\ (\bibinfo  {publisher} {Clarendon Press},\
  \bibinfo {address} {Oxford},\ \bibinfo {year} {1979})\BibitemShut {NoStop}%
\end{thebibliography}%

\end{document}